%% file: EXO-16-048_temp.tex
\pdfoutput=1

\documentclass[11pt,twoside,a4paper,cmspaper,final,collab]{cms-tdr}
\def\svnVersion{462350P}\def\cmsCernNoTag{CERN-EP-2017-294}\def\cmsCernDate{\today}\def\cmsMessage{Published in Physical Review D as \href{http://dx.doi.org/10.1103/PhysRevD.97.092005}{\doi{10.1103/PhysRevD.97.092005.}}}

\begin{document}\cmsNoteHeader{EXO-16-048}

\hyphenation{had-ron-i-za-tion}
\hyphenation{cal-or-i-me-ter}
\hyphenation{de-vices}
\RCS$HeadURL: svn+ssh://svn.cern.ch/reps/tdr2/papers/EXO-16-048/trunk/EXO-16-048.tex $
\RCS$Id: EXO-16-048.tex 453153 2018-03-28 15:15:15Z alverson $

\ifthenelse{\boolean{cms@external}}{\newcommand{\PV}{\ensuremath{V}\xspace}}{\newcommand{\PV}{\ensuremath{\mathrm{V}}\xspace}}
\newcommand{\Zmm}{\ensuremath{\PZ\to\PGm^+\PGm^-}}
\newcommand{\Zee}{\ensuremath{\PZ\to \Pe^+\Pe^-}}
\newcommand{\Zll}{\ensuremath{\PZ\to\ell\ell}}
\newcommand{\Zvv}{\ensuremath{\PZ\to\PGn\PGn}}
\newcommand{\Wlv}{\ensuremath{\PW\to \ell\PGn}}
\newcommand{\Wmn}{\ensuremath{\PW\to \PGm\PGn}}
\newcommand{\Wen}{\ensuremath{\PW\to \Pe\PGn}}
\newcommand{\Zmmjets}{{$\PZ(\PGm\PGm){+}\textrm{jets}$}\xspace}
\newcommand{\Zeejets}{{$\PZ(\Pe\Pe){+}\textrm{jets}$}\xspace}
\newcommand{\Zlljets}{{$\PZ(\ell\ell){+}\textrm{jets}$}\xspace}
\newcommand{\Zjets}{{$\PZ{+}\textrm{jets}$}\xspace}
\newcommand{\Wjets}{{$\PW{+}\textrm{jets}$}\xspace}
\newcommand{\Zvvjets}{{$\PZ(\PGn\PGn){+}\textrm{jets}$}\xspace}
\newcommand{\Wlvjets}{{$\PW(\ell\PGn){+}\textrm{jets}$}\xspace}
\newcommand{\Wmvjets}{{$\PW(\PGm\PGn){+}\textrm{jets}$}\xspace}
\newcommand{\Wmnjets}{{$\PW(\PGm\PGn){+}\textrm{jets}$}\xspace}
\newcommand{\Wevjets}{{$\PW(\Pe\PGn){+}\textrm{jets}$}\xspace}
\newcommand{\phojets}{{$\gamma{+}\textrm{jets}$}\xspace}
\newcommand{\brhiggs}{\ensuremath{0.62}}
\newcommand{\higgsbr}{\ensuremath{0.62}}
\newcommand{\higgsbrobs}{\ensuremath{0.53}}
\newcommand{\cchiggsbr}{\ensuremath{0.92}}
\newcommand{\Et}{\ensuremath{E_\mathrm{T}}}
\newcommand{\mt}{\ensuremath{M_\mathrm{T}}}
\newcommand{\met}{\ensuremath{\Et^{\mathrm{miss}}}}
\newcommand{\Ht}{\ensuremath{H_\mathrm{T}}}
\newcommand{\sieie}{\ensuremath{\sigma_{i\eta i\eta}} }
\newcommand{\vmet}{\ensuremath{\vec{E}_\mathrm{T}}^{\text{miss}}\xspace}
\newcommand\numberthis{\addtocounter{equation}{1}\tag{\theequation}}
\newcommand{\mettrig}{\ensuremath{E_{\mathrm{T, trig}}^{\mathrm{miss}}}}
\newcommand{\ptmisstrig}{\ensuremath{p_{\mathrm{T, trig}}^{\mathrm{miss}}}}
\newcommand{\mhttrig}{\ensuremath{H_{\mathrm{T, trig}}^{\mathrm{miss}}}}
\newcommand{\brhinv}{\ensuremath{\mathcal{B}(\mathrm{H}\rightarrow \mathrm{inv.})}}
\newcommand{\ptvecjet}{\ensuremath{{\vec p}_{\mathrm{T}}^{\kern1pt\text{jet}}}\xspace}
\newcommand{\pthat}{\ensuremath{\hat{p}_{\mathrm{T}}}\xspace}

\newcommand{\x}{\ensuremath{\phantom{0}}}
\newcolumntype{.}{D{,}{\,\pm\,}{3}}
\ifthenelse{\boolean{cms@external}}{\providecommand{\CL}{C.L.\xspace}}{\providecommand{\CL}{CL\xspace}}
\ifthenelse{\boolean{cms@external}}{\providecommand{\NA}{\ensuremath{\cdots}\xspace}}{\providecommand{\NA}{\ensuremath{\text{---}}\xspace}}
\newlength\cmsFigWidth
\ifthenelse{\boolean{cms@external}}{\setlength\cmsFigWidth{0.85\columnwidth}}{\setlength\cmsFigWidth{0.45\textwidth}}
\ifthenelse{\boolean{cms@external}}{\providecommand{\cmsLeft}{top\xspace}}{\providecommand{\cmsLeft}{left\xspace}}
\ifthenelse{\boolean{cms@external}}{\providecommand{\cmsRight}{bottom\xspace}}{\providecommand{\cmsRight}{right\xspace}}

\cmsNoteHeader{EXO-16-048}

\title{Search for new physics in final states with an energetic jet or a hadronically decaying \texorpdfstring{$\PW$}{W} or \texorpdfstring{$\PZ$}{Z} boson and transverse momentum imbalance at \texorpdfstring{$\sqrt{s} = 13\TeV$}{sqrt(s) = 13 TeV}}

\date{\today}

\abstract{
A search for new physics using events containing an imbalance in transverse momentum and one or more energetic jets arising from initial-state radiation or the hadronic decay of $\PW$ or $\PZ$ bosons is presented. A data sample of proton-proton collisions at $\sqrt{s}=13\TeV$, collected with the CMS detector at the LHC and corresponding to an integrated luminosity of 35.9\fbinv, is used. The observed data are found to be in agreement with the expectation from standard model processes. The results are interpreted as limits on the dark matter production cross section in simplified models with vector, axial-vector, scalar, and pseudoscalar mediators. Interpretations in the context of fermion portal and nonthermal dark matter models are also provided. In addition, the results are interpreted in terms of invisible decays of the Higgs boson and set stringent limits on the fundamental Planck scale in the Arkani-Hamed, Dimopoulos, and Dvali model with large extra spatial dimensions.}

\hypersetup{%
pdfauthor={CMS Collaboration},%
pdftitle={Search for new physics in final states with an energetic jet or a hadronically decaying W or Z boson and transverse momentum imbalance at sqrt(s) = 13 TeV},%
pdfsubject={CMS},%
pdfkeywords={CMS, physics, exotica, dark matter, monojet}}

\maketitle

\section{Introduction}

Several astrophysical observations \cite{Bertone:2004pz,Feng:2010gw,Porter:2011nv} provide compelling evidence for the existence of
dark matter (DM), a type of matter not accounted for in the standard model (SM).
To date only gravitational interactions of DM have been observed and it remains unknown
if DM has a particle origin and could interact with ordinary matter via SM processes. However,
many theoretical models have been proposed in which DM and SM
particles interact with sufficient strength that DM may be directly produced with
observable rates in high energy collisions at the CERN LHC.
While the DM particles would remain undetected, they may recoil
with large transverse momentum (\pt) against other detectable particles
resulting in an overall visible \pt imbalance in a collision event. This type of event topology is
rarely produced in SM processes and therefore enables a highly sensitive search for DM. Similar event topologies
are predicted by other extensions of the SM, such as the Arkani-Hamed, Dimopoulos, and Dvali (ADD)
model ~\cite{bib:ADD1,ADDPRD,Antoniadis,ADDGiudice,ADDPeskin} of large extra spatial dimensions (EDs).

This paper describes a search for new physics resulting in final
states with one or more energetic jets and an imbalance in \pt
due to undetected particles. The jets are the result of the fragmentation and hadronization of
quarks or gluons, which may be produced directly in the hard scattering process as initial-state radiation or
as the decay products of a vector boson $\PV$ ($\PW$ or $\PZ$).
These final states are commonly referred to as `monojet' and `mono-$\PV$'.
Several searches have been performed at the LHC using the
monojet and mono-$\PV$ channels~\cite{Aad:2013oja,Khachatryan:2014rra,Aad:2015zva,Khachatryan:2016mdm,Aaboud:2016tnv,paper-exo-037,Aaboud:2017phn}.
This analysis makes use of a data sample of proton-proton
(pp) collisions at $\sqrt{s}=13\TeV$ collected with the
CMS detector at the LHC, corresponding to an integrated
luminosity of 35.9\fbinv. This sample is approximately
three times larger than the one used in Ref.~\cite{paper-exo-037}.
The analysis strategy is similar to that of previous CMS searches, and simultaneously
employs event categories to target both the monojet and mono-$\PV$ final states.
In an improvement compared to previous searches, in this paper revised theoretical predictions and uncertainties
for \phojets, \Zjets, and \Wjets processes based on recommendations of Ref.~\cite{DMTheory} are used.
In addition to interpretations in the context of simplified DM models~\cite{Busoni:2013lha,Buchmueller:2013dya,Buchmueller:2014yoa},
in this paper the results are further studied in the context of the fermion portal (FP) dark matter model~\cite{fermionportal_ref1},
the light nonthermal DM model~\cite{bib:NonThDM_paper,Allahverdi:2013mza},
and the ADD model.

In many simplified DM models, DM particles are assumed to be Dirac fermions that interact with SM particles through
a spin-1 or spin-0  mediator~\cite{Alwall:2008ag,Goodman:2011jq,Alves:2011wf,An:2012va,An:2012ue,DiFranzo:2013vra,Buchmueller:2013dya,fermionportal_ref1,Bai:2014osa,An:2013xka,Abdallah:2014hon,Malik:2014ggr,Harris:2014hga,Buckley:2014fba,Haisch:2015ioa,Harris:2015kda,Carpenter:2012rg,Bell:2012rg}.
These interactions are classified into four different types, depending on whether
the mediator is a vector, axial-vector, scalar, or pseudoscalar particle.
The spin-0 mediators are assumed to couple to the SM particles via Yukawa couplings.
The SM Higgs boson is a specific example of a scalar mediator that may couple to the DM particles.
Combined results of the direct searches for invisible Higgs bosons have been presented by both the ATLAS and CMS
Collaborations, which respectively obtain observed upper limits of 0.25 and 0.24 on the Higgs boson invisible
branching fraction, $\brhinv$, at 95\% confidence level (\CL)~\cite{Aad:2015pla,Khachatryan:2016whc}.

\begin{figure*}[htb]
\begin{center}
\includegraphics[scale=0.28]{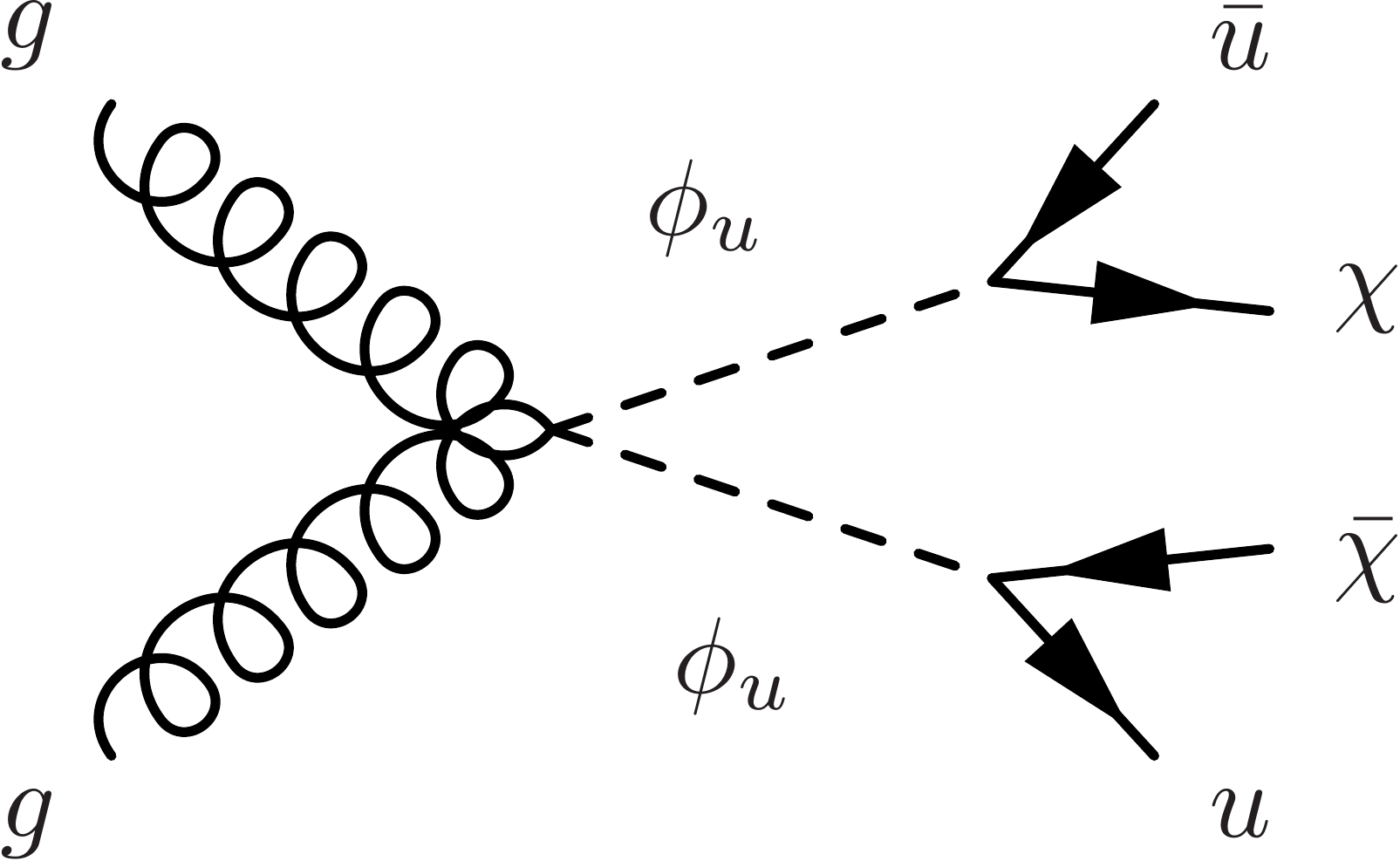}\hspace{0.6cm}
\includegraphics[scale=0.30]{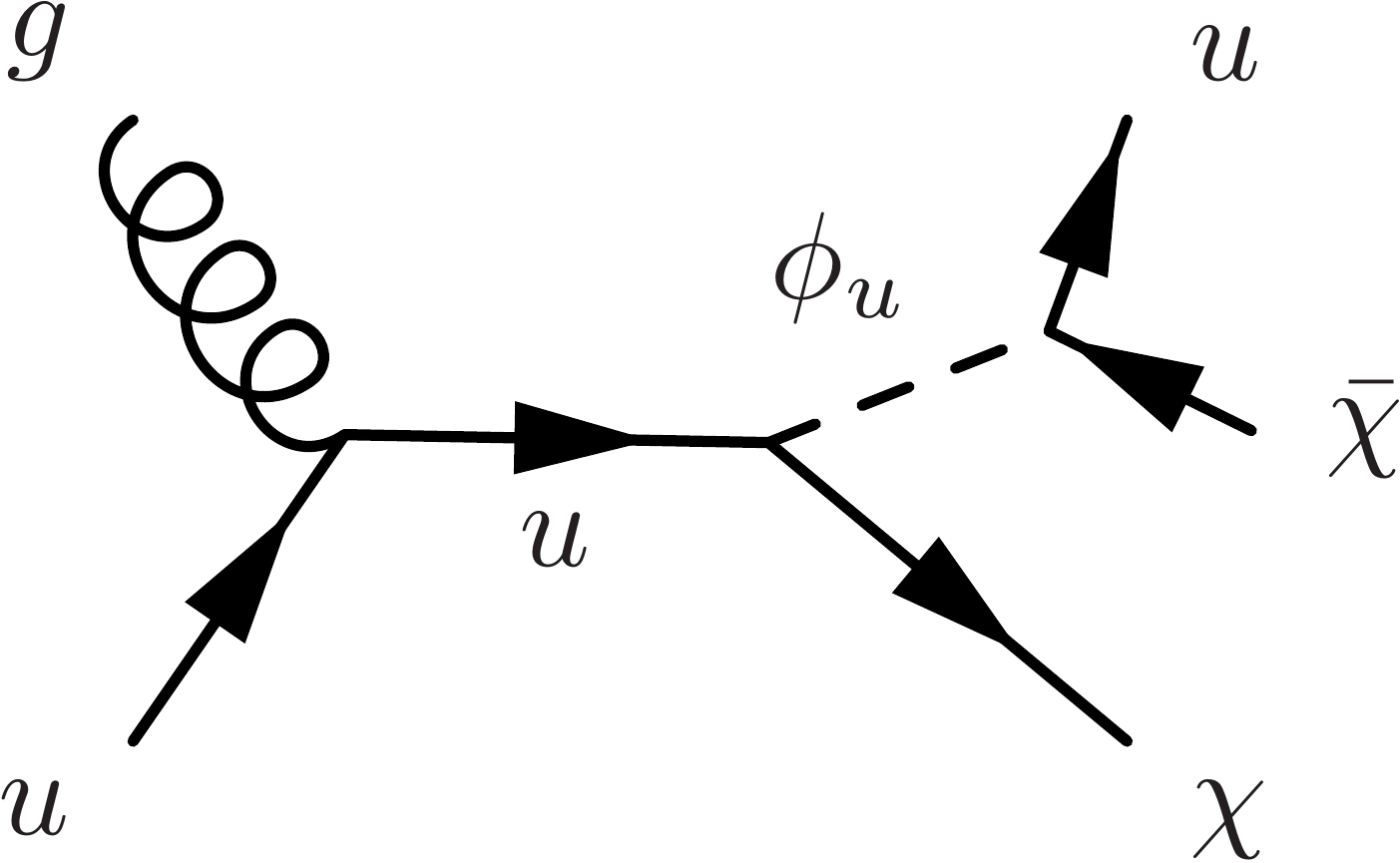}\hspace{0.6cm}
\includegraphics[scale=0.28]{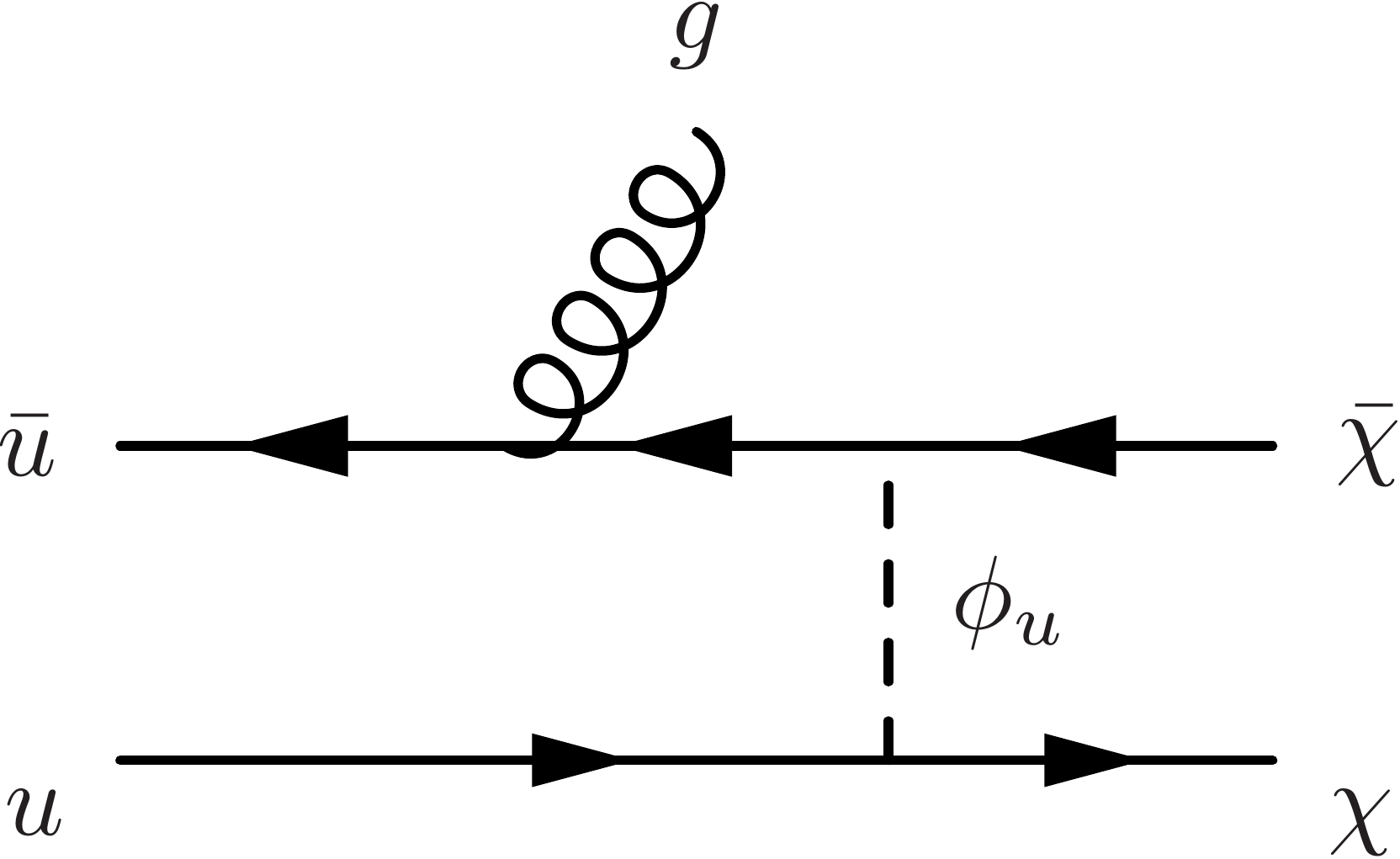}
\caption{Examples of Feynman diagrams of the main production mechanisms at the LHC of DM particles in association with a quark or gluon
in the fermion portal model providing multijet (left) and monojet (middle, right) signatures.}
\label{fig:fermionportal_FeynDiag}
\end{center}
\end{figure*}

In the FP dark matter model~\cite{fermionportal_ref1}, the
DM particle, assumed to be either a Dirac or Majorana fermion,
couples to a color-triplet scalar mediator ($\phi_{\rm{u}}$) and an SM fermion.
In the investigated model, the DM candidate is assumed to couple only to up-type
quarks, with a coupling strength parameter $\lambda_{\rm{u}}=1$.
In this model, the mediators couple to quarks and the DM candidate, and may be
singly produced in association with a DM particle.
This associated production yields
a monojet signature, while pair production of mediators can be
observed in multijet final states with significant \pt imbalance,
as shown in Fig.~\ref{fig:fermionportal_FeynDiag}.

\begin{figure}[htb]
\begin{center}
\includegraphics[scale=0.3]{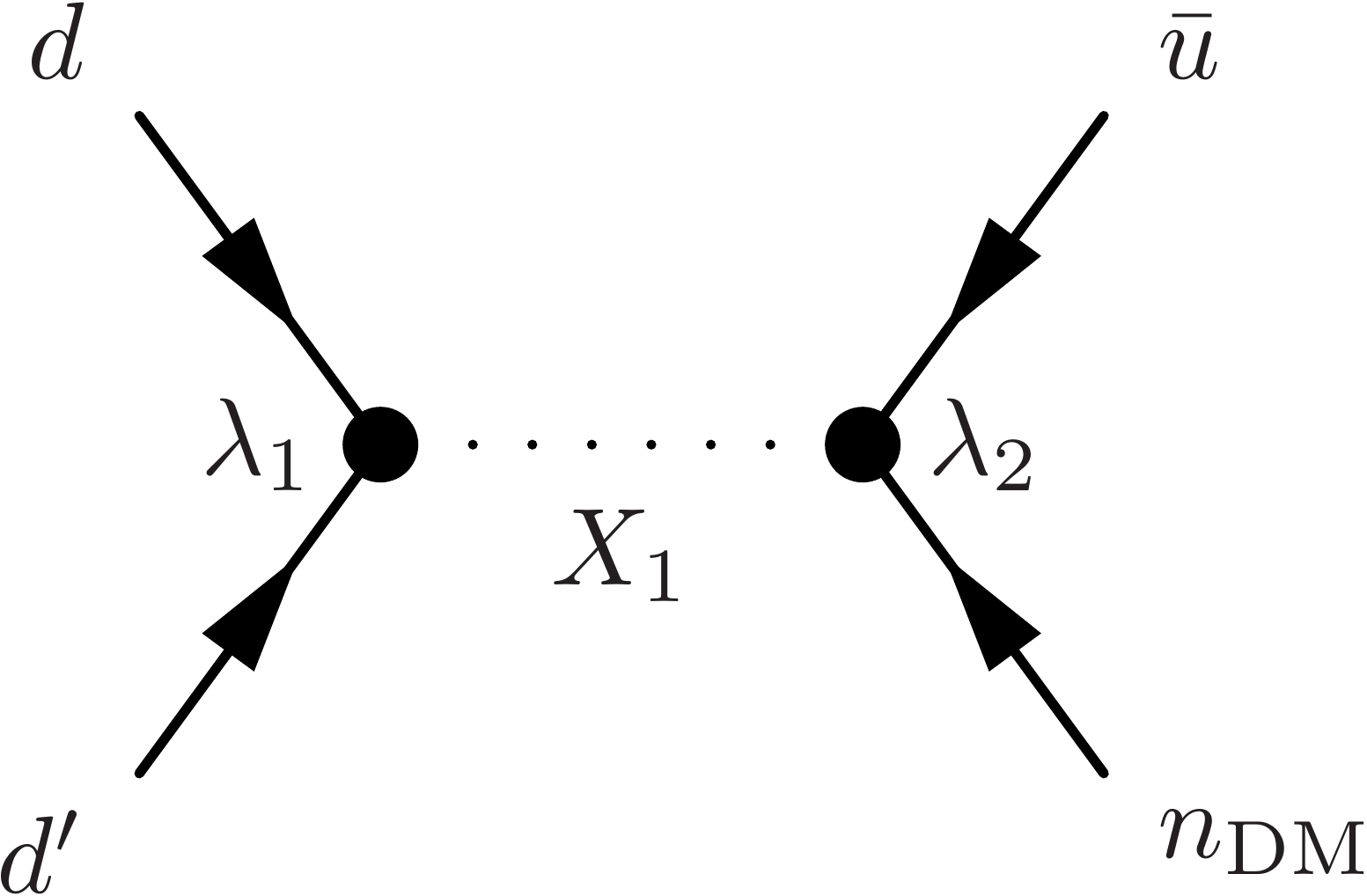}
\caption{Example of Feynman diagram of the main production mechanism at the
LHC of DM particles in the nonthermal model
resulting in the monojet final state.
In this diagram, $\rm{d}$ and $\rm{d^{'}}$ represent different down-type quark generations.
}
\label{fig:nonthermal_diagram}
\end{center}
\end{figure}

The light nonthermal DM model~\cite{bib:NonThDM_paper,Allahverdi:2013mza} is a minimal extension of the SM where the
DM particle is a Majorana fermion ($\rm{n_{DM}}$) that interacts with the up-type quarks via a colored scalar mediator
($\rm{X_{1}}$) with a coupling strength parameter $\lambda_2$.
This new colored mediator also interacts with the down-type quarks
with a coupling strength parameter $\lambda_1$.
Baryon number is not conserved in interactions of such mediators, and therefore the nonthermal
DM model could explain both the baryon abundance and the DM content of the universe.
The DM particle mass in this model must be nearly degenerate with the proton mass to
ensure the stability of both the proton and the DM particle.
Thus, the latter can be singly produced at the LHC, as shown in Fig.~\ref{fig:nonthermal_diagram}.
This leads to a final state that includes large \pt imbalance and an energetic
jet, whose \pt distribution is a Jacobian peak at half the $\rm{X_{1}}$ mass.

\begin{figure*}[htb]
\begin{center}
\includegraphics[scale=0.30]{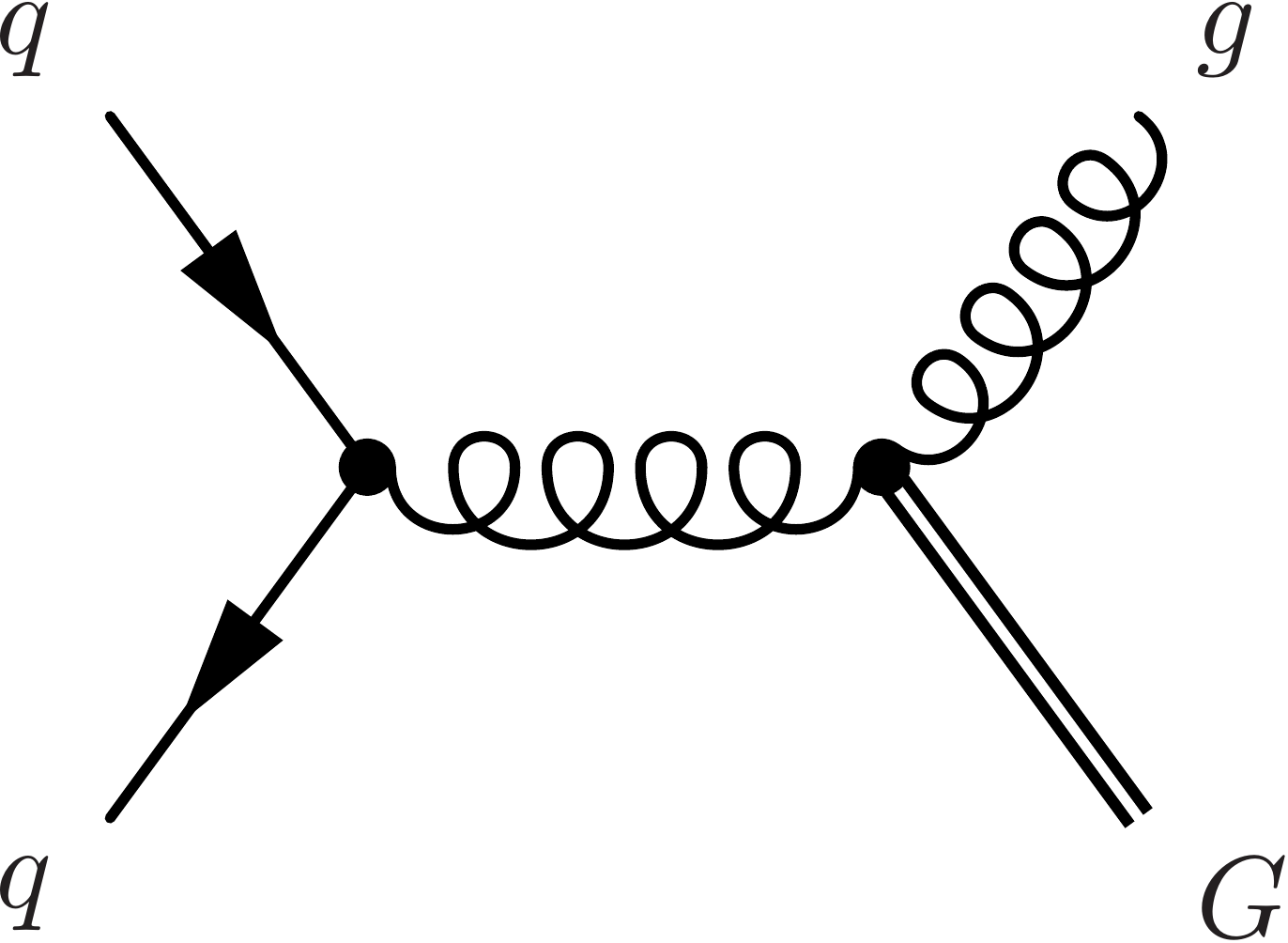}\hspace{1cm}
\includegraphics[scale=0.30]{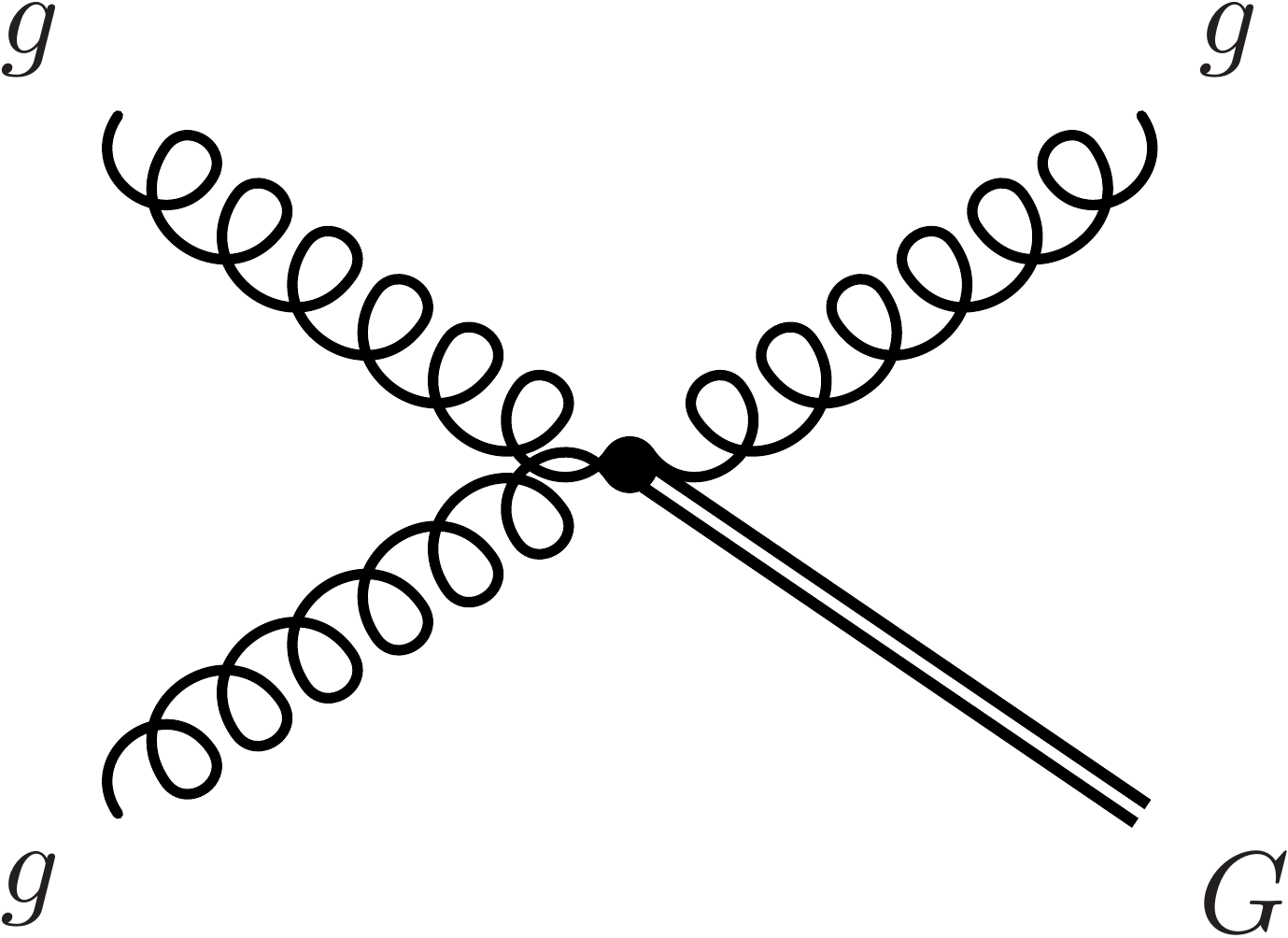}\hspace{1cm}
\includegraphics[scale=0.30]{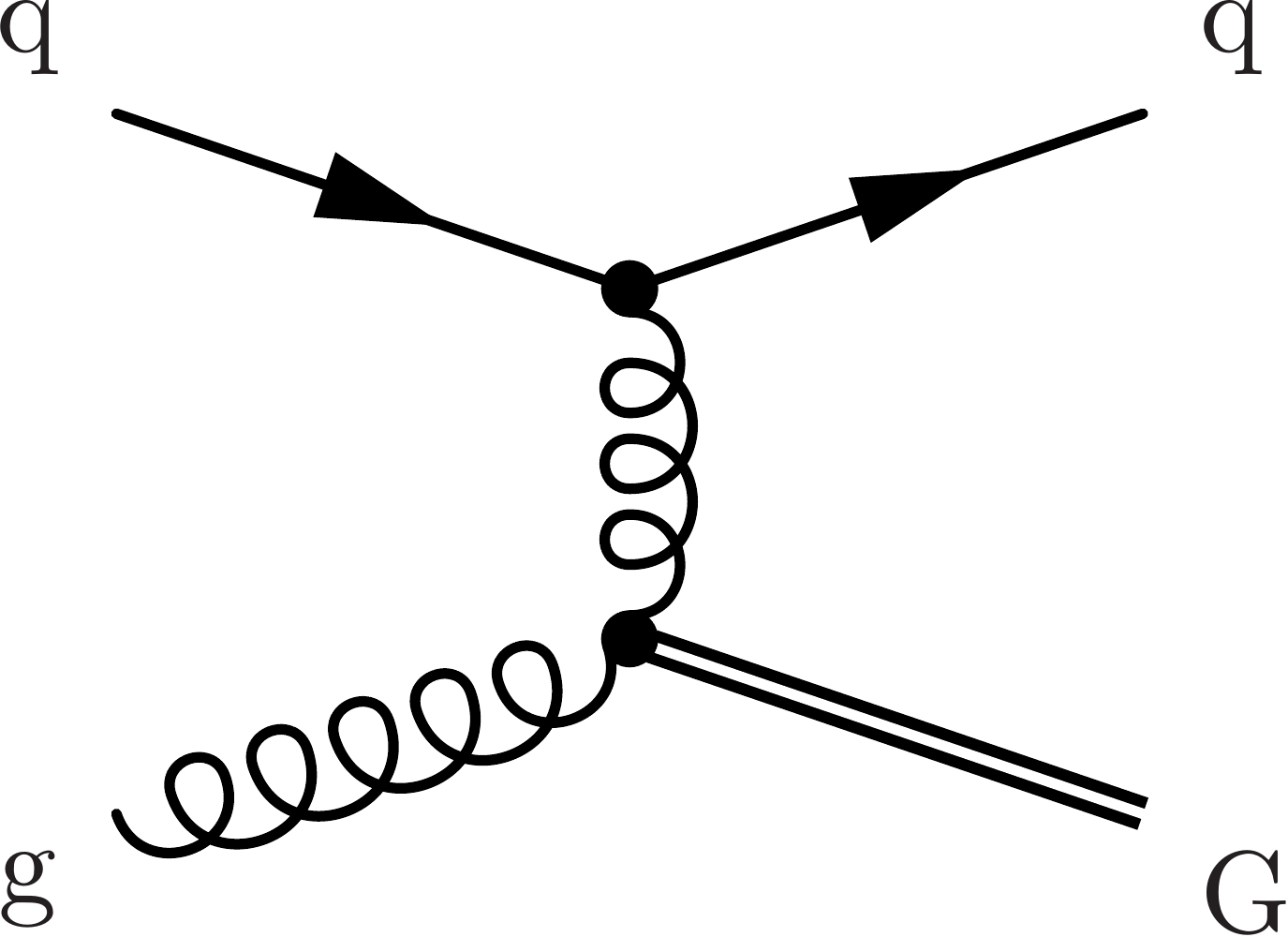}
\caption{Examples of Feynman diagrams of the main production mechanisms of gravitons at the LHC that provide monojet signatures in the ADD model.}
\label{fig:addfeyn}
\end{center}
\end{figure*}

The ADD model of EDs offers an explanation of the large difference
between the electroweak unification scale and the Planck scale ($\Mpl$),
at which gravity becomes as strong as the SM interactions.
In the simplest ADD model, a number ($n$) of EDs are introduced and are compactified on an
$n$-dimensional torus of common radius $R$.
In this framework, the SM particles and their interactions are
confined to the ordinary $3{+}1$ space-time dimensions, while
gravity is free to propagate through the
entire multidimensional space. The strength of the gravitational force
in $3{+}1$ dimensions is effectively
diluted. The fundamental Planck scale  $\MD$ of this 4+$n$-dimensional theory is related to the apparent
four-dimensional Planck scale according to $\Mpl^2\approx \MD^{n+2} R^{n}$.
The production of gravitons (G) is expected to be greatly enhanced by the increased phase space available in the EDs.
Once produced in proton-proton collisions, the graviton escapes undetected into the EDs and its presence must
be inferred from an overall \pt imbalance in the collision event, again leading to a monojet signature,
as shown in Fig.~\ref{fig:addfeyn}.

For all models, the signal extraction is performed using the distribution of the \pt
imbalance in each event category.
In the context of simplified DM models, the results of the search are reported in terms of excluded values of the masses of
the mediator and of the DM particles.
In the context of the FP and nonthermal DM models, the
results of the search are reported in terms of excluded values of the mass of the mediator particle, and
either the DM particle mass or the strength of the coupling between the mediator and the DM or SM particles.
The case of a Higgs boson decaying to invisible (e.g., DM) particles is also considered, and
the results are reported in terms of upper limits on the branching fraction to invisible particles of the Higgs boson
with a mass of 125\GeV~\cite{Aad:2012tfa,Chatrchyan:2012xdj,Chatrchyan:2013lba}, assuming SM production cross sections ($\sigma_{\textrm{SM}}$).
In the ADD model, the results are reported in terms of limits on the
fundamental Planck scale as a function of the number of extra spatial dimensions.

This paper is organized as follows. A brief overview of the CMS detector
and a description of the event reconstruction is given in Section~\ref{sec:cms}.
Information about the event simulation is provided in Section~\ref{sec:simulation} and
the event selection
is provided in Section~\ref{sec:selection}.
Section~\ref{sec:bkg} details the background estimation strategy used in the
analysis. Finally, the results of the search are described in Section~\ref{sec:results}
and summarized in Section~\ref{sec:summary}.

\section{The CMS detector and event reconstruction}
\label{sec:cms}
The central feature of the CMS apparatus is a superconducting
solenoid of 6\unit{m} internal diameter, providing a magnetic field of 3.8\unit{T}.
Within the solenoid volume are a silicon pixel and strip tracker, a lead
tungstate crystal electromagnetic calorimeter (ECAL), and a brass and scintillator
hadron calorimeter (HCAL), each composed of a barrel and two endcap sections.
Forward calorimeters extend the pseudorapidity ($\eta$) coverage provided by the barrel and
endcap detectors. Muons are detected in gas-ionization chambers embedded in the
steel flux-return yoke outside the solenoid. A more detailed description of the CMS
detector, together with a definition of the coordinate system used and the relevant
kinematic variables, can be found in Ref.~\cite{Chatrchyan:2008aa}.

The CMS particle-flow (PF) event algorithm~\cite{CMS-PRF-14-001} reconstructs and identifies each
individual particle with an optimized combination of information from the various elements
of the detector. The energy of photons is directly obtained from the ECAL measurement,
corrected for zero-suppression effects.
The energy of muons is
obtained from the curvature of the corresponding track.
The energy of electrons is determined from a combination
of the electron momentum at the primary interaction vertex as determined by the tracker, the
energy of the corresponding ECAL cluster, and the energy sum of all bremsstrahlung photons
spatially compatible with originating from the electron track.
The energy of charged hadrons
is determined from a combination of their momentum measured in the tracker and the
matching ECAL and HCAL energy deposits, corrected for zero-suppression effects and for the
response function of the calorimeters to hadronic showers. Finally, the energy of neutral
hadrons is obtained from the corresponding corrected ECAL and HCAL energy.

The missing transverse momentum vector ($\ptvecmiss$) is computed as the negative vector
sum of the transverse momenta ($\ptvec$) of all the PF candidates in an event, and its magnitude
is denoted as \ptmiss. Hadronic jets are reconstructed by clustering PF candidates
using the infrared and collinear safe anti-\kt algorithm~\cite{Cacciari:2008gp}. Jets clustered
with distance parameters of 0.4 and 0.8 are referred to as AK4 and AK8 jets, respectively.
The reconstructed vertex with the largest value of summed physics-object $\pt^2$ is taken to
be the primary $\Pp\Pp$ interaction vertex. The physics objects are those returned by a
jet finding algorithm~\cite{Cacciari:2008gp,Cacciari:2011ma} applied to all charged PF candidates
associated with the vertex, plus the corresponding associated \ptmiss.

Jet momentum is determined as the vector sum of all particle momenta in the jet,
and is found from simulation to be within 5 to 10\% of the true momentum over the full \pt
spectrum and detector acceptance. An offset correction is applied to jet energies to take
into account the contribution from additional proton-proton interactions within the
same or nearby bunch crossings (pileup). Jet energy corrections are derived from simulation,
and are confirmed with {\it in situ} measurements of the energy balance in dijet, multijet,
$\gamma$+jet, and leptonic $\PZ$+jet events~\cite{Khachatryan:2016kdb}. Additional selection criteria
are applied to each event to remove spurious jet-like features originating from isolated noise
patterns in certain HCAL regions. Such corrections and selections are also propagated to
the \ptmiss calculation~\cite{CMS-PAS-JME-16-004,Khachatryan:2014gga}.

Muons within the geometrical acceptance of $|\eta| < 2.4$ are reconstructed
by combining information from the silicon tracker and the muon system~\cite{Chatrchyan:2012xi}.
The muons are required to pass a set of quality criteria based on the number of
spatial points measured in the tracker and in the muon system, the fit quality
of the muon track, and its consistency with the primary vertex of the event.
The isolation requirements for muons are
based on the sum of the energies of the PF candidates originating from the primary vertex
within a cone of $\Delta R < 0.4$ around the muon direction, excluding the muons and electrons
from the sum. The muon isolation variable
is corrected for pileup effects by subtracting half of the \pt sum of the
charged particles that are inside the isolation cone and
not associated with the primary vertex. In this paper, `loose' muons are
selected with an average efficiency of 98\% and are used as a condition to veto the events,
whereas `tight' muons are selected with an average efficiency of 95\% and are
used to tag the events in the control samples.

Electrons within the geometrical acceptance of $|\eta| < 2.5$
are reconstructed by associating tracks reconstructed
in the silicon detector with clusters of energy in the ECAL~\cite{Khachatryan:2015hwa}.
Well-identified electron candidates are required to satisfy
additional identification criteria based on the shower
shape of the energy deposit in the ECAL and the consistency of the
electron track with the primary vertex~\cite{TRK-11-001}. Electron candidates
that are identified as coming from photon conversions in
the detector material are removed. The isolation
requirements are separated from electron identification, and are
based on the sum of the energies of the PF candidates originating from the primary vertex within a
cone of $\Delta R < 0.3$ around the electron direction, excluding the muons and electrons
from the sum. The mean
energy deposit in the isolation cone of the electron coming from
pileup is estimated following the method described in Ref.~\cite{Khachatryan:2015hwa}
and subtracted from the isolation sum. In this paper, `loose' electrons
are selected with an average efficiency of 95\% and are used as a condition to veto the events,
whereas `tight' electrons with an average efficiency of 70\% are used to select the events
in the control samples.

Photon candidates are reconstructed from energy deposits in the ECAL using algorithms
that constrain the clusters to the size and shape expected from a photon~\cite{CMS:EGM-14-001}.
The identification of the candidates is based on shower-shape and isolation variables.
For a photon to be considered to be isolated, scalar \pt sums of PF candidates originating from the primary vertex,
excluding the muons and electrons within a cone of $\Delta R < 0.3$
around the photon candidate, are required to be below the bounds defined. Only the PF candidates
that do not overlap with the electromagnetic shower of the candidate photon are included in the isolation sums.
In this paper, `loose' photon candidates are required to be reconstructed within $|\eta| < 2.5$,
whereas `tight' photon candidates used are required to be reconstructed in the ECAL barrel
($|\eta| < 1.44$). The tight photon candidates are also required to pass identification and isolation
criteria that ensure an efficiency of 80\% in selecting prompt photons, and a sample
purity of 95\% for the control samples.

Hadronically decaying $\tau$ lepton candidates detected within $|\eta| < 2.3$ are required to pass identification criteria
using the hadron-plus-strips algorithm~\cite{Khachatryan:2015dfa}. The algorithm
identifies a jet as a hadronically decaying $\tau$ lepton candidate if a subset of the
particles assigned to the jet is consistent with the decay products of a $\tau$ candidate.
In addition, $\tau$ candidates are required to be isolated from other activity in the
event. The isolation requirement is computed by summing the \pt of the PF
charged and PF photon candidates within an isolation cone of $\Delta R = 0.5$ and $0.3$, respectively,
around the $\tau$ candidate direction. A more detailed description of the isolation requirement can be found in Ref.~\cite{Khachatryan:2015dfa}.

\section{Simulated samples} \label{sec:simulation}

To model the SM backgrounds, simulated Monte Carlo (MC) samples are produced for the $\PZ$+jets, $\PW$+jets, $\gamma$+jets,
and quantum chromodynamics (QCD) multijet processes at leading order (LO) using the \MGvATNLO 2.2.2~\cite{Alwall:2014hca}
generator and are generated with up to four additional partons in the matrix element calculations.
The samples for the \ttbar and single top quark background processes are
produced at next-to-leading order (NLO) using \POWHEG 2.0 and 1.0, respectively~\cite{Oleari:2010nx,Alioli:2009je},
and the set of diboson ($\PW\PW$, $\PW\PZ$, $\PZ\PZ$) samples is produced at LO with \PYTHIA 8.205~\cite{Sjostrand:2014zea}.

Vector and axial-vector monojet and mono-$\PV$ dark matter signals are simulated at NLO
using  the DMSIMP models~\cite{Backovic:2015soa,Mattelaer:2015haa} with  the \MGvATNLO generator.
Both scalar and pseudoscalar monojet and mono-V production contain gluon-initiated loop processes. In the case of
mono-V signals, no direct couplings of the mediator to vector bosons are considered. All samples are generated at LO with one
additional parton in the matrix element calculations, taking into account finite top quark mass effects and using
the \MGvATNLO generator in conjunction with the \textsc{DMSIMP} models.

The SM Higgs boson signal events produced through vector boson fusion and gluon fusion
are generated using the \POWHEG generator~\cite{Alioli:2008tz,Nason:2009ai}; for each sample the cross
section is normalized to the  next-to-NLO (NNLO) and next-to-NNLO, respectively.  The SM Higgs boson
production in association with $\PW$ or $\PZ$ bosons is simulated at LO using the \textsc{JHUGenerator} 5.2.5
generator~\cite{Anderson:2013afp} and normalized to the NNLO cross section.

The ADD ED signal is simulated at LO in QCD using the \PYTHIA generator, requiring $\pthat > 80\GeV$,
where $\pthat$ denotes the transverse momentum of the outgoing parton in the parton-parton center-of-mass frame.
The \PYTHIA truncation setting is used to suppress the cross section by a factor of ${\MD}^{4}/\hat{s}^2$ for
$\hat{s} > {\MD}^2$, where $\hat{s}$ is the center-of-mass energy of the
incoming partons, to ensure validity of the effective field theory.

Lastly, both the FP dark matter signal and the nonthermal DM signal models are simulated
at LO using the \MGvATNLO generator.
In the FP dark matter signal model, the coupling strength parameter is fixed
to be $\lambda_{\rm{u}}=1$ while, in the
nonthermal DM signal model, the mass of the DM particle is fixed to the
proton mass to assure the stability of both the proton and the
DM particle. In this latter model, coupling ranges of 0.01--1.5 for $\lambda_1$
and 0.01--2.0 for $\lambda_2$ are considered, to ensure the mediator width is less
than about 30\% of its mass.

The MC samples produced using \MGvATNLO, \POWHEG, and \textsc{JHUGenerator}
generators are interfaced with \PYTHIA using the CUETP8M1 tune~\cite{Khachatryan:2015pea}
for the fragmentation, hadronization, and underlying event description.
In the case of the \MGvATNLO samples, jets from the matrix element calculations
are matched to the parton shower description following the MLM~\cite{Mangano:2006rw} (FxFx~\cite{Frederix:2012ps})
prescription to match jets from matrix element calculations and parton shower description for LO (NLO) samples.
The NNPDF 3.0~\cite{Ball:2014uwa}
parton distribution functions (PDFs) are used in all generated samples.
The propagation of all final-state particles through the CMS detector
are simulated with \GEANT4~\cite{Agostinelli:2002hh}.
The simulated events include the effects of pileup, with the multiplicity of reconstructed primary
vertices matching that in data. The average number of pileup interactions per proton bunch crossing
is found to be 23 for the data sample used in this analysis~\cite{CMS:2017sdi}.

\section{Event selection}\label{sec:selection}

Signal region events are selected using triggers with thresholds of  110 or 120\GeV
on both \ptmisstrig~and \mhttrig, depending on the data taking period.
The \ptmisstrig~corresponds to the magnitude of the vector
\ptvec sum of all the PF candidates reconstructed at the trigger level, while the \mhttrig~is computed
as the magnitude of the vector \ptvec sum of jets with $\pt > 20\GeV$ and $|\eta| < 5.0$
reconstructed at the trigger level. The energy fraction attributed to neutral hadrons in these
jets is required to be smaller than 0.9. This requirement suppresses anomalous events with jets
originating from detector noise. To be able to use the same triggers for selecting events
in the muon control samples used for background prediction, muon candidates are not included in the
\ptmisstrig~nor \mhttrig~computation. The trigger efficiency is measured to be 97\% for events passing
the analysis selection for $\ptmiss > 250\GeV$ and becomes fully efficient for events with $\ptmiss > 350\GeV$.

Candidate events are required to have $\ptmiss > 250\GeV$. In the monojet category, the highest $p_{\rm T}$ (leading) AK4 jet in
the event is required to have $ \pt > 100\GeV$ and $|\eta| < 2.4$, whereas in the mono-$\PV$ category, the
leading AK8 jet is required to have $ \pt > 250\GeV$ and $|\eta| < 2.4$. In both categories, the leading
jet is also required to have at least 10\% of its energy coming from charged particles and less than 80\%
of its energy attributed to neutral hadrons. This selection helps to remove events originating from
beam-induced backgrounds. In addition, the analysis employs various event filters to reduce events
with large misreconstructed \ptmiss~\cite{CMS-PAS-JME-16-004} originating from noncollision backgrounds.

The main background processes in this search are the \Zvvjets and \Wlvjets processes.
The \Zvvjets process is an irreducible background and constitutes the
largest background in the search. In contrast, the background from \Wlvjets is suppressed by
imposing a veto on events containing one or more loose muons or
electrons with $ \pt >10\GeV$, or $\tau$ leptons with $\pt>18\GeV$.
Events that contain a loose, isolated photon with $\pt>15\GeV$ and $|\eta| < 2.5$
are also vetoed.
This helps to suppress electroweak (EW) backgrounds in which a photon is radiated from the initial state.
To reduce the contamination from top quark backgrounds, events are rejected if they contain a b-tagged jet
with $\pt > 20\GeV$ and $|\eta| < 2.4$. These jets are identified using the combined secondary vertex
algorithm (CSVv2)~\cite{Chatrchyan:2012jua,CMS-PAS-BTV-15-001}, adopting a working point corresponding to correctly
identifying a jet originating from a bottom quark with a probability of 80\% and misidentifying a
jet originating from a charm quark (light-flavor jet) with a probability of 40 (10)\%.
Lastly, QCD multijet background with \ETm arising from mismeasurements of the jet momenta
is suppressed by requiring the minimum
azimuthal angle between the \ptvecmiss direction and each of the first four leading jets with
\pt greater than 30\GeV to be larger than 0.5 radians.

To select an event in the mono-$\PV$ category, a leading AK8 jet is
identified as a jet arising from hadronic decays of Lorentz-boosted $\PW$ or $\PZ$ bosons.
Such jets typically have an invariant mass, computed from the momenta of jet's constituents, between 65 and 105\GeV~\cite{2017wyc}.
The mass of the leading AK8 jet is computed after pruning based on the technique~\cite{Ellis:2009me,Ellis:2009su}
involving reclustering the constituents of the jet using the Cambridge--Aachen algorithm~\cite{Dokshitzer:1997in}
and removing the soft and wide-angle contributions to jets in every recombination step.
The pruning algorithm is controlled by a soft threshold parameter $\rm{z_{cut}} = 0.1$ and an angular separation threshold of
$\Delta R > m_{\rm jet}/p_{\rm T}^{\rm jet}$. This technique yields improved jet mass resolution
owing to reduced effects coming from
the underlying event and pileup.
The $N$-subjettiness variable $\tau_N$~\cite{Thaler:2010tr}
is also employed to further isolate jets arising from hadronic decays of $\PW$ or $\PZ$ bosons.
This observable measures the distribution of jet constituents relative to candidate subjet axes in order to quantify
how well the jet can be divided into $N$ subjets. Therefore, the ratio of the `2-subjettiness' to
the `1-subjettiness' ($\tau_2 / \tau_1$) has excellent capability for distinguishing jets
originating from boosted vector bosons from jets originating from light quarks and gluons.
The pruned jet mass and $N$-subjettiness requirements, whose use if referred to as $\PV$ tagging, result in a
70\% efficiency for tagging jets originating from $\PV$ bosons and a 5\% probability of misidentifying a jet as a $\PV$ jet.
Events that do not qualify for the mono-$\PV$ category are assigned to the monojet category. The common selection
requirements for both signal categories are summarized in Table~\ref{tab:selection},
while the category-specific selection requirements are reported in Table~\ref{tab:category}.

\begin{table*}[htb]
\topcaption{Summary of the common selection requirements for mono-$\PV$ and monojet categories.}
\begin{center}
\renewcommand{\arraystretch}{1}
\ifthenelse{\boolean{cms@external}}{\footnotesize}{\resizebox{\textwidth}{!}}
{
\begin{scotch}{l c c}
Variable                           & Selection                       & Target background \\
\hline
Muon (electron) veto               & $\pt > 10\GeV,~|\eta| < 2.4 (2.5)$  & \Zlljets,~\Wlvjets \\
$\tau$ lepton veto                 & $\pt > 18\GeV,~|\eta| < 2.3$        & \Zlljets,~\Wlvjets  \\
Photon veto                        & $\pt > 15\GeV,~|\eta| < 2.5$        & \phojets \\
Bottom jet veto                    & CSVv2 $< 0.8484$, $\pt > 15\GeV,~|\eta| < 2.4$ &  Top quark\\
$\ptmiss$                          & ${>} 250\GeV$                          & QCD, top quark, \Zlljets \\
$\Delta\phi$($\ptvecjet$,$\ptvecmiss$)   &  $ {>} 0.5$ radians               & QCD \\
Leading AK4 jet $\pt$ and $\eta$   & ${>} 100\GeV$ and $ |\eta| < 2.4$      & All \\
\end{scotch}
}
\label{tab:selection}
\end{center}
\end{table*}

\begin{table}[htb]
\topcaption{Summary of the selection requirements for the mono-$\PV$ category.
Events that fail the mono-$\PV$ selection are assigned to the monojet category.}
\begin{center}
\renewcommand{\arraystretch}{1}
\ifthenelse{\boolean{cms@external}}{\footnotesize}{}
{
\begin{scotch}{l c}
Leading AK8 jet                                    & Mono-$\PV$ selection             \\
\hline
$\pt$ and $\eta$            & $ {>} 250\GeV $ and $ |\eta| < 2.4 $   \\
$\tau_2 / \tau_1$           & $ {<} 0.6 $                             \\
Mass ($m_{\mathrm{jet}}$)   & $65 < m_{\mathrm{jet}} < 105\GeV$    \\
\end{scotch}
}
\label{tab:category}
\end{center}
\end{table}

\section{Background estimation} \label{sec:bkg}

The largest background contributions, from \Zvvjets and \Wlvjets processes,
are estimated using data from five mutually exclusive control samples selected
from dimuon, dielectron, single-muon, single-electron, and \phojets final states
as explained below. The hadronic recoil \pt is used as a proxy for \ptmiss in these
control samples, and is defined by excluding identified leptons or photons from
the \ptmiss calculation.

\subsection{Control sample selection}

Dimuon and single-muon control sample events are selected using full signal region
criteria with the exception of the muon veto.
Events in the dimuon control sample
are selected requiring leading (subleading) muon \pt greater than 20 (10)\GeV and
an invariant mass in the range 60 to 120\GeV, compatible with a $\PZ$ boson decay.
Events are vetoed if there is an additional loose muon or electron with $\pt > 10\GeV$.
In the single-muon control sample, exactly one tightly identified, isolated muon with $\pt > 20\GeV$
is required. No additional loose muons or electrons with $\pt > 10\GeV$ are allowed.
In addition, the transverse mass (\mt) of the muon-$\vec p_{\rm T}^{\rm miss}$ system is required to be less
than 160\GeV and is computed as $\mt = \sqrt{\smash[b]{2\ptmiss \pt^{\mu} (1 - \mathrm{cos}\Delta\phi)}}$,
where $\pt^{\mu}$ is the \pt of the muon, and $\Delta\phi$ is the angle between $\ptvec^{\mu}$ and $\ptvecmiss$.

Dielectron and single-electron control sample events are selected with
an isolated single-electron trigger with a \pt threshold of 27\GeV.
In boosted \Zeejets events, the two electrons produced in the decay
typically have so little separation such that their tracks are included in each
other's isolation cones.
Therefore, to recover efficiency in selecting
high-\pt $Z$ candidates at the trigger level, a nonisolated single-electron
trigger with a \pt threshold of 105\GeV is used.
Events in the dielectron control sample are required to contain exactly two oppositely charged
electrons with leading (trailing) electron \pt greater than 40 (10)\GeV. Similar to the dimuon control
sample case, the invariant mass of the dielectron system is required to be between 60 and 120\GeV
to be consistent with a $\PZ$ boson decay. The events in the single-electron control sample are required to
contain exactly one tightly identified and isolated electron with $\pt > 40\GeV$.
In addition, the contamination from QCD multijet events in this control sample is suppressed by
requiring $\ptmiss > 50\GeV$ and $\mt < 160\GeV$.

Lastly, the \phojets control sample is selected using events with one high-\pt photon
collected using single-photon triggers with \pt thresholds of 165 or 175\GeV, depending
on the data taking conditions. The photon is required to have $\pt > 175\GeV$ and to
pass tight identification and isolation criteria, to ensure a high trigger efficiency of 98\%.

\subsection{Signal extraction}\label{sec:constraints}

A binned likelihood fit to the data as presented in Ref.~\cite{paper-exo-037}
is performed simultaneously in the five different control samples and in the signal
region, for events selected in
both the monojet and mono-$\PV$ categories, to estimate the \Zvvjets and \Wlvjets rate
in each \ptmiss bin. In this likelihood, the expected numbers of \Zvvjets events in each
bin of \ptmiss are the free parameters of the fit. Transfer factors, derived from simulation,
are used to link the yields of the \Zlljets,~\Wlvjets and \phojets processes in the control
regions with the \Zvvjets and \Wlvjets background estimates in the signal region.
These transfer factors are defined as the ratio of expected yields of the target process
in the signal region and the process being measured in the control sample.

To estimate the \Wlvjets background in the signal region, the transfer factors between
the \Wmvjets and \Wevjets event yields in the single-lepton control samples and
the estimates of the \Wlvjets  background in the signal region are constructed. These transfer factors take into account
the impact of lepton acceptances and efficiencies, lepton veto efficiencies, and
the difference in the trigger efficiencies in the case of the single-electron control sample.

The \Zvv~background prediction in the signal region is connected to the yields of \Zmm~and \Zee~events
in the dilepton control samples. The associated transfer factors account for the differences in the
branching ratio of $\PZ$ bosons to charged leptons relative to neutrinos
and the impact of lepton acceptance and selection
efficiencies. In the case of dielectron events, the transfer factor also takes into account the
difference in the trigger efficiencies. The resulting constraint on the \Zvvjets process from the dilepton
control samples is limited by the statistical uncertainty in the dilepton control samples
because of the large difference in branching fractions between $\PZ$ boson decays to neutrinos
and $\PZ$ boson decays to muons and electrons.

The \phojets control sample is also used to predict the \Zvvjets process in the signal region through a
transfer factor, which accounts for the difference in the cross sections of the \phojets and \Zvvjets processes,
the effect of acceptance and efficiency of identifying photons along with the difference in the efficiencies of
the photon and \ptmiss triggers. The addition of the \phojets control sample mitigates the impact of the limited
statistical power of the dilepton constraint, because of the larger production cross section of \phojets process compared to that of \Zvvjets process.

Finally, a transfer factor is also defined to connect the \Zvvjets and \Wlvjets background yields
in the signal region, to further benefit from the larger statistical power that
the \Wlvjets background provides, making it possible to experimentally
constrain \Zvvjets production at high \ptmiss.

These transfer factors rely on an accurate prediction of the
ratio of \Zjets, \Wjets, and \phojets cross sections. Therefore, LO simulations for these
processes are corrected using boson \pt-dependent NLO QCD K-factors derived using
\MGvATNLO. They are also corrected using \pt-dependent higher-order
EW corrections extracted from theoretical
calculations~~\cite{Denner:2009gj,Denner:2011vu,Denner:2012ts,Kuhn:2005gv,Kallweit:2014xda,Kallweit:2015dum}.
The higher-order corrections are found to improve the data-to-simulation agreement for both the
absolute prediction of the individual \Zjets, \Wjets, and \phojets processes, and their respective ratios.

The remaining backgrounds that contribute to the total event yield in the signal region
are much smaller than those from \Zvvjets and \Wlvjets processes. These smaller backgrounds
include QCD multijet events which are measured from data using a $\Delta\phi$ extrapolation
method~\cite{Collaboration:2011ida,paper-exo-037}, and top quark and diboson processes, which are
obtained directly from simulation.

\subsection{Systematic uncertainties}

Systematic uncertainties in the transfer factors are
modeled as constrained nuisance parameters and include both
experimental and theoretical uncertainties
in the \phojets to \Zjets and \Wjets to \Zjets differential cross section ratios.

Theoretical uncertainties in $\PV$-jets and \phojets processes include effects from QCD and EW higher-order
corrections along with PDF modeling uncertainty. To estimate the theoretical uncertainty
in the $\PV$-jets and \phojets ratios due to QCD and EW higher-order effects as well as their correlations across
the processes and \pt bins, the recommendations of Ref.~\cite{DMTheory} are employed,
as detailed in the following explanation.

Three separate sources of uncertainty associated with QCD higher order corrections are used.
One of the uncertainties considered comes from the variations around the central
renormalization and factorization scale choice. It is evaluated by taking the differences in the NLO cross
section as a function of boson \pt after changing the renormalization and factorization scales by a factor of two and a factor
of one-half with respect to the default value. These constant scale variations mainly affect the
overall normalization of the boson \pt distributions and therefore underestimate the shape uncertainties
that play an important role in the extrapolation of low-\pt measurements to high-\pt.
A second, conservative shape uncertainty derived from altered boson \pt spectra is used to
supplement the scale uncertainties and account for the \pt dependence of the uncertainties.
The modeling of the correlations between
the processes assumes a close similarity of QCD effects between all $\PV$-jets and \phojets processes.
However, the QCD effects in \phojets production could differ compared to the case of \Zjets and \Wjets productions.
In order to account for this variation, a third uncertainty is computed based on the
difference of the known QCD K-factors
of the \Wjets and \phojets processes with respect to \Zjets production.
All QCD uncertainties are
correlated across the \Zjets, \Wjets, \phojets processes, and also correlated across
the bins of the hadronic recoil~\pt.

For the $\PV$-jets and \phojets processes, nNLO EW corrections are applied, which correspond to full
NLO EW corrections~\cite{Denner:2009gj,Denner:2011vu,Denner:2012ts,Kallweit:2015dum}
supplemented by two-loop Sudakov EW logarithms~\cite{Kuhn:2004em,Kuhn:2005gv,Kuhn:2005az,Kuhn:2007cv}.
We also considered three separate sources of uncertainty arising from the
following: pure EW higher-order corrections failing to cover the effects of
unknown Sudakov logarithms in the perturbative expansion beyond NNLO, missing
NNLO effects that are not
included in the nNLO EW calculations, and the difference between the next-to-leading logarithmic
(NLL) Sudakov approximation
at two-loop and simple exponentiation of the full NLO EW correction.
The variations due to the effect of unknown Sudakov
logs are correlated across the \Zjets, \Wjets, and \phojets processes and are also
correlated across the bins of hadronic recoil \pt. On the other hand, the other
two sources of EW uncertainties
are treated as uncorrelated across the $\PV$-jet and \phojets processes, and an independent
nuisance parameter is used for each process.

A recommendation that includes a factorized approach to partially include
mixed QCD-EW corrections is outlined in Ref.~\cite{DMTheory}. An
additional uncertainty is introduced to account for the difference between
the corrections done in the multiplicative and the additive approaches, to account
for the non-factorized mixed EW-QCD effects.

The summary of the aforementioned theoretical uncertainties including their magnitude and correlation is outlined in Table~\ref{tab:sys}.

\begin{table*}[htb]
\topcaption{Theoretical uncertainties considered in the $\PV$-jets and \phojets processes, and their ratios. The correlation between each process and between the \pt bins are described.}
\begin{center}
\renewcommand{\arraystretch}{1}
\ifthenelse{\boolean{cms@external}}{\footnotesize}{\resizebox{\textwidth}{!}}
{
\begin{scotch}{lll}
Uncertainty source                           & Process (magnitude)                                                                                     & Correlation                    \\
\hline \\[-1.5ex]
Fact. \& renorm. scales (QCD) & \begin{tabular}[c]{@{}l@{}}\Zvv/\Wlv~(0.1~--~0.5\%) \\ \Zvv/\phojets (0.2~--~0.5\%)  \end{tabular} & \begin{tabular}[c]{@{}l@{}}Correlated between processes; \\ and in \pt \end{tabular}\\[2.5ex]
\pt shape dependence (QCD)                   & \begin{tabular}[c]{@{}l@{}}\Zvv/\Wlv~(0.4~--~0.1\%)\\ \Zvv/\phojets (0.1~--~0.2\%) \end{tabular} & \begin{tabular}[c]{@{}l@{}}Correlated between processes; \\ and in \pt \end{tabular}\\[2.5ex]
Process dependence (QCD)                     & \begin{tabular}[c]{@{}l@{}}\Zvv/\Wlv~(0.4~--~1.5\%)\\ \Zvv/\phojets (1.5~--~3.0\%)  \end{tabular} & \begin{tabular}[c]{@{}l@{}}Correlated between processes; \\ and in \pt \end{tabular}\\[2.5ex]
Effects of unknown Sudakov logs (EW)        & \begin{tabular}[c]{@{}l@{}}\Zvv/\Wlv~(0~--~0.5\%) \\ \Zvv/\phojets (0.1~--~1.5\%) \end{tabular} & \begin{tabular}[c]{@{}l@{}}Correlated between processes; \\ and in \pt \end{tabular}\\[2.5ex]
Missing NNLO effects (EW)                   & \begin{tabular}[c]{@{}l@{}}\Zvv~(0.2~--~3.0\%) \\ \Wlv~(0.4~--~4.5\%)\\ \phojets (0.1~--~1.0\%)\end{tabular}  & \begin{tabular}[c]{@{}l@{}}Uncorrelated between processes; \\ correlated in \pt \end{tabular} \\[4ex]
Effects of NLL Sudakov approx. (EW)   & \begin{tabular}[c]{@{}l@{}}\Zvv~(0.2~--~4.0\%) \\ \Wlv~(0~--~1.0\%)\\ \phojets (0.1~--~3.0\%)\end{tabular}   &  \begin{tabular}[c]{@{}l@{}}Uncorrelated between processes; \\ correlated in \pt \end{tabular} \\[4ex]
Unfactorized mixed QCD-EW corrections        & \begin{tabular}[c]{@{}l@{}}\Zvv/\Wlv~(0.15~--~0.3\%)\\ \Zvv/\phojets ($<$0.1\%)\end{tabular}    & \begin{tabular}[c]{@{}l@{}}Correlated between processes; \\ and in \pt \end{tabular}\\[2.5ex]
PDF                                          &  \begin{tabular}[c]{@{}l@{}}\Zvv/\Wlv~(0~--~0.3\%)\\ \Zvv/\phojets (0~--~0.6\%)\end{tabular}       & \begin{tabular}[c]{@{}l@{}}Correlated between processes; \\ and in \pt \end{tabular}\\[2ex]
\end{scotch}
}
\label{tab:sys}
\end{center}
\end{table*}

Experimental uncertainties including the reconstruction efficiency
(1\% per muon or electron) and the selection efficiencies of leptons
(1\% per muon and 2\% per electron), photons (2\%), and hadronically
decaying $\tau$ leptons (5\%), are also incorporated.
These reconstruction and selection efficiencies further translate into an uncertainty in the lepton veto efficiency of 3\%.
Uncertainties in the purity of photons
in the~\phojets control sample (2\%), and in the efficiency of the electron (2\%),
photon (2\%), and \ptmiss (1--4\%) triggers, are included and are fully correlated
across all the bins of hadronic recoil \pt and \ptmiss. The uncertainty in the efficiency of the b jet
veto is estimated to be 6\% (2)\% for the contribution of the top quark (diboson) background.

The uncertainty in the efficiency of the $\PV$
tagging requirements is estimated to be 9\% in the mono-$\PV$ category.
The uncertainty in the modeling of \ptmiss in simulation~\cite{Khachatryan:2014gga}
is estimated to be 4\% and is dominated by the uncertainty in the jet energy scale.

A systematic uncertainty of 10\% is included for the top quark background
associated with the modeling of the top quark \pt distribution in simulation~\cite{Khachatryan:2016mnb}.
In addition, systematic uncertainties of 10 and 20\% are included in the normalizations of the
top quark~\cite{Khachatryan:2015uqb} and diboson backgrounds~\cite{Khachatryan:2016txa,Khachatryan:2016tgp},
respectively, to account for the uncertainties in their cross sections in the relevant
kinematic phase space. Lastly, the uncertainty in the QCD multijet background estimate
is found to be between 50--150\% due to the variations of the jet response and the
statistical uncertainty of the extrapolation factors.

\subsection{Control sample validation}

An important cross-check of the application of \pt-dependent NLO QCD and EW corrections
is represented by the agreement between data and simulation in the ratio of
\Zjets events to both \phojets events
and \Wjets events in the control samples, as a function of hadronic recoil~\pt.

\begin{figure*}[htbp]
\centering
\includegraphics[width=\cmsFigWidth]{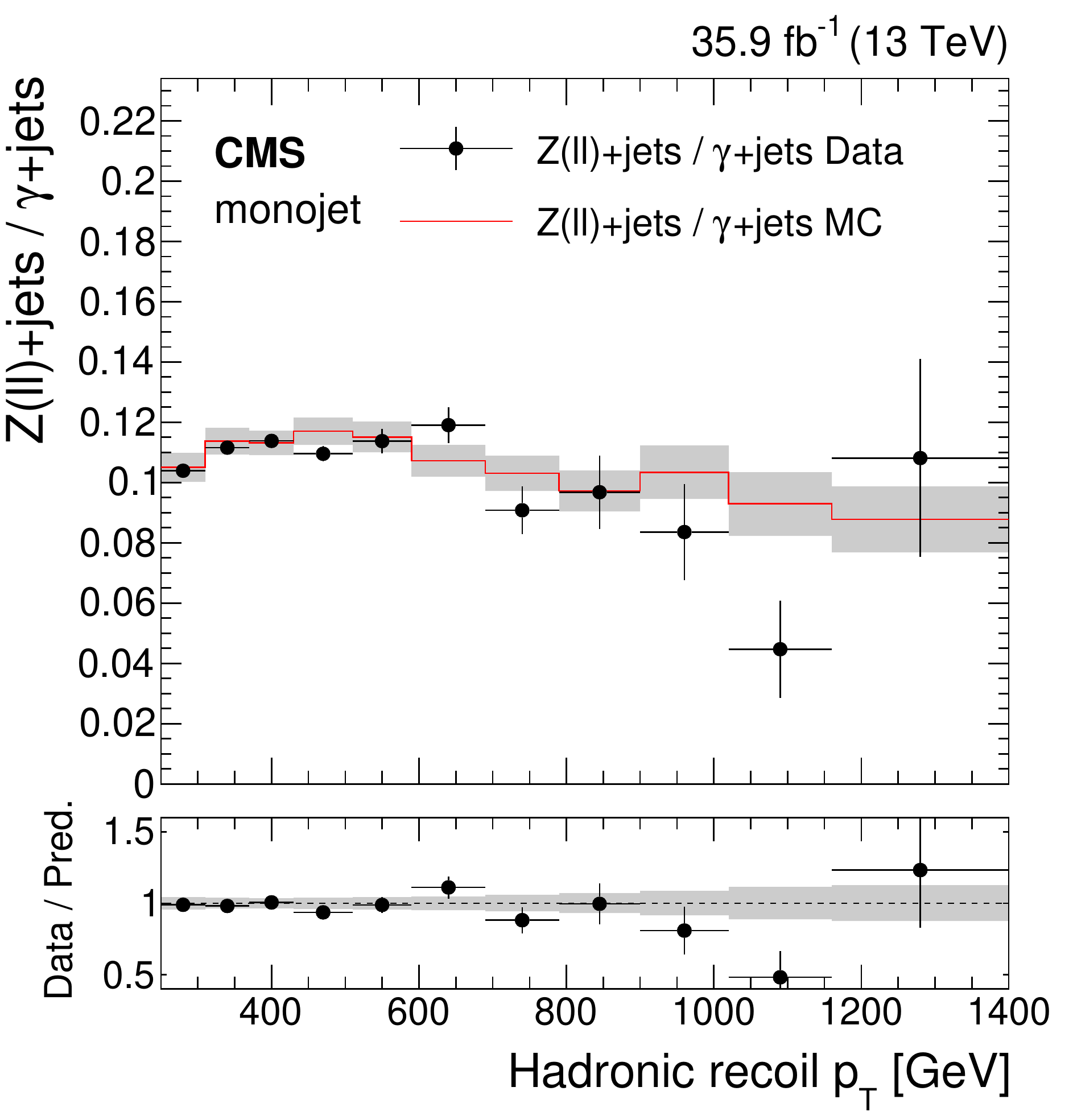}\hfil
\includegraphics[width=\cmsFigWidth]{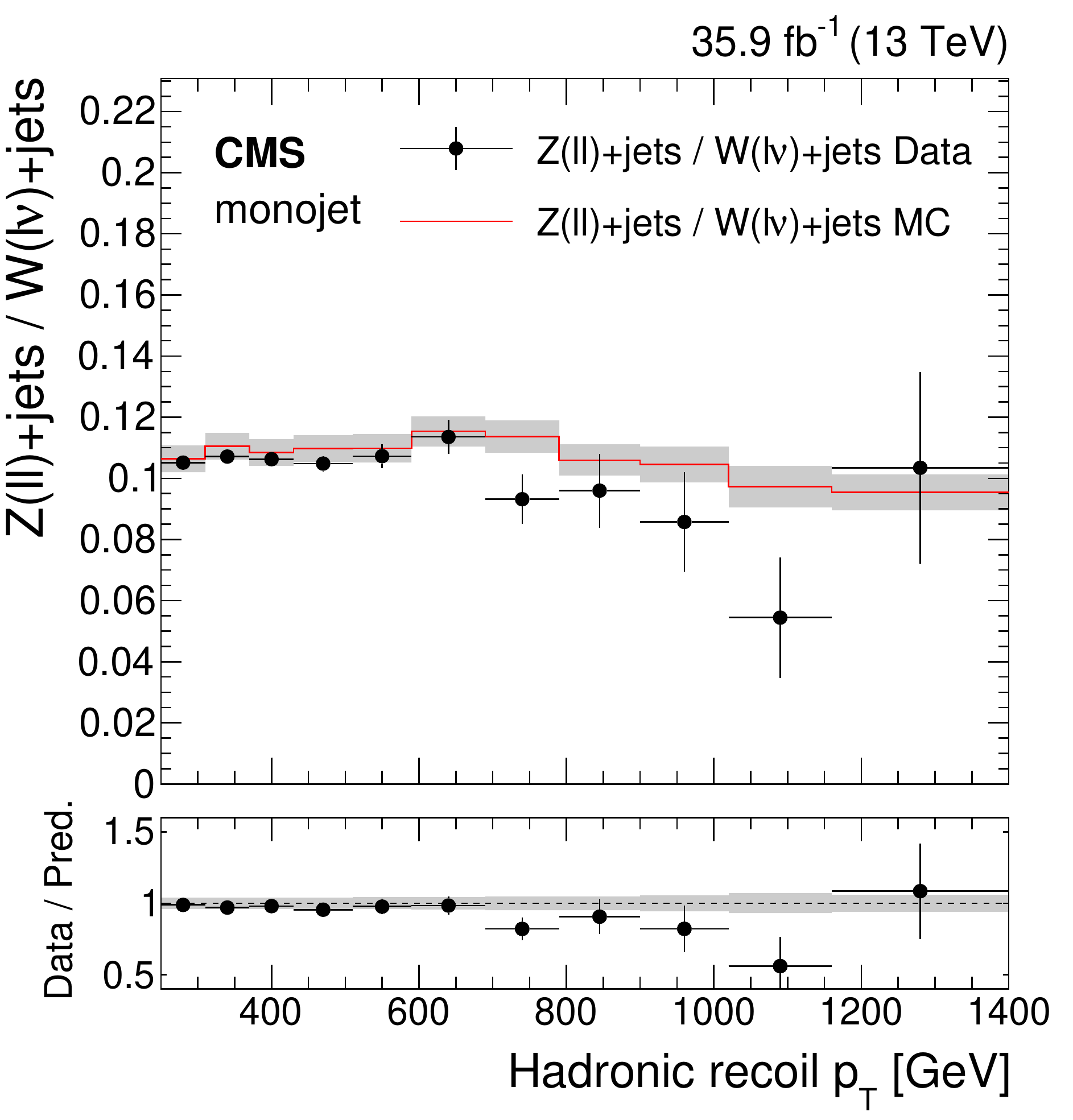}
\\
\includegraphics[width=\cmsFigWidth]{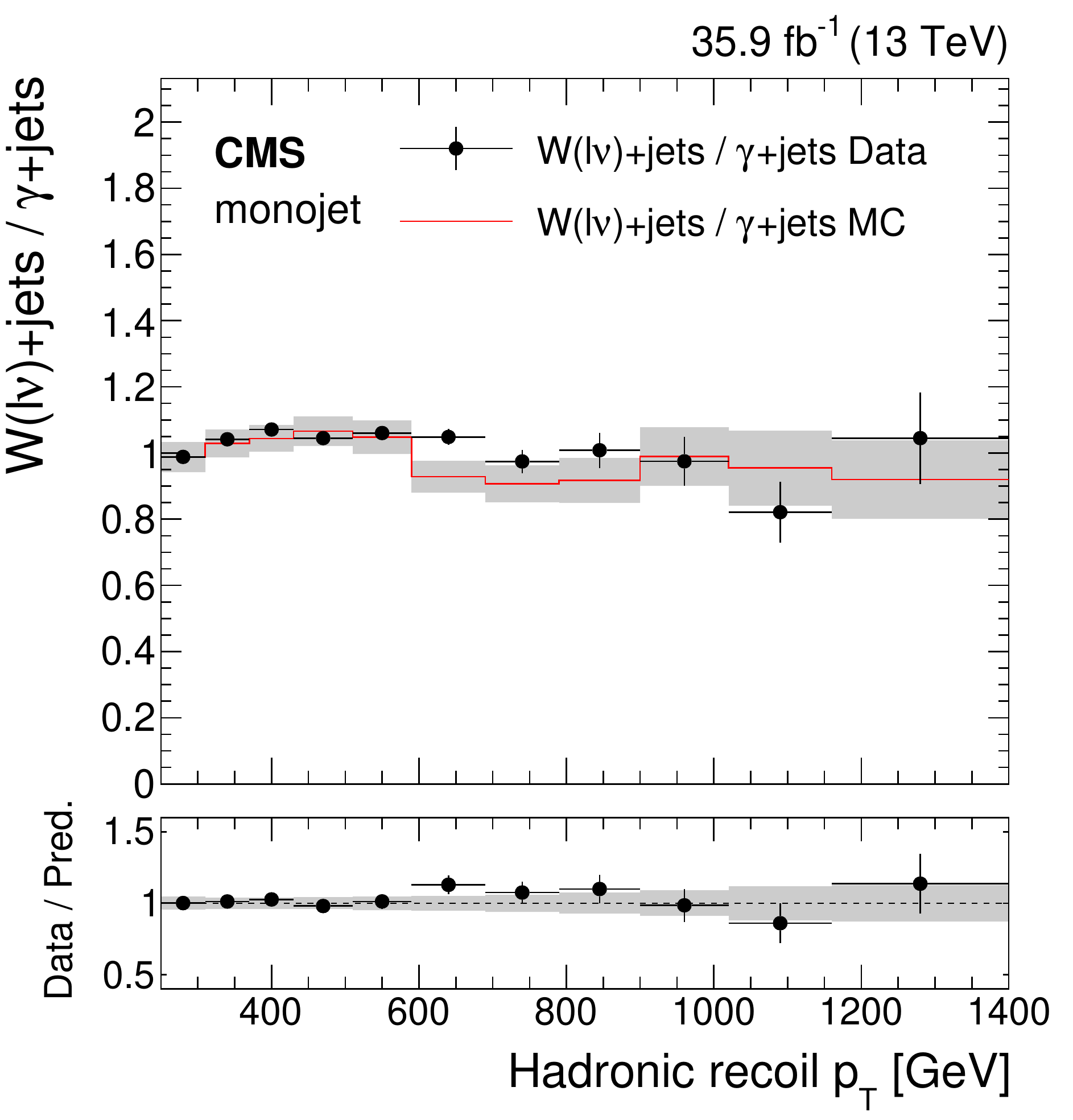}
\caption{Comparison between data and MC simulation for the  $\PZ(\ell\ell)$/\phojets,
$\PZ(\ell\ell)$/$\PW(\ell\nu)$, and $\PW(\ell\nu)$/\phojets ratios as a function
of the hadronic recoil in the monojet category.
In the lower panels, ratios of data with the pre-fit background prediction are shown.
The gray bands include both the pre-fit systematic uncertainties and the statistical uncertainty in the simulation.}
\label{fig:Ratio}
\end{figure*}

Figure~\ref{fig:Ratio} shows the ratio
between \Zlljets and \phojets (left), \Zlljets and \Wlvjets (middle),
and the one between \Wlvjets/\phojets processes (right) as a function of the recoil for events selected in the monojet category.
While we do not explicitly use a \Wlvjets/\phojets constraint in the analysis, the two cross sections
are connected through the \Zjets/\phojets and \Zjets/\Wjets constraints that are explained in Section~\ref{sec:constraints}. Therefore, it is instructive to examine
the data-MC comparison of the \Wlvjets/\phojets ratio.
Good agreement is observed between data and simulation after the application of the NLO corrections as shown in Fig.~\ref{fig:Ratio}.
The ratio between \Zmmjets and \phojets, \Zmmjets and \Wmvjets and the one between \Wmvjets/\phojets processes as a function of the boson \pt is
also studied and the results can be seen in Fig.~\ref{fig:Ratio_pt} in Section~\ref{sec:app}.

\begin{figure*}[hbtp]\begin{center}
\includegraphics[width=\cmsFigWidth]{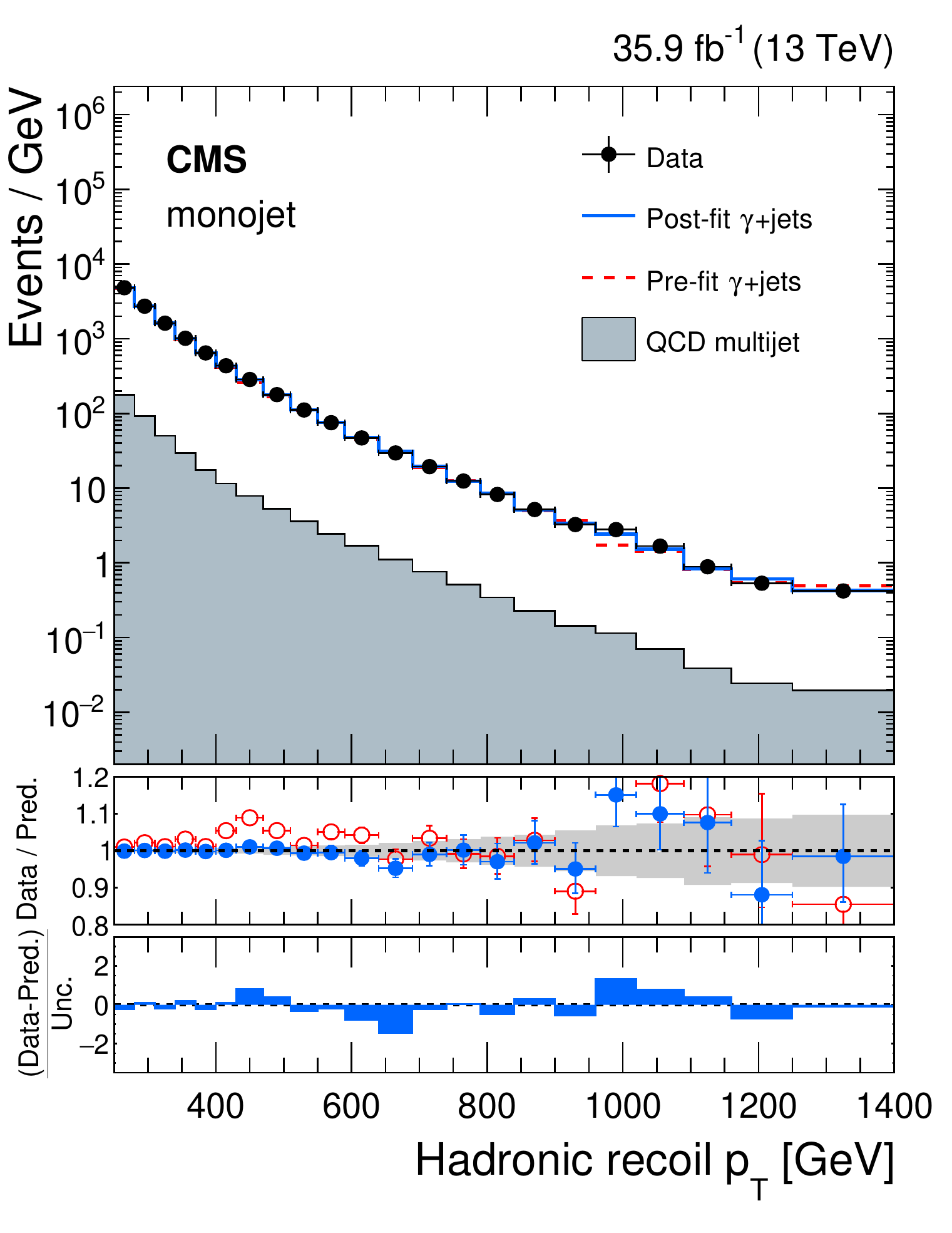}\hfil
\includegraphics[width=\cmsFigWidth]{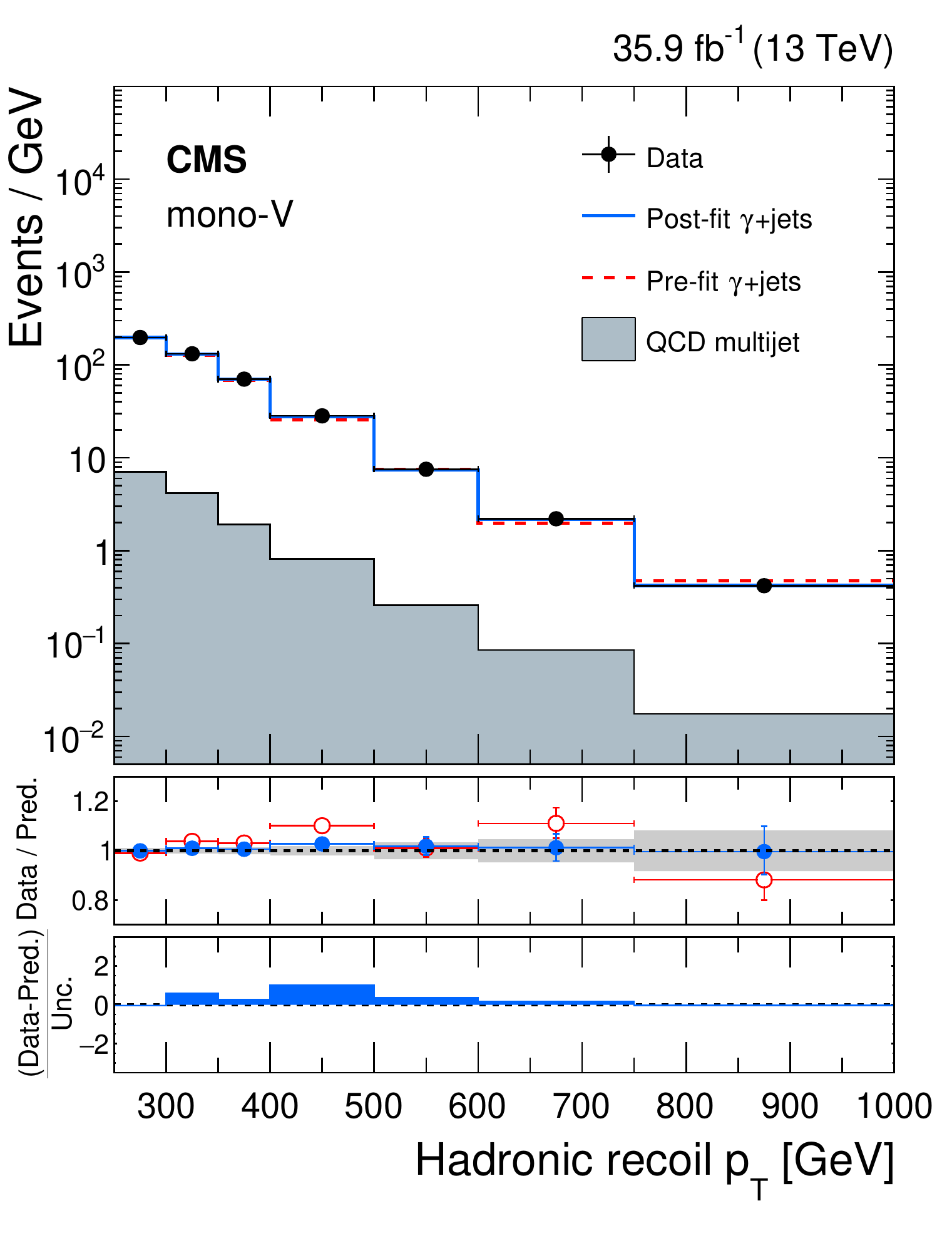}
\caption{
Comparison between data and MC simulation in the \phojets control sample
before and after performing the simultaneous fit across all the control samples and the signal region
assuming the absence of any signal. The left plot shows the monojet category and the right plot shows the
mono-$\PV$ category. The hadronic recoil \pt in \phojets events is used as a proxy for \ptmiss in the signal region.
The last bin includes all events with hadronic recoil \pt larger than 1250 (750)\GeV in the monojet (mono-$\PV$) category.
In the lower panels, ratios of data with the pre-fit background
prediction (red open points) and post-fit background
prediction (blue full points) are shown for both the monojet and mono-$\PV$ categories.
The gray band in the lower panel indicates the post-fit uncertainty
after combining all the systematic uncertainties. Finally, the distribution of the pulls, defined as the
difference between data and the post-fit background prediction relative to the quadrature sum of the
post-fit uncertainty in the prediction and statistical uncertainty in data, is shown in the lowest panel.
}
\label{fig:gamCR}\end{center}\end{figure*}

\begin{figure*}[hbtp]\begin{center}
\includegraphics[width=\cmsFigWidth]{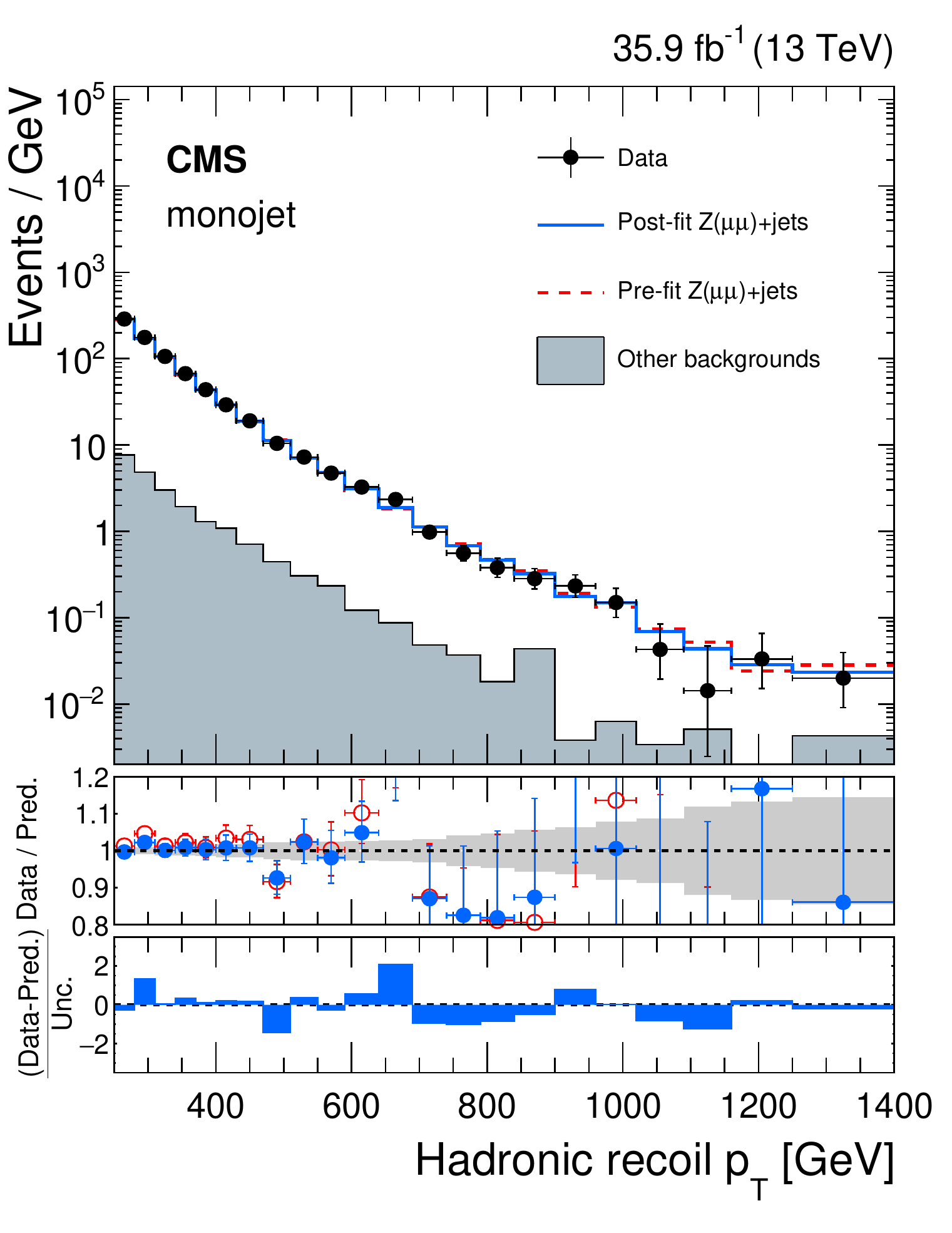}\hfil
\includegraphics[width=\cmsFigWidth]{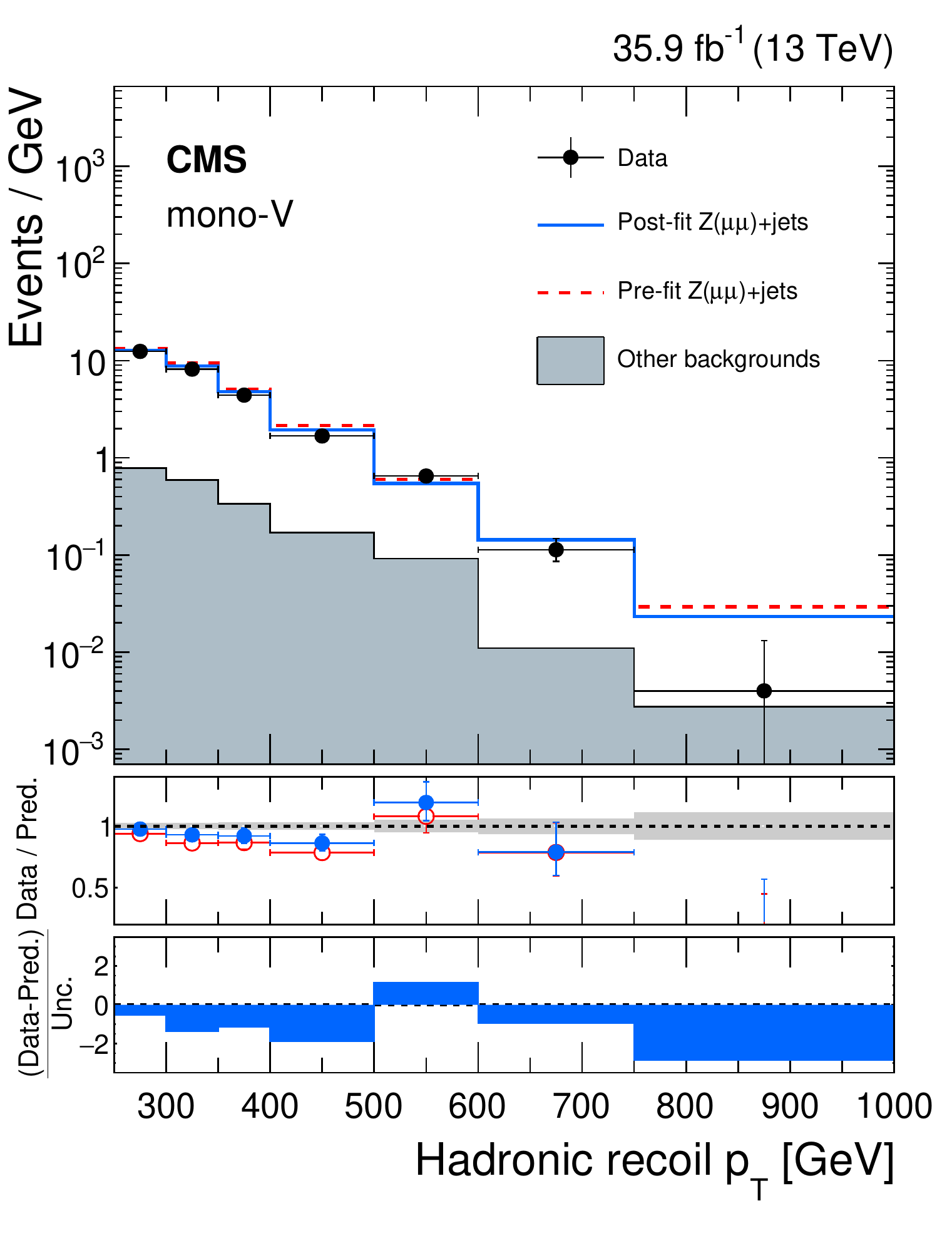}
\\
\includegraphics[width=\cmsFigWidth]{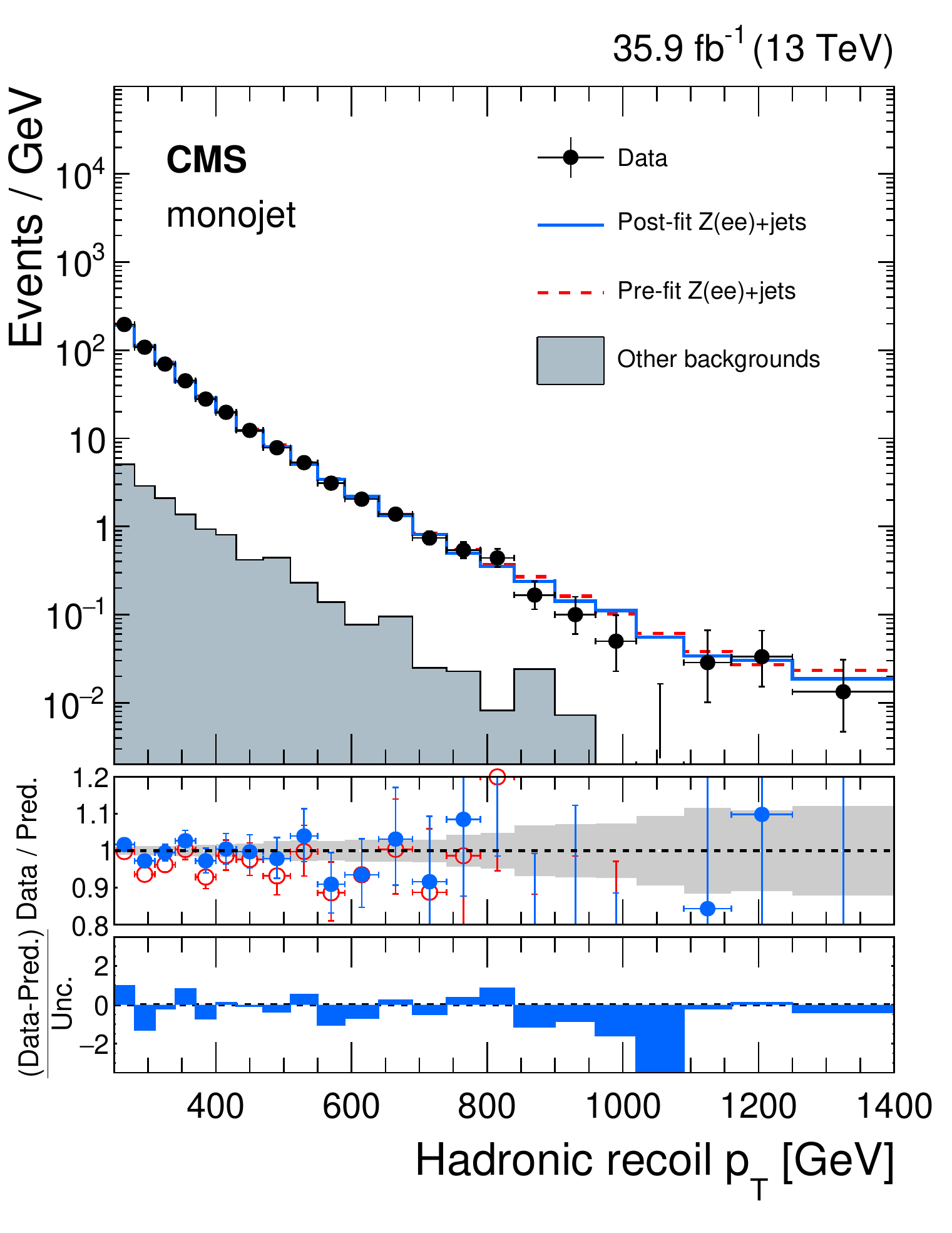}\hfil
\includegraphics[width=\cmsFigWidth]{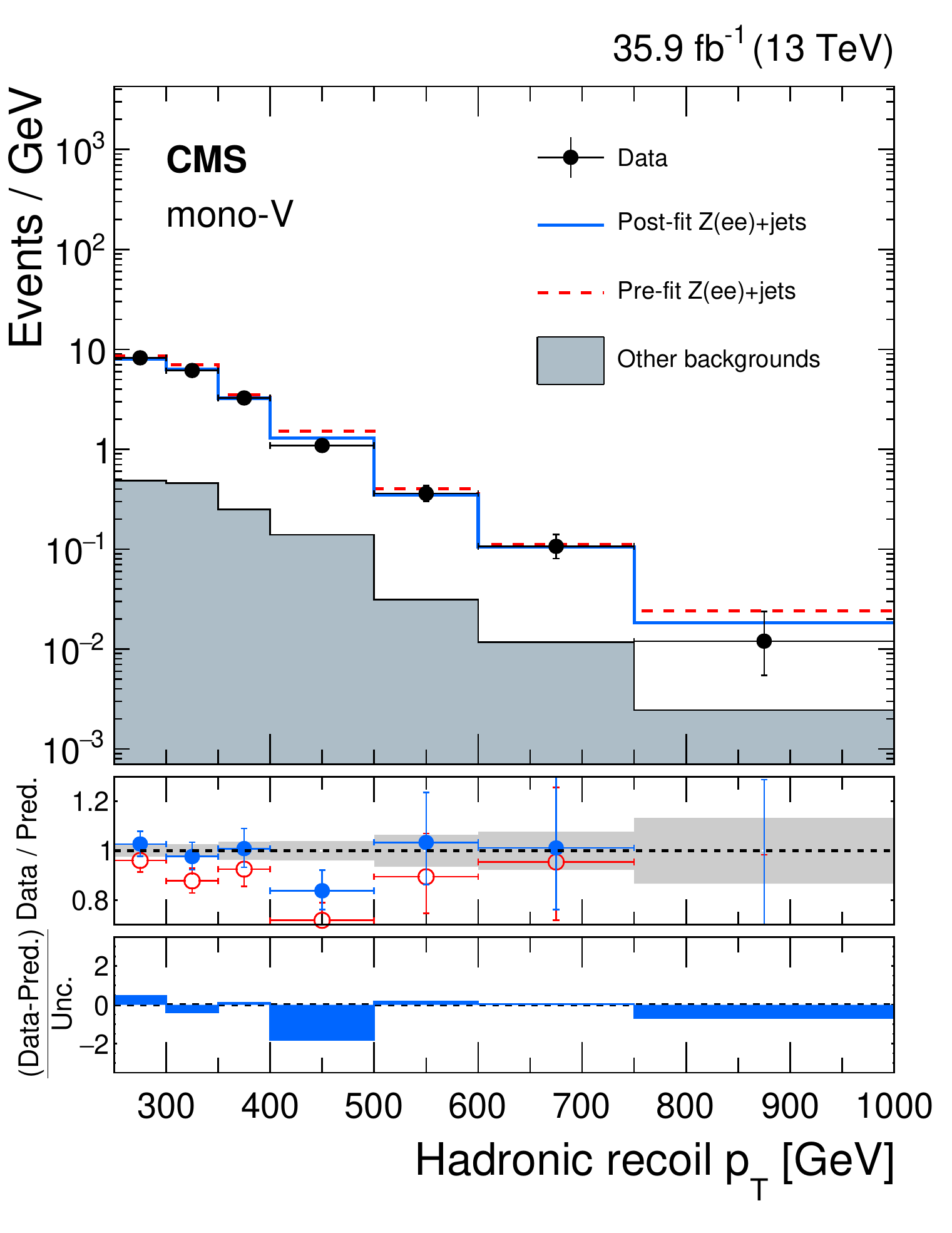}
\caption{
Comparison between data and MC simulation in the  dimuon (upper row) and dielectron  (lower row)
control samples before and after
performing the simultaneous fit across all the control samples and the signal region
assuming the absence of any signal. Plots correspond to the monojet (left) and mono-$\PV$ (right)
categories, respectively, in the dilepton control sample.
The hadronic recoil \pt in dilepton events is used as a proxy for \ptmiss in the signal region.
The other backgrounds include top quark, diboson, and \Wjets processes.
The description of the lower panels is the same as in Fig.~\ref{fig:gamCR}.
}
\label{fig:zmmCR}\end{center}\end{figure*}

\begin{figure*}[hbtp]\begin{center}
\includegraphics[width=\cmsFigWidth]{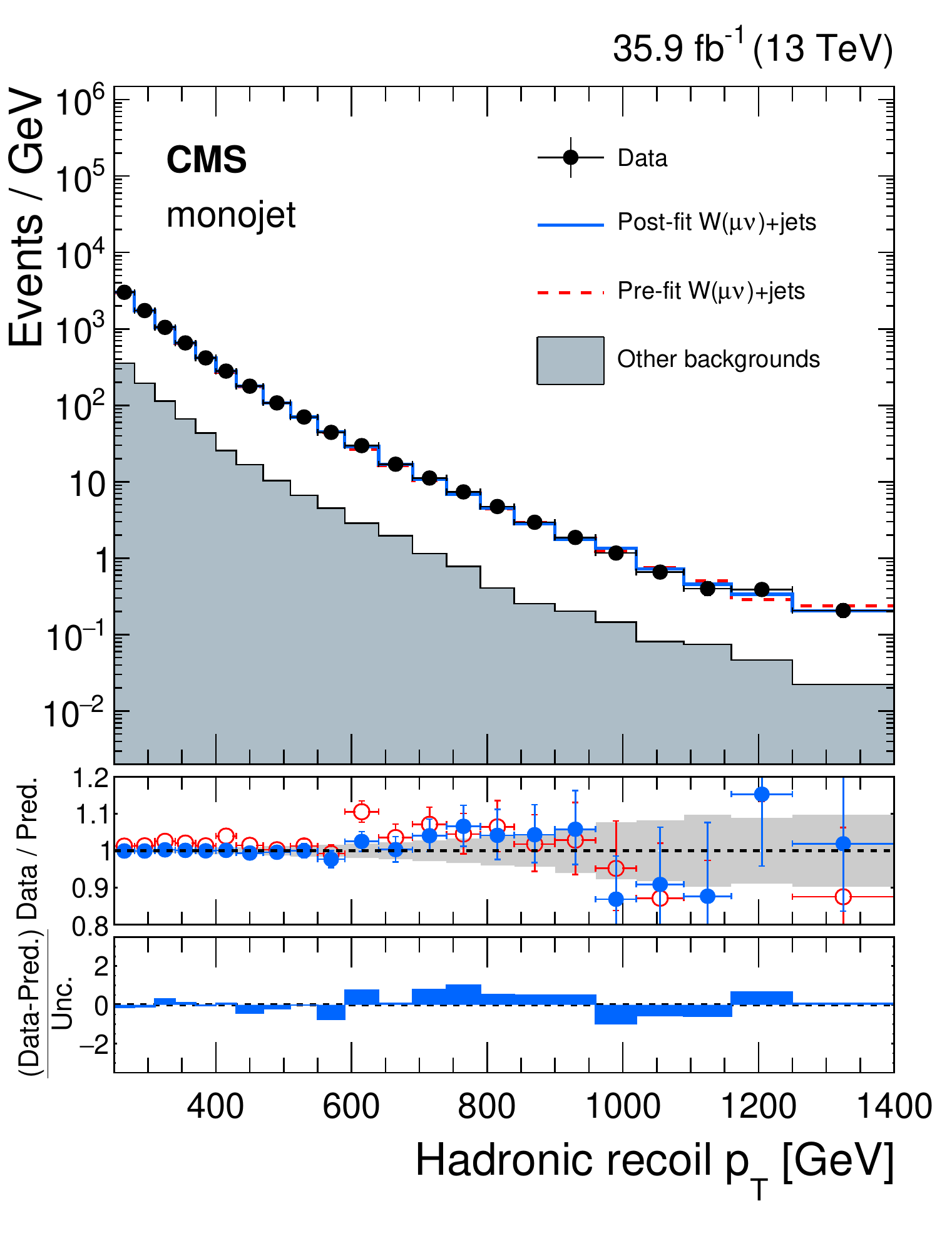}\hfil
\includegraphics[width=\cmsFigWidth]{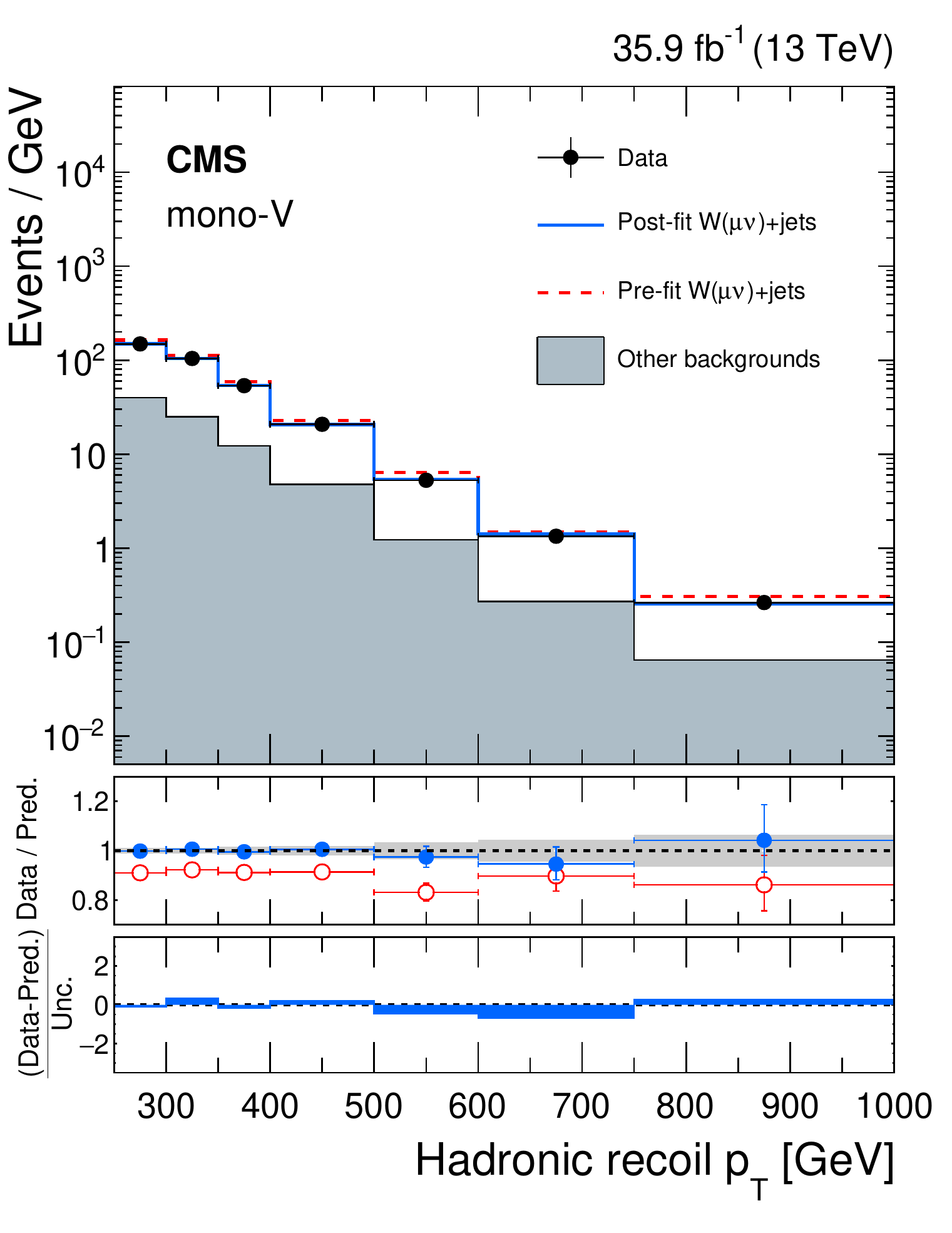}
\\
\includegraphics[width=\cmsFigWidth]{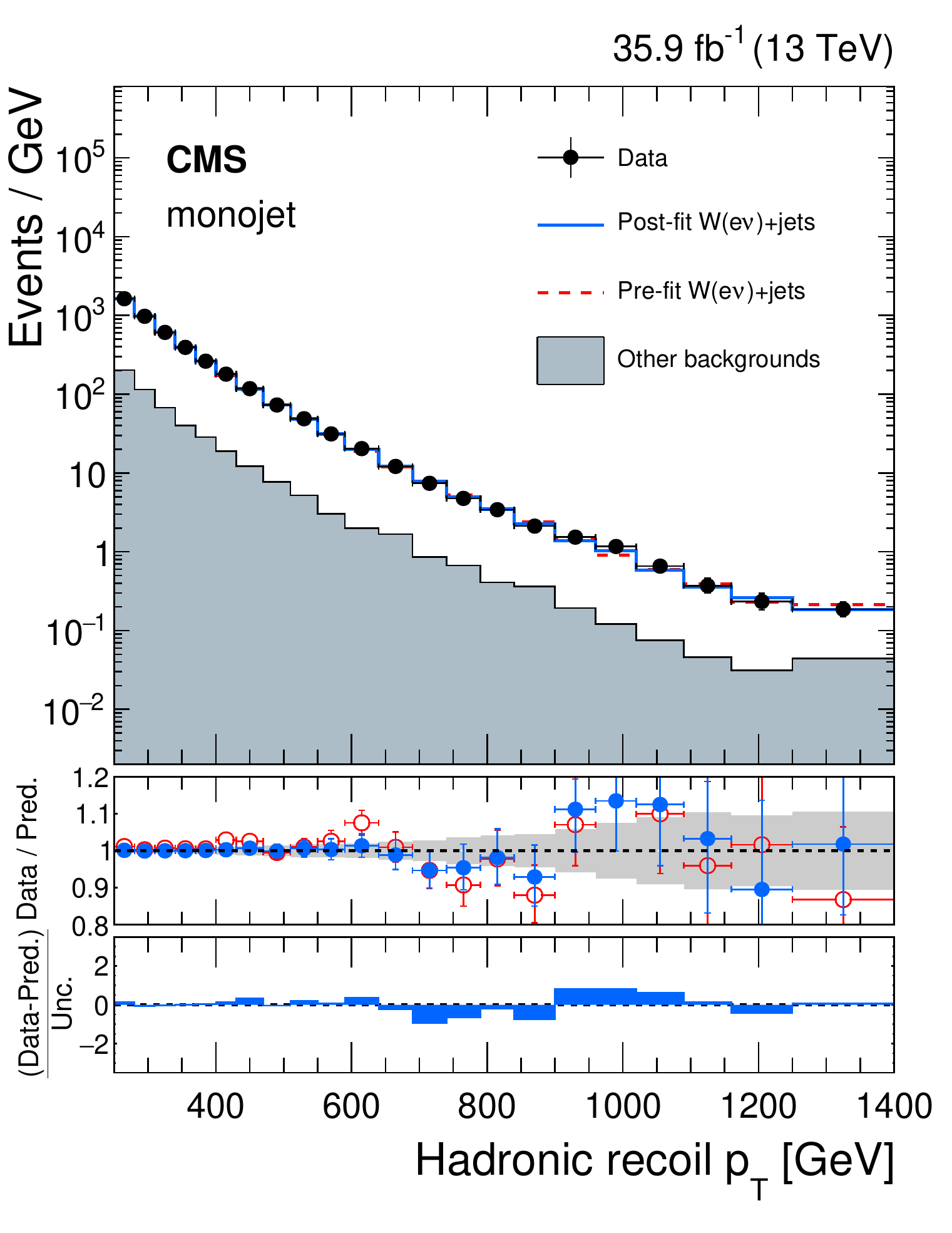}\hfil
\includegraphics[width=\cmsFigWidth]{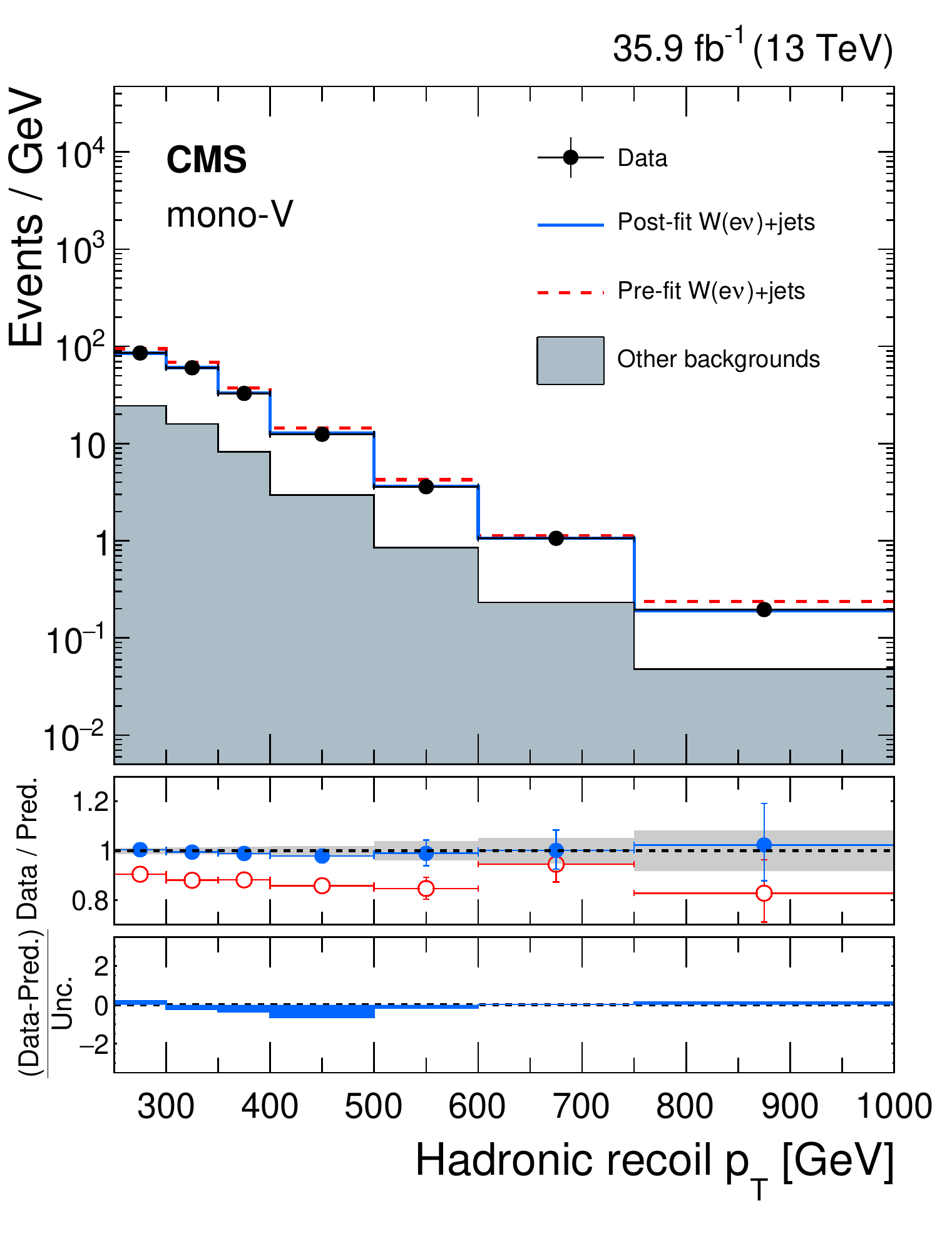}
\caption{
Comparison between data and MC simulation in the single-muon (upper row) and single-electron (lower row)
control samples before and
after performing the simultaneous fit across all the control samples and the signal region
assuming the absence of any signal. Plots correspond to the monojet (left) and mono-$\PV$ (right) categories,
respectively, in the single-lepton control samples.
The hadronic recoil \pt in single-lepton events is used as a proxy for \ptmiss in the signal region.
The other backgrounds include top quark, diboson, and QCD multijet processes.
The description of the lower panels is the same as in Fig.~\ref{fig:gamCR}.
}
\label{fig:wmnCR}\end{center}\end{figure*}

Figures~\ref{fig:gamCR}--\ref{fig:wmnCR} show the results of the combined fit in all control samples and the signal region.
Data in the control samples are compared to the pre-fit predictions from simulation and
the post-fit estimates obtained after performing the fit.
The control samples with larger yields dominate the fit results.
A normalization difference of 7\% is observed in the pre-fit distributions for the mono-V category in the single-lepton and dilepton control regions. The sources of the differences are identified to be the modeling of the pruned mass variable and the large theoretical uncertainties in the diboson and top quark backgrounds, which are the leading backgrounds in these regions. The normalization difference is found to be fully mitigated by the fitting procedure.

\section{Results and interpretation}\label{sec:results}

\begin{figure*}[!hbt]
\begin{center}
\includegraphics[width=\cmsFigWidth]{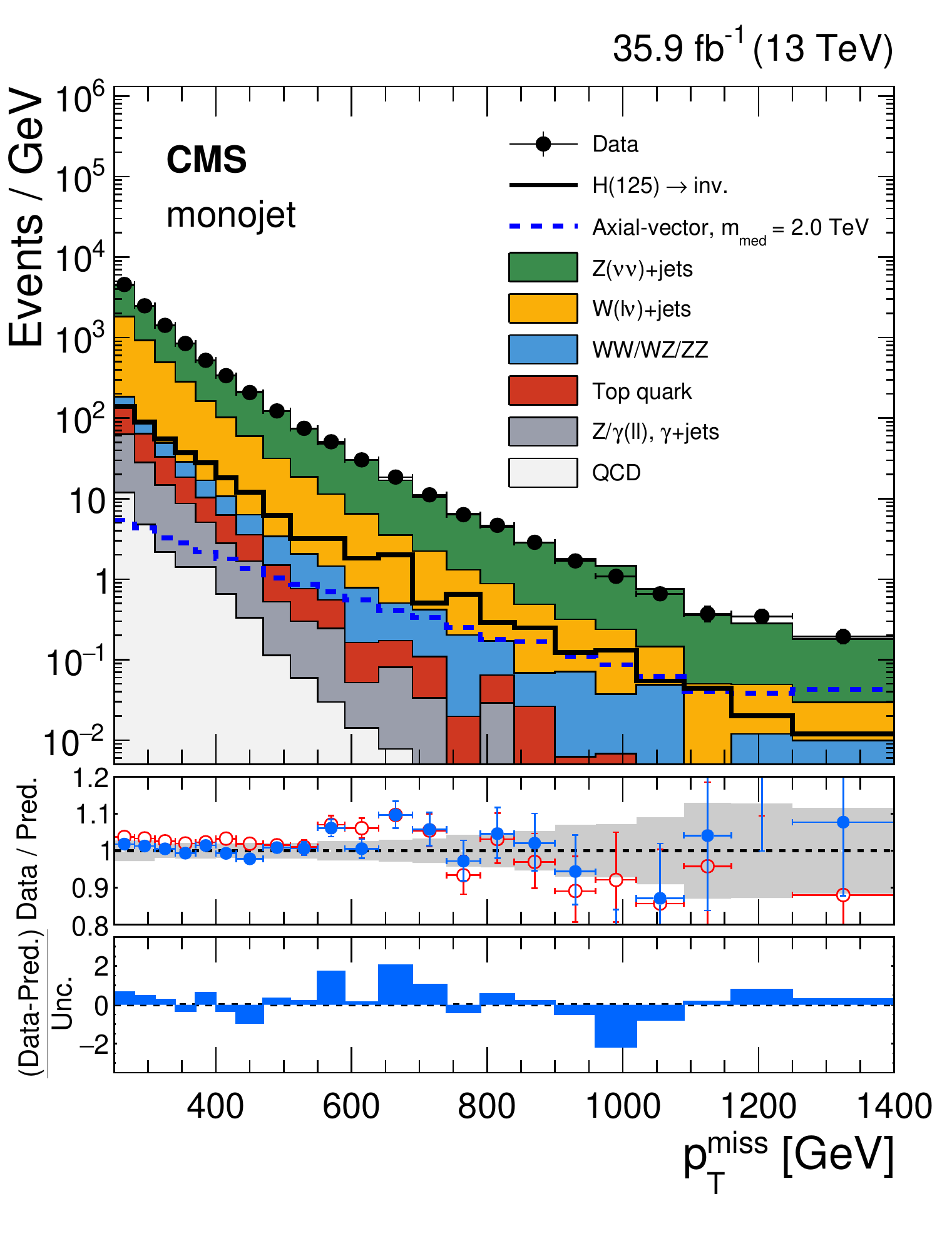}\hfil
\includegraphics[width=\cmsFigWidth]{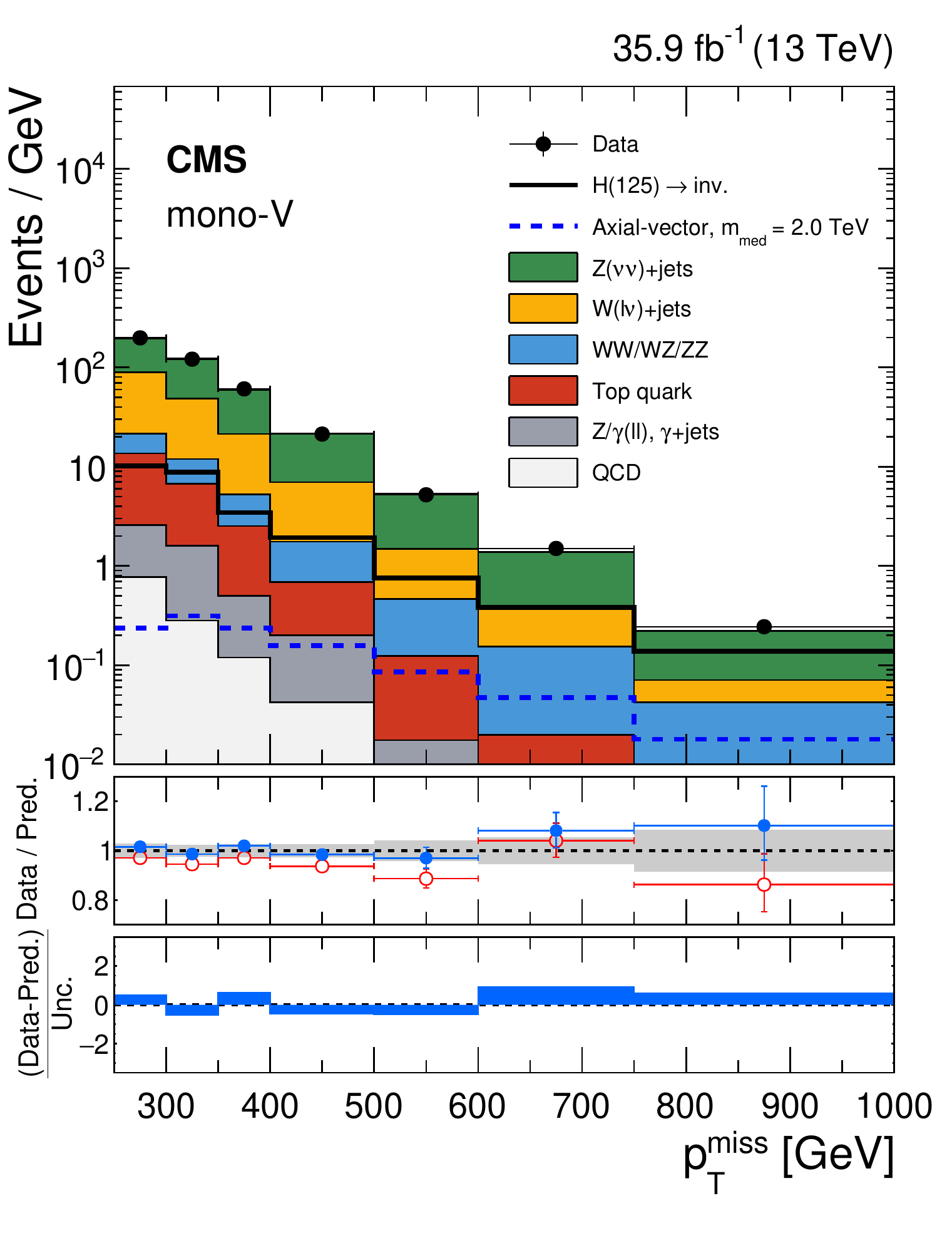}
\caption{
Observed \ptmiss distribution in the monojet (left) and mono-$\PV$ (right)
signal regions compared with the post-fit background expectations for various SM processes.
The last bin includes all events with $\ptmiss > 1250\, (750)\GeV$ for the monojet (mono-$\PV$) category.
The expected background distributions are evaluated after performing a
combined fit to the data in all the control samples, not including the signal region.
Expected signal distributions for the 125\GeV Higgs boson decaying exclusively to
invisible particles, and a 2\TeV axial-vector mediator decaying to 1\GeV DM particles, are overlaid.
The description of the lower panels is the same as in Fig.~\ref{fig:gamCR}.
}
\label{fig:moneyplots_SRmask}
\end{center}
\end{figure*}

The search is performed by extracting the signal through a combined fit of the signal
and control regions.
Figure~\ref{fig:moneyplots_SRmask} shows the comparison between data and the post-fit
background predictions in the signal region in the monojet and mono-$\PV$ categories, where
the background prediction is obtained from a combined fit performed in all control regions,
excluding the signal region.
Expected signal distributions for the 125\GeV Higgs boson decaying exclusively
to invisible particles, and a 2\TeV axial-vector mediator decaying to 1\GeV DM particles,
are overlaid. Data are found to be in agreement with the SM prediction.

The expected yields in each bin of \ptmiss for all SM backgrounds, after the fit to the data in the
control regions, are given in Tables~\ref{tab:yield_monojet_SR_mask} and~\ref{tab:yield_monov_SR_mask}
for the monojet and mono-$\PV$ signal regions, respectively. The correlations between the predicted background
yields across all the \ptmiss bins in the two signal regions are shown in Figs.~\ref{fig:correlation_matrix_monojet}
and~\ref{fig:correlation_matrix_monov} in Section~\ref{sec:app}. The expected yields together with the correlations can
be used with the simplified likelihood approach detailed in Ref.~\cite{CMS-NOTE-2017-001} to reinterpret the results
for models not studied in this paper.

\begin{table*}[htb]
\topcaption{Expected event yields in each \ptmiss bin for various background processes in the monojet signal region. The background yields and the corresponding uncertainties are obtained after performing a combined fit to data in all the control samples, excluding data in the signal region. The other backgrounds include QCD multijet and \phojets processes.The expected signal contribution for a 2\TeV axial-vector mediator decaying to 1\GeV DM particles and the observed event yields in the monojet signal region are also reported.}
\begin{center}
\renewcommand{\arraystretch}{1}
\ifthenelse{\boolean{cms@external}}{\footnotesize}{\resizebox{\textwidth}{!}}
{
\begin{scotch}{c*{7}{.}c}
$\ptmiss$ (GeV) &
\multicolumn{1}{c}{Signal} &
\multicolumn{1}{c}{$\PZ(\nu\nu)$+jets} &
\multicolumn{1}{c}{$\PW(\ell\nu)$+jets} &
\multicolumn{1}{c}{Top quark} &
\multicolumn{1}{c}{Diboson} &
\multicolumn{1}{c}{Other} &
\multicolumn{1}{c}{Total bkg.} &
Data \\
\hline
250-280   & 162 , 3       & 79700 , 2300  & 49200 , 1400  & 2360 , 200 & 1380 , 220    & 1890	, 240  & 134500 , 3700   & 136865     \\
280-310   & 130 , 3       & 45800 , 1300  & 24950 , 730   & 1184 , 99  & 770 , 120    & 840 , 110  & 73400 , 2000   & 74340      \\
310-340   & 97.8 , 2.4     & 27480 , 560   & 13380 , 260   & 551 , 53  & 469	 , 77     & 445	, 63   & 42320 , 810	 & 42540       \\
340-370   & 84.8 , 2.1     & 17020 , 350   & 7610 , 150   & 292 , 28  & 301	 , 51     & 260	, 39   & 25490 , 490	 & 25316       \\
370-400   & 65.2 , 1.9     & 10560 , 220   & 4361 , 91    & 157 , 17  & 198	 , 33     & 152	, 26   & 15430 , 310	 & 15653       \\
400-430   & 53.5 , 1.8     & 7110 , 130   & 2730 , 47    & 104 , 12  & 133	 , 23     & 84	, 15   & 10160 , 170	 & 10092	\\
430-470   & 53.9 , 1.8     & 6110 , 100   & 2123 , 37    & 75.2 , 7.9 & 110	 , 19     & 67	, 11   & 8480 , 140	 & 8298 	\\
470-510   & 41.4 , 1.5     & 3601 , 75    & 1128 , 22    & 38.6 , 5.3	& 75	 , 12     & 21.0	, 3.9  & 4865 , 95	 & 4906 	\\
510-550   & 34.3 , 1.4     & 2229 , 39    & 658 , 12    & 18.5 , 3.3	& 51.7 , 9.5    & 12	, 2.4  & 2970 , 49	 & 2987 	\\
550-590   & 28.1 , 1.2     & 1458 , 27    & 398 , 8     & 12.3 , 2.6	& 35.9 , 7.1    & 9.7	, 1.9  & 1915 , 33	 & 2032 	\\
590-640   & 27.5 , 1.2     & 1182 , 26    & 284 , 7     & 5.5 , 1.4	& 30.9 , 5.7    & 2.6	, 0.7  & 1506 , 32	 & 1514 	\\
640-690   & 20.4 , 1.1     & 667 , 15    & 151 , 4     & 4.6 , 1.7	& 16.7 , 3.9    & 4.0	, 0.8  & 844 , 18	 & 926         \\
690-740   & 16.6 , 0.9     & 415 , 12    & 90.4 , 3.0   & 3.8 , 1.5	& 15.6 , 3.6    & 1.7	, 0.4  & 526 , 14	 & 557  	\\
740-790   & 12.5 , 0.8     & 259 , 9.6   & 55.2 , 2.3   & 0.8 , 0.5 & 9.14 , 2.3    & 0.2	, 0.1  & 325 , 12	 & 316  	\\
790-840   & 8.94 , 0.72    & 178 , 7.1   & 35.3 , 1.7   & 1.7 , 0.8 & 5.35 , 1.7    & 1.4	, 0.3  & 223 , 9	 & 233  	\\
840-900   & 10.1 , 0.7     & 139 , 6.2   & 25.2 , 1.3   & 1.5 , 1.2 & 2.52 , 1.05   & 0.04	, 0.03 & 169 , 8	 & 172  	\\
900-960   & 6.62 , 0.61    & 88.1 , 4.9   & 14.7 , 0.9   & 0.3 , 0.3 & 3.88 , 1.42   & 0.03	, 0.02 & 107 , 6	 & 101  	\\
\x960-1020  & 5.19 , 0.54    & 73.8 , 4.7   & 12.0 , 0.8   & 0.4 , 0.3 & 1.83 , 0.92   & 0.02	, 0.01 & 88.1 , 5.3	 & 65		 \\
1020-1090 & 4.35 , 0.52    & 42.6 , 3.1   & 6.7 , 0.6   & 0.0 , 0.0 & 3.42 , 1.33   & 0.01	, 0.01 & 52.8 , 3.9	 & 46		 \\
1090-1160 & 2.84 , 0.43    & 21.5 , 2.1   & 3.5 , 0.4   & 0.0 , 0.0 & 0.00 , 0.00   & 0.01	, 0.00 & 25.0 , 2.5	 & 26		 \\
1160-1250 & 3.44 , 0.38    & 21.0 , 2.2   & 3.3 , 0.4   & 0.0 , 0.0 & 1.07 , 0.69   & 0.01	, 0.00 & 25.5 , 2.6	 & 31		 \\
$>$1250  & 6.39 , 0.58    & 22.5 , 2.4   & 2.9 , 0.3   & 0.0 , 0.0 & 1.49 , 0.91   & 0.01	, 0.00 & 26.9 , 2.8	 & 29		 \\
\end{scotch}
}
\label{tab:yield_monojet_SR_mask}
\end{center}
\end{table*}

\begin{table*}[htb]
\topcaption{Expected event yields in each \ptmiss bin for various background processes in the mono-$\PV$ signal region. The background yields and the corresponding uncertainties are obtained after performing a combined fit to data in all the control samples, but excluding data in the signal region. The other backgrounds include QCD multijet and \phojets processes. The expected signal contribution for a 2\TeV axial-vector mediator decaying to 1\GeV DM particles and the observed event yields in the mono-$\PV$ signal region are also reported.}
\begin{center}
\renewcommand{\arraystretch}{1}
\ifthenelse{\boolean{cms@external}}{\footnotesize}{\resizebox{\textwidth}{!}}
{
\begin{scotch}{c*{7}{.}{c}}
$\ptmiss$ (GeV) &
\multicolumn{1}{c}{Signal} &
\multicolumn{1}{c}{$\PZ(\nu\nu)$+jets} &
\multicolumn{1}{c}{$\PW(\ell\nu)$+jets} &
\multicolumn{1}{c}{Top quark} &
\multicolumn{1}{c}{Diboson} &
\multicolumn{1}{c}{Other} &
\multicolumn{1}{c}{Total bkg.} &
Data \\
\hline
250-300      & 11.7 , 0.6   & 5300 , 170   & 3390 , 120    & 553 , 54   & 396 , 69  & 128 , 25	& 9770 , 290	& 9929    \\
300-350      & 15.7 , 0.7   & 3720 , 98	 & 1823 , 53     & 257 , 27   & 261 , 46  & 79.8 , 13	& 6140 , 140	& 6057    \\
350-400      & 11.8 , 0.6   & 1911 , 59	 & 808 , 28     & 101 , 12   & 134 , 25  & 25.0 , 4.8	& 2982 , 79	& 3041     \\
400-500      & 15.8 , 0.7   & 1468 , 45	 & 521 , 15     & 48.8 , 5.7  & 107 , 20  & 20.0 , 3.6	& 2165 , 55	& 2131      \\
500-600      & 8.59 , 0.56  & 388 , 18	 & 103.0 , 5.1    & 10.7 , 1.9  & 33.8 , 7.0 & 1.76 , 0.53	& 537	, 23	& 521        \\
600-750      & 7.04 , 0.47  & 151.0 , 9.9   & 33.4 , 2.3    & 1.9 , 1.1  & 20.2 , 4.5 & 1.05 , 0.25	& 208	, 11	& 225         \\
$>$750      & 4.48 , 0.40  & 37.7 , 3.7   & 7.09 , 0.69   & 0.28 , 0.25 & 10.2 , 2.3 & 0.06 , 0.03	& 55.3 , 4.6	& 61          \\
\end{scotch}
}
\label{tab:yield_monov_SR_mask}
\end{center}
\end{table*}

\begin{figure*}[!htb]
\begin{center}
\includegraphics[width=\cmsFigWidth]{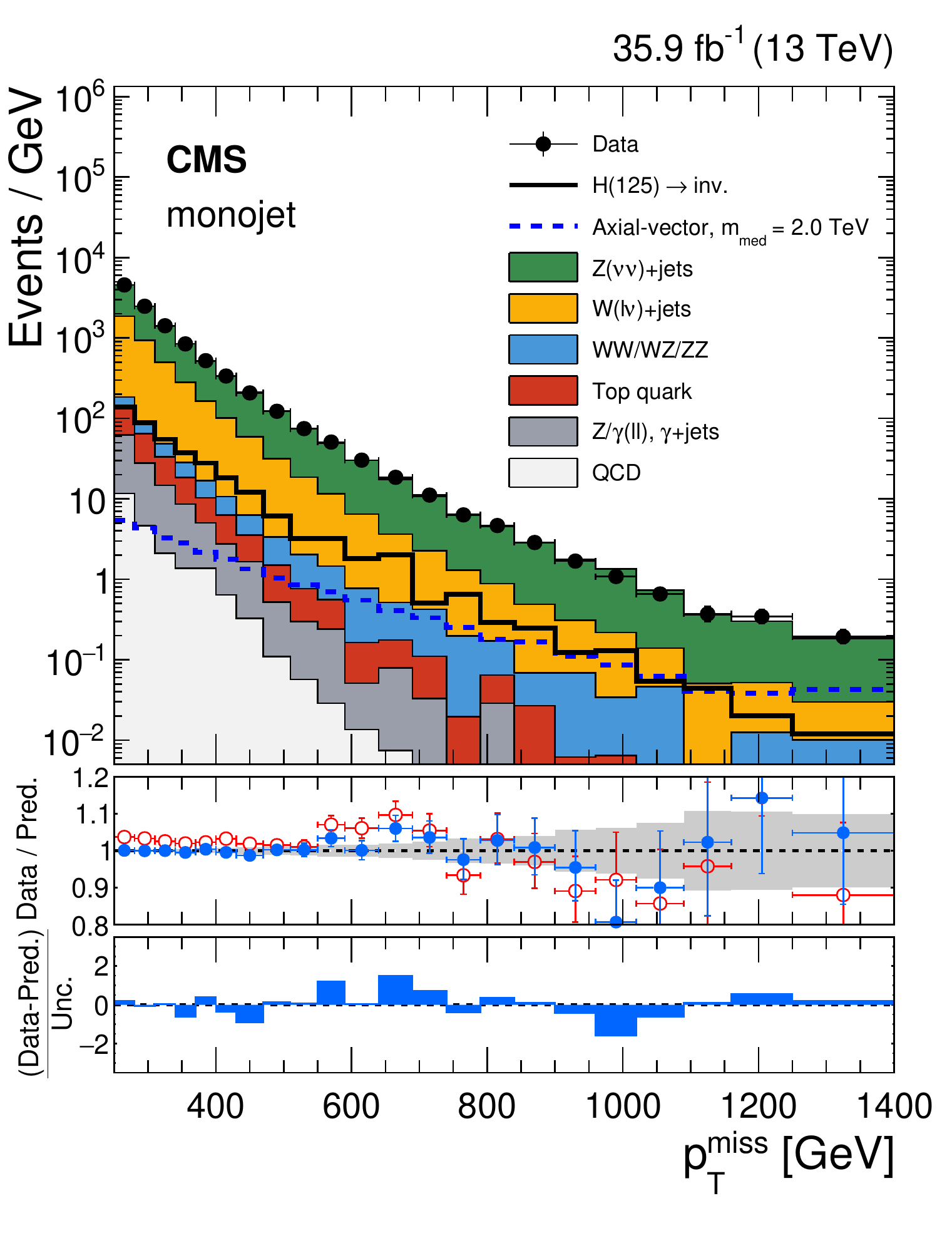}\hfil
\includegraphics[width=\cmsFigWidth]{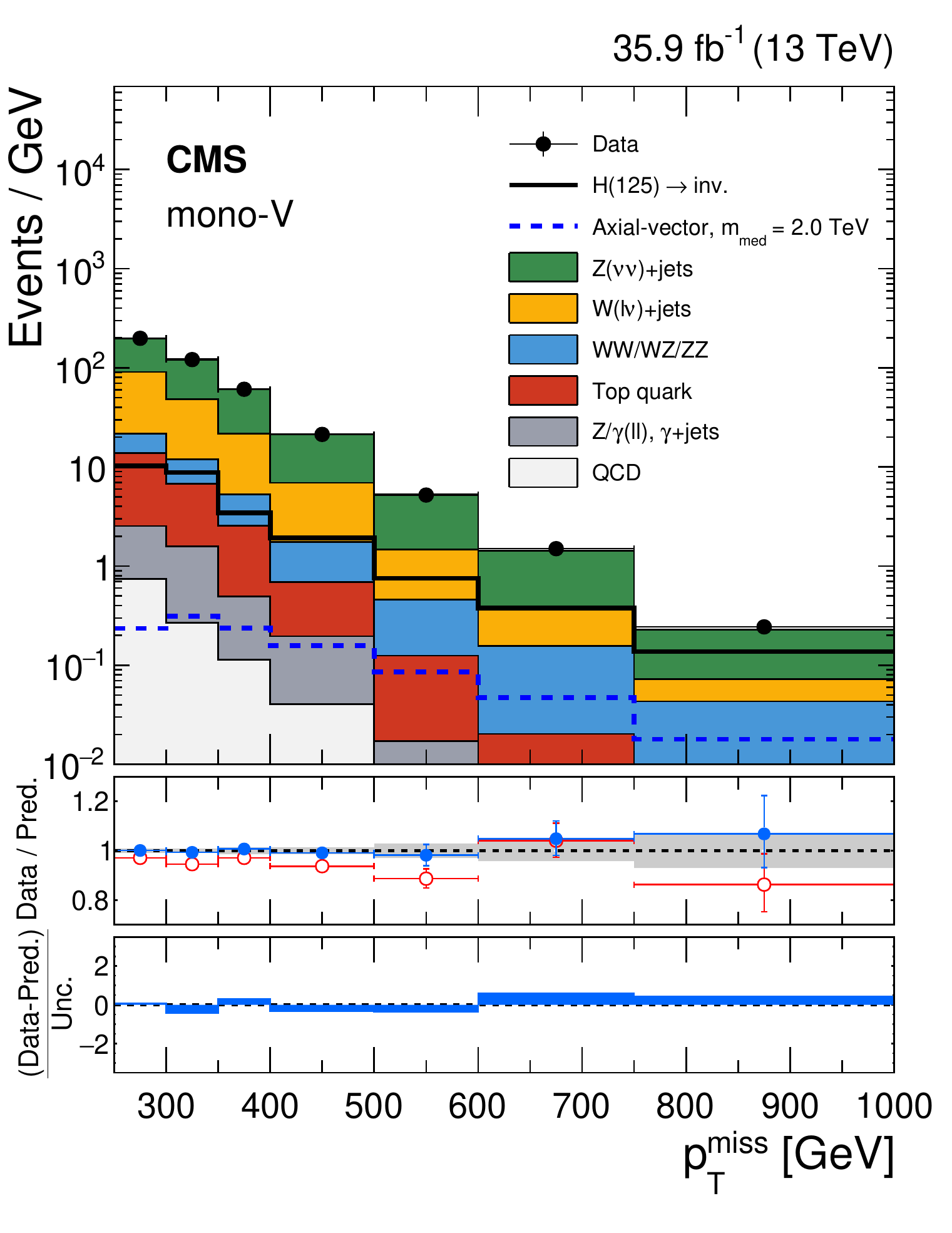}
\caption{
Observed \ptmiss distribution in the monojet (left) and mono-$\PV$ (right) signal regions
compared with the post-fit background expectations for various SM processes.
The last bin includes all events with $\ptmiss >  1250\, (750)\GeV$ for the monojet (mono-$\PV$) category.
The expected background distributions are evaluated after performing a combined
fit to the data in all the control samples, as well as in the signal region.
The fit is performed assuming the absence of any signal. Expected signal distributions
for the 125\GeV Higgs boson decaying exclusively to invisible particles, and a 2\TeV
axial-vector mediator decaying to 1\GeV DM particles, are overlaid.
The description of the lower panels is the same as in Fig.~\ref{fig:gamCR}.
}
\label{fig:moneyplots}
\end{center}
\end{figure*}

Figure~\ref{fig:moneyplots} shows a comparison between data and the post-fit background
predictions in the signal region in the monojet and mono-$\PV$ categories, where the fit
is performed under the background-only hypothesis including signal region events
in the likelihood. The limits on the production cross section of the various models
described below is set after comparing this fit with an alternative one assuming
the presence of signal.

\subsection{Dark matter interpretation}

The results are interpreted in terms of simplified $s$-channel DM models assuming a  vector, axial-vector,
scalar, or pseudoscalar mediator decaying into a pair of fermionic DM particles.
The coupling of the mediators to the DM is assumed to be unity for all four types of mediators.
The spin-0 particles are assumed to couple to the quarks with a coupling strength ($g_{\mathrm{q}}$) of 1.
In the case of the spin-1 mediators, $g_{\mathrm{q}}$ is taken to be 0.25. The choice of all the signal model parameters follows the recommendations
from Ref.~\cite{Abercrombie:2015wmb}. Uncertainties of 20 and 30\% are assigned to the inclusive
signal cross section in the case of the spin-1 and spin-0 mediators, respectively.
These estimates include the renormalization and factorization scale uncertainties, as well as the PDF uncertainty.

\begin{figure*}[hbtp]
\begin{center}
\includegraphics[width=\cmsFigWidth]{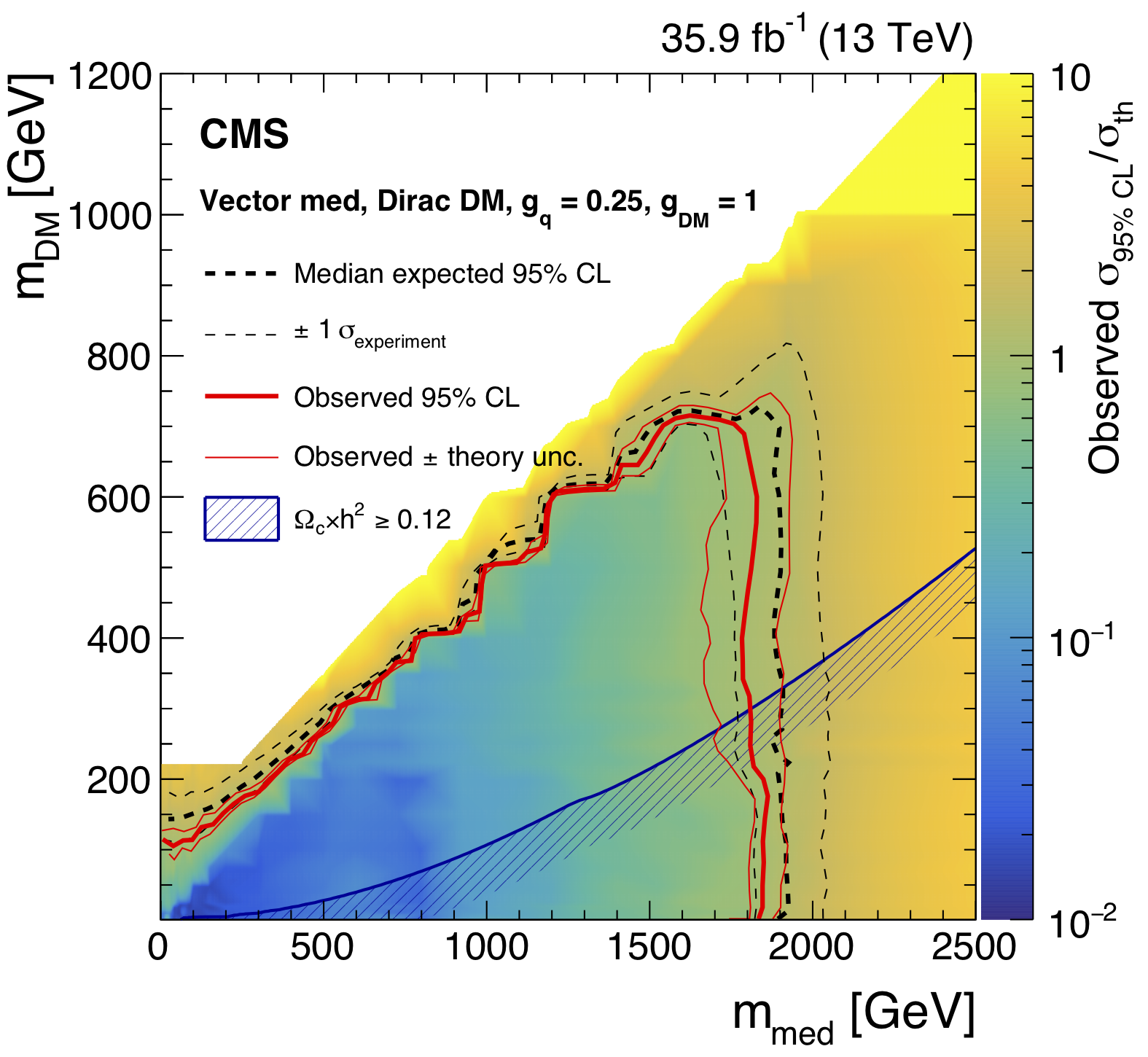}\hfil
\includegraphics[width=\cmsFigWidth]{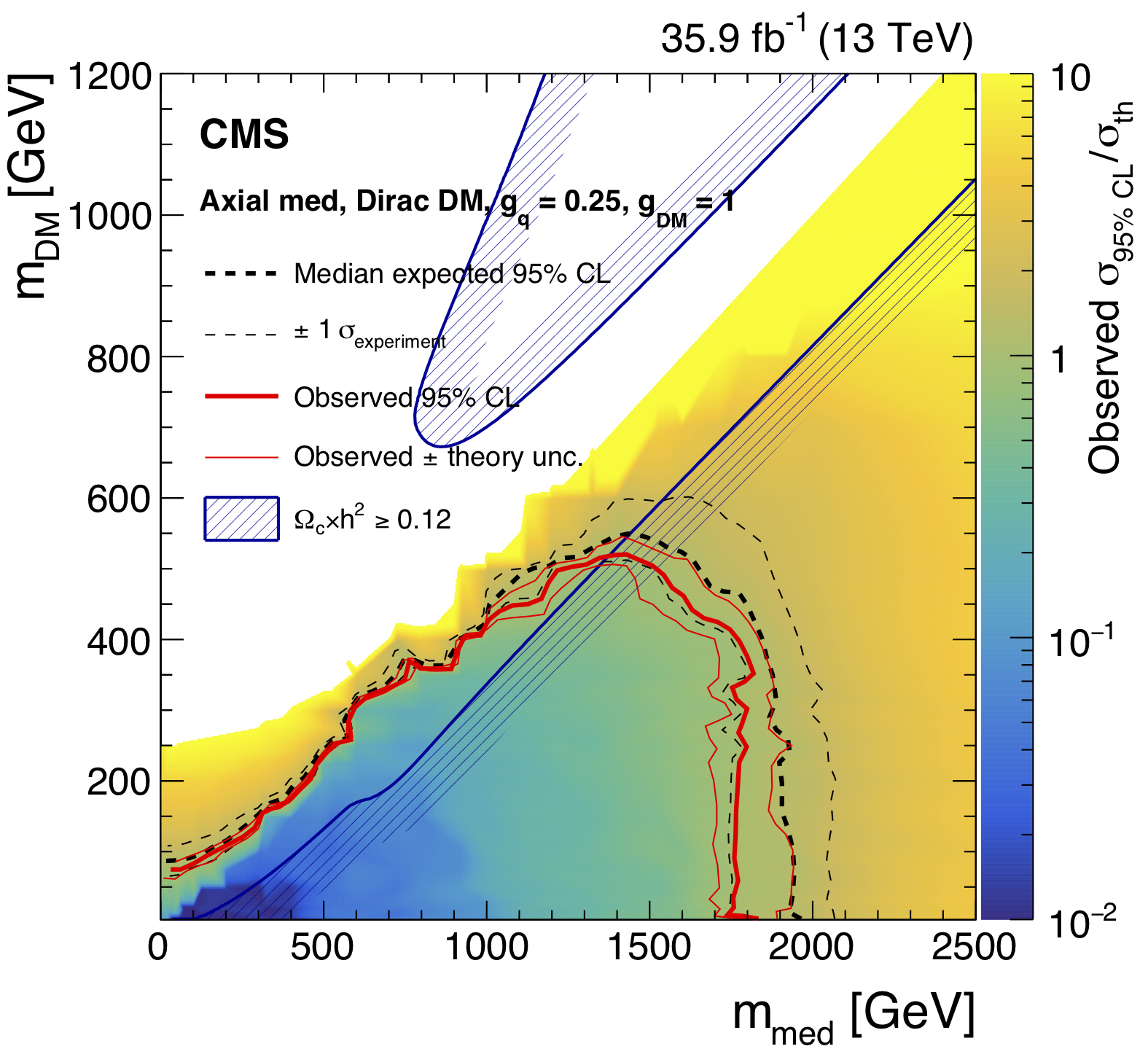}
\caption{
Exclusion limits at 95\%\,\CL on $\mu=\sigma/\sigma_{\textrm{th}}$ in the
$m_{\textrm{med}}$-$m_{\textrm{DM}}$ plane assuming vector (left) and axial-vector (right) mediators.
The solid (dotted) red (black) line shows the contour for the observed (expected) exclusion.
The solid contours around the observed limit and the dashed contours
around the expected limit represent one standard deviation due to theoretical uncertainties
in the signal cross section and the combination of the statistical
and experimental systematic uncertainties, respectively. Constraints from the
Planck satellite
experiment~\cite{Ade:2015xua} are shown as dark blue contours;
in the shaded area DM is overabundant.
}
\label{fig:scan_spin1}
\end{center}
\end{figure*}

\begin{figure*}[!h]
\begin{center}
\includegraphics[width=\cmsFigWidth]{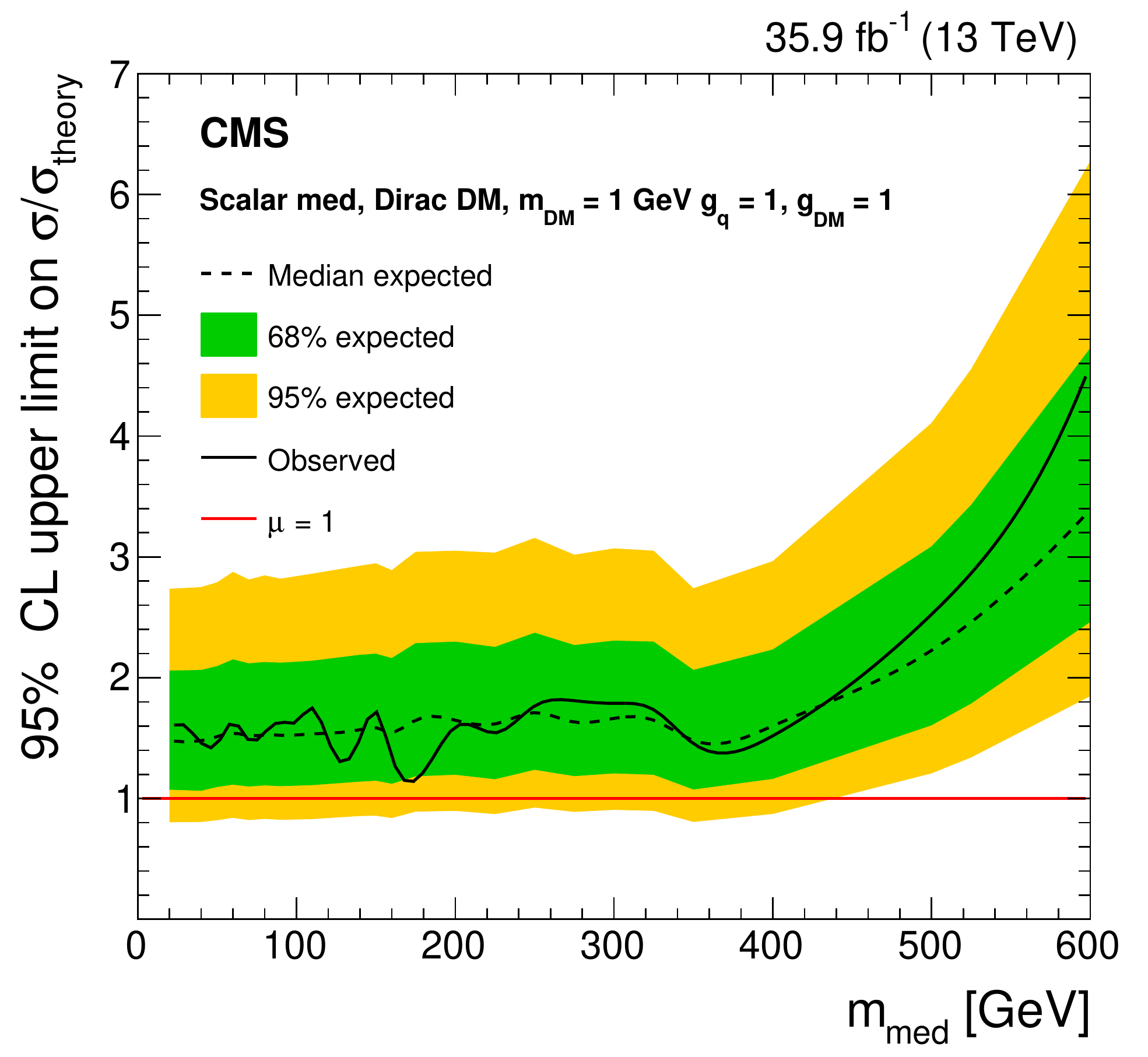}\hfil
\includegraphics[width=\cmsFigWidth]{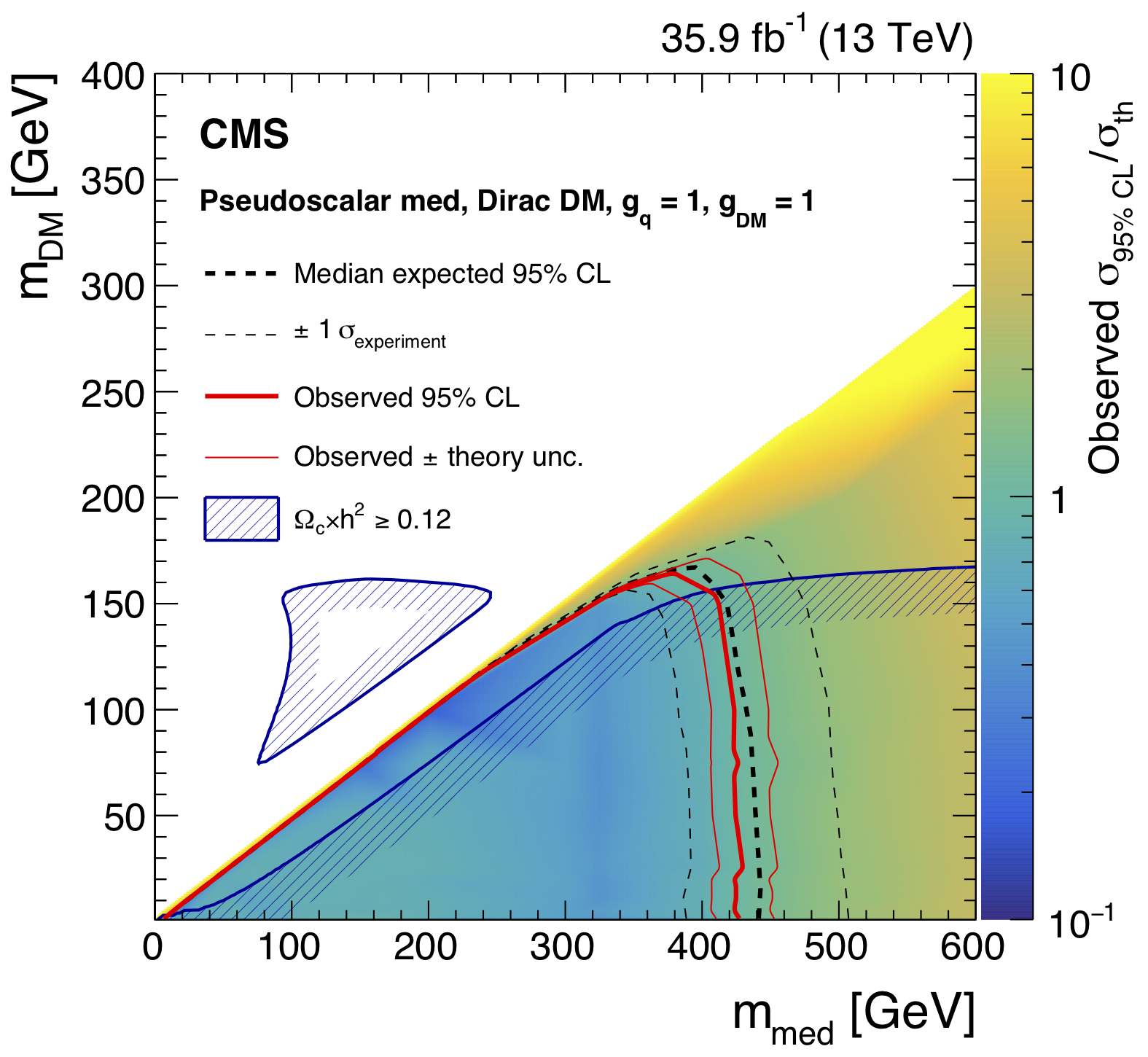}
\caption{
Expected (dotted black line) and observed (solid black line) 95\%\,\CL upper limits on the
signal strength $\mu=\sigma/\sigma_{\textrm{th}}$  as a function of the mediator mass
for the scalar mediators (left) for $m_{\textrm{DM}}=1\GeV$. The horizontal red line denotes $\mu = 1 $.
Exclusion limits at 95\%\,\CL on $\mu=\sigma/\sigma_{\textrm{th}}$ in the
$m_{\textrm{med}}$-$m_{\textrm{DM}}$ plane assuming pseudoscalar mediators (right).
The solid (dashed) red (back) line shows the contours for the observed (expected) exclusion.
Constraints from the Planck satellite experiment~\cite{Ade:2015xua} are shown with the
dark blue contours; in the shaded area DM is overabundant.
}
\label{fig:scan_spin0}
\end{center}
\end{figure*}

Upper limits are computed at 95\%\,\CL on the ratio of the measured signal cross section to the predicted
one, denoted by $\mu=\sigma/\sigma_{\textrm{th}}$, with the \CL$_{\mathrm{s}}$
method~\cite{Junk:1999kv,Read:2002av}, using the asymptotic approximation~\cite{Cowan:2010js}.
Limits are obtained as a function of the mediator mass ($m_{\textrm{med}}$) and the DM mass ($m_{\textrm{DM}}$).
Figure~\ref{fig:scan_spin1} shows the exclusion contours in the $m_{\textrm{med}}$-$m_{\textrm{DM}}$ plane
for the vector and axial-vector mediators.
Mediator masses up to 1.8\TeV, and DM masses up to 700 and 500\GeV are excluded for the vector and axial-vector models,
respectively.
Figure~\ref{fig:scan_spin0} shows the limits for the scalar mediators as a function of the mediator
mass, for a fixed DM mass of 1\GeV and the exclusion contours in the $m_{\textrm{med}}$-$m_{\textrm{DM}}$ plane for
pseudoscalar mediators, respectively. Pseudoscalar mediator (dark matter) masses up to 400 (150)\GeV are excluded at 95\%\,\CL.
A direct comparison of the results for simplified DM models of this paper,
to the one presented in Ref.~\cite{paper-exo-037}
can be seen in Fig.~\ref{fig:scan_comparision} and Fig.~\ref{fig:scan_spin1_dmf} in Section~\ref{sec:app}.

The results for vector, axial-vector, and pseudoscalar mediators are compared to constraints
from the observed cosmological relic density of DM as determined from measurements of the
cosmic microwave background by the Planck satellite experiment~\cite{Ade:2015xua}.
The expected DM abundance is estimated, separately for each model, using the thermal freeze-out
mechanism implemented in the {\sc MadDM}~\cite{Backovic:2013dpa} framework and  compared to
the observed cold DM density $\Omega_c h^2=0.12$~\cite{Ade:2013zuv}, where $\Omega_c$
is the DM relic abundance and $h$ is the Hubble constant.

\begin{figure*}[hbtp]
\begin{center}
\includegraphics[width=\cmsFigWidth]{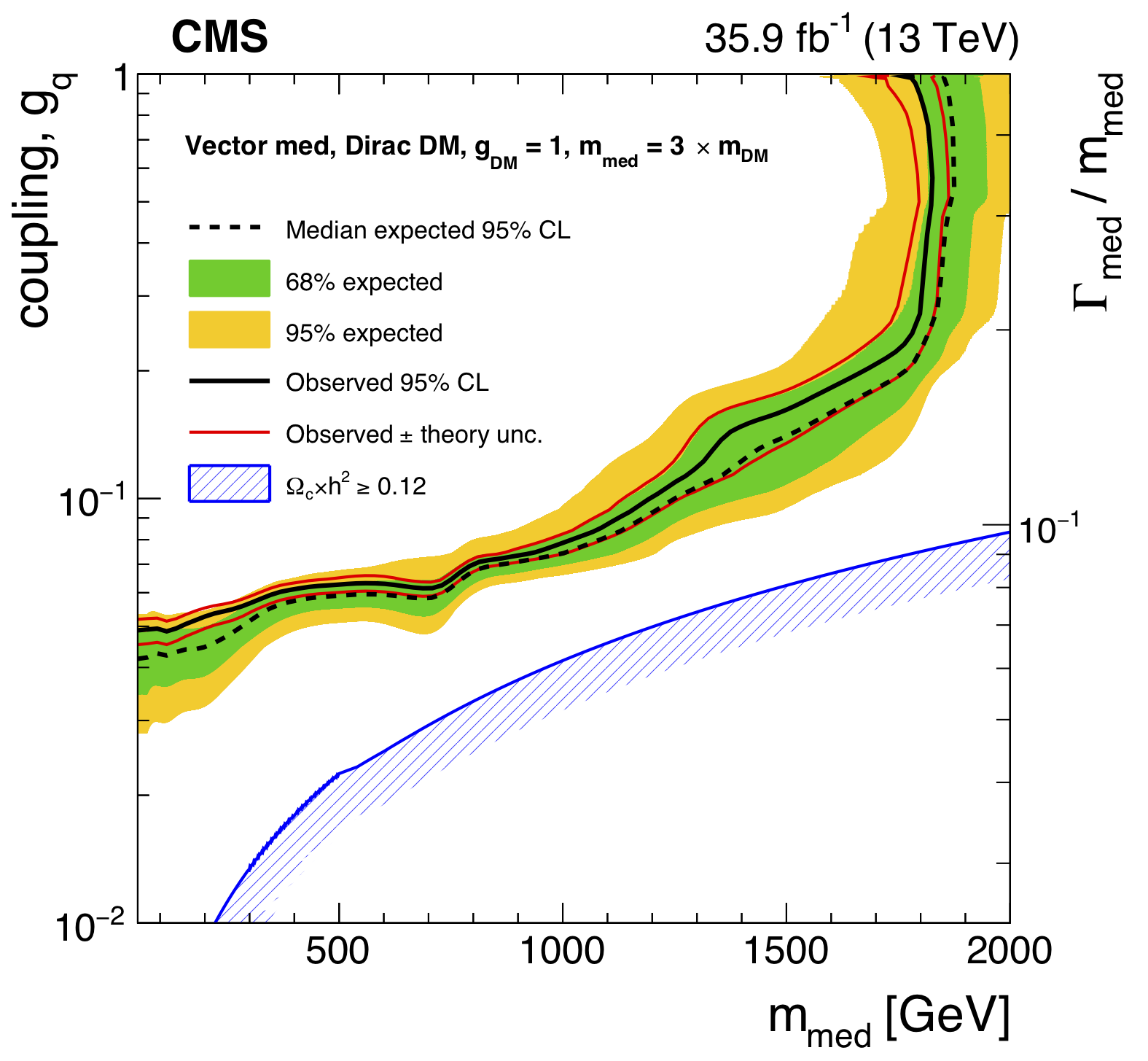}\hfil
\includegraphics[width=\cmsFigWidth]{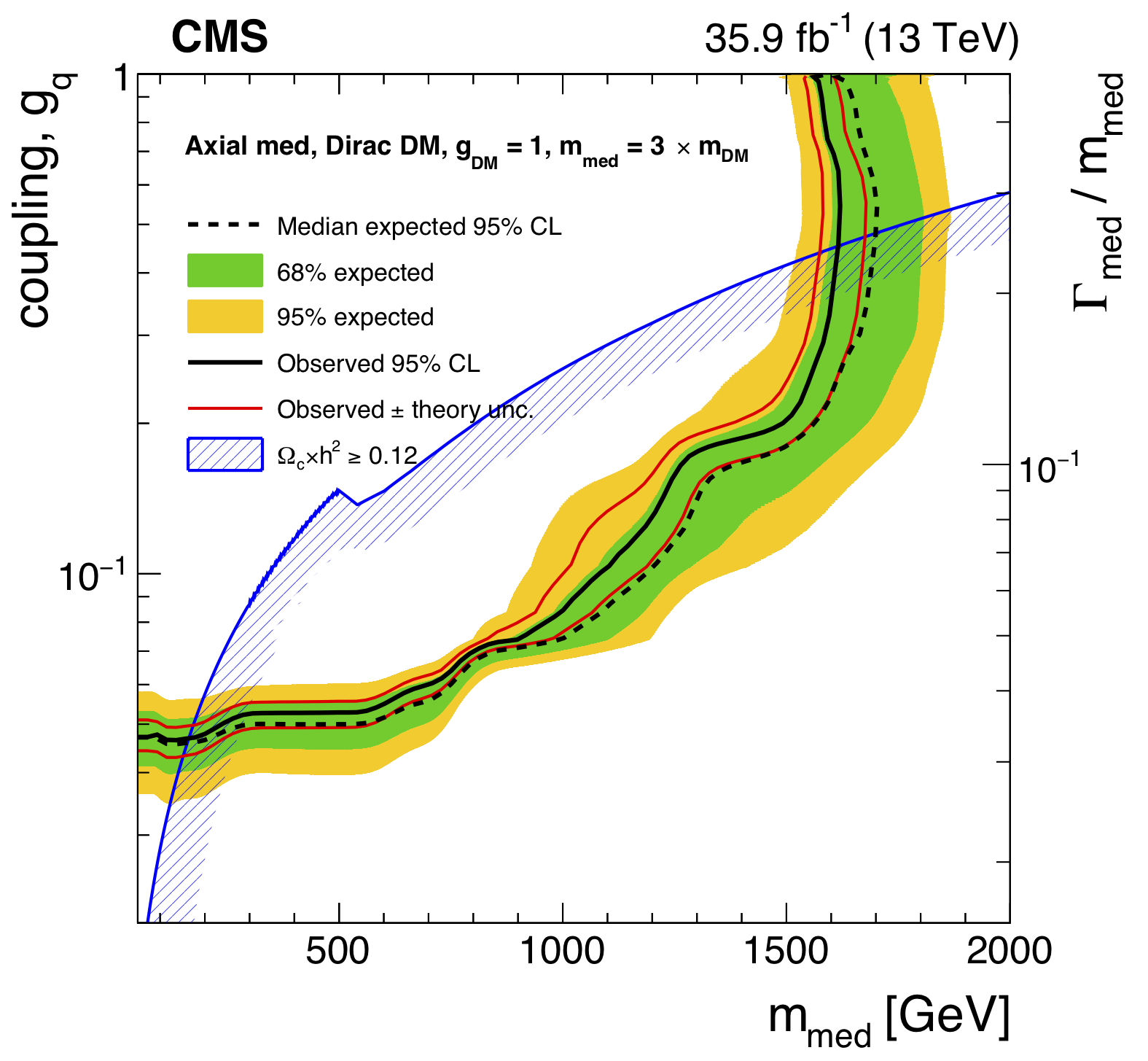}
\caption{
Exclusion limits at 95\%\,\CL on $\mu=\sigma/\sigma_{\textrm{th}}$ in the
$m_{\textrm{med}}$-$g_{\mathrm{q}}$ plane assuming vector (left) and axial-vector (right) mediators.
The widths shown on the axis correspond to mediator masses above 400\GeV, where the top quark decay channel is fully open.
For the mediator masses below the top quark decay channel threshold the width is 9\% less.
The solid (dotted) black line shows the contour for the observed (expected) exclusion.
The solid red contours around the observed limit represent one standard deviation due to theoretical uncertainties
in the signal cross section. Constraints from the Planck satellite experiment~\cite{Ade:2015xua}  are shown as dark blue contours;
in the shaded area DM is overabundant.
}
\label{fig:scan_spin1_gq}
\end{center}
\end{figure*}

In addition to scanning the $m_{\textrm{med}}$-$m_{\textrm{DM}}$ plane, for a fixed $g_{\mathrm{q}}$ value, the analysis interprets the results
in the $m_{\textrm{med}}$-$g_{\mathrm{q}}$ plane for a fixed ratio of $m_{\textrm{med}}$/$m_{\textrm{DM}}=3$. The ratio is chosen to ensure a valid relic abundance solution for every allowed  $g_{\mathrm{q}}$ value scanned
for a spin-1 simplified model. Quark couplings down to 0.05 for mediator masses at 50\GeV are excluded for the spin-1 simplified models
as shown in Fig.~\ref{fig:scan_spin1_gq}.

\begin{figure*}[hbtp]
\begin{center}
\includegraphics[width=\cmsFigWidth]{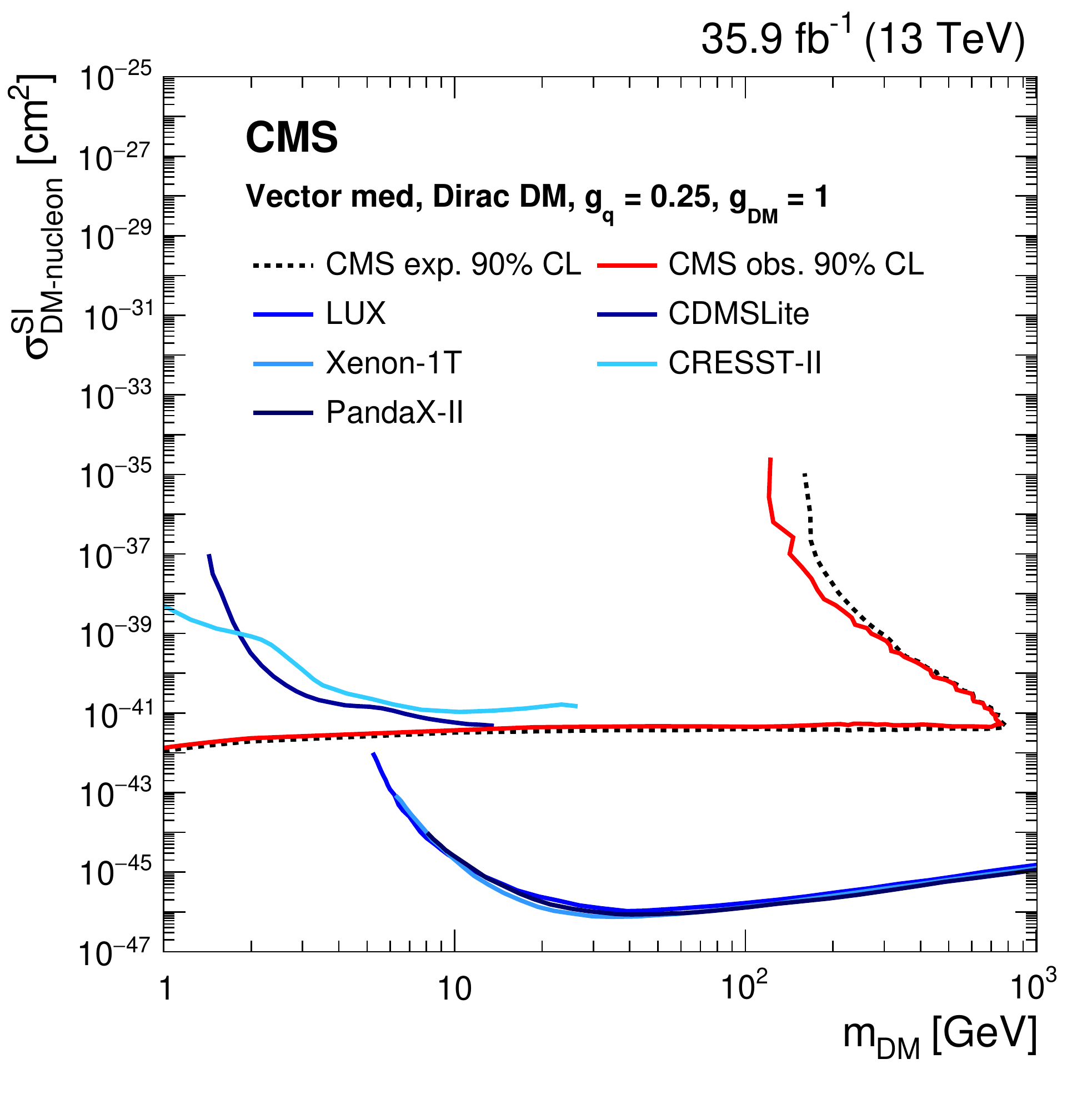}\hfil
\includegraphics[width=\cmsFigWidth]{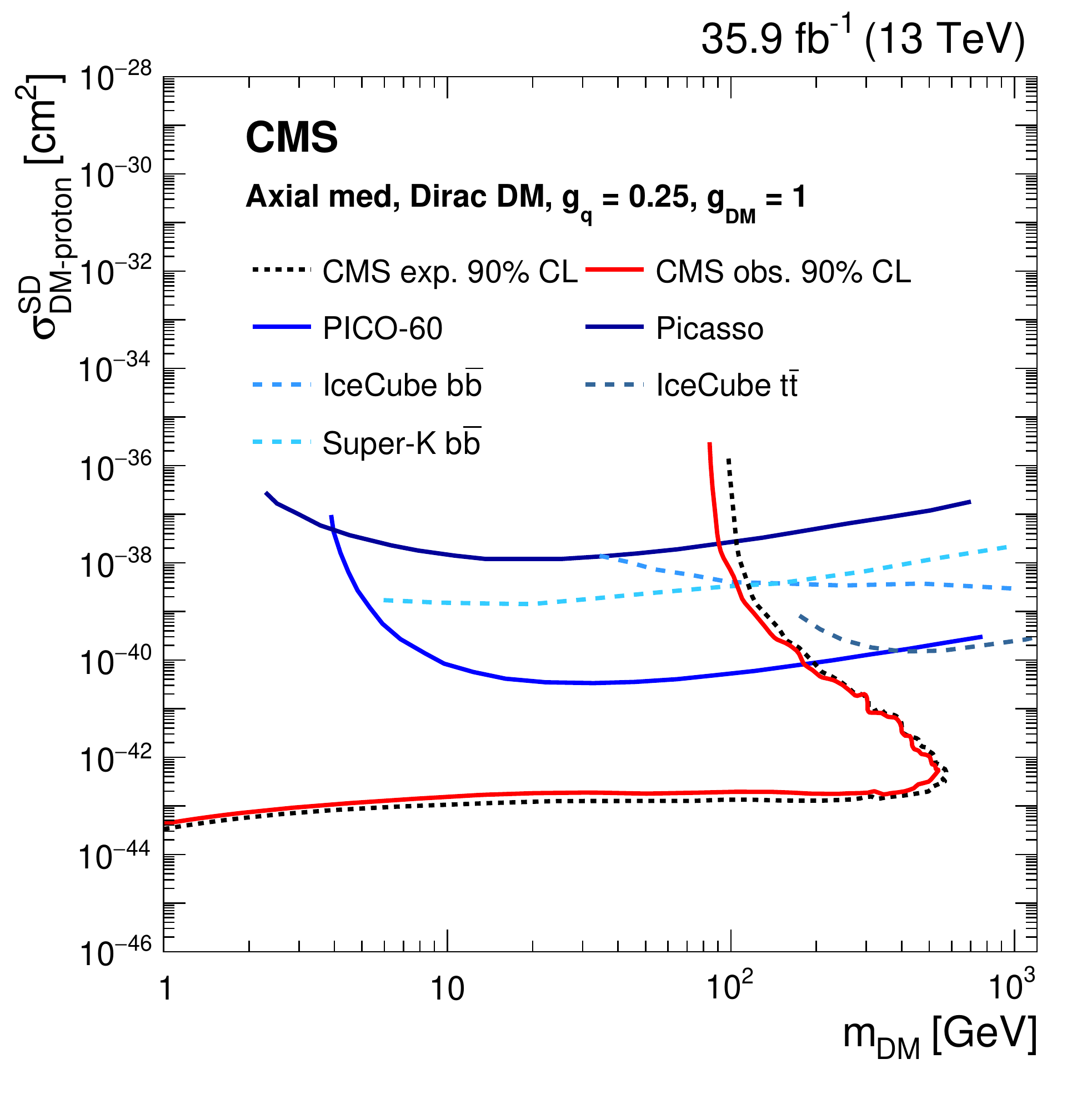}
\caption{
Exclusion limits at 90\%\CL in the $m_{\textrm{DM}}$ vs. $\sigma_{\textrm{SI/SD}}$ plane for vector
(left) and axial-vector (right) mediator models.
The solid red (dotted black) line shows the contour for the
observed (expected) exclusion in this search.
Limits from CDMSLite~\cite{Agnese:2015nto}, LUX~\cite{Akerib:2016vxi},
XENON-1T~\cite{Aprile:2017iyp}, PANDAX-II~\cite{Cui:2017nnn}, and
CRESST-II~\cite{cresst} are shown for the vector mediator.
Limits from Picasso~\cite{Behnke:2016lsk}, PICO-60~\cite{Amole:2015pla}, IceCube~\cite{Aartsen:2016exj}, and
Super-Kamiokande \cite{Choi:2015ara}
are shown for the axial-vector mediator. }
\label{fig:nucleon}
\end{center}
\end{figure*}

\begin{figure*}[hbtp]
\centering
\includegraphics[width=\cmsFigWidth]{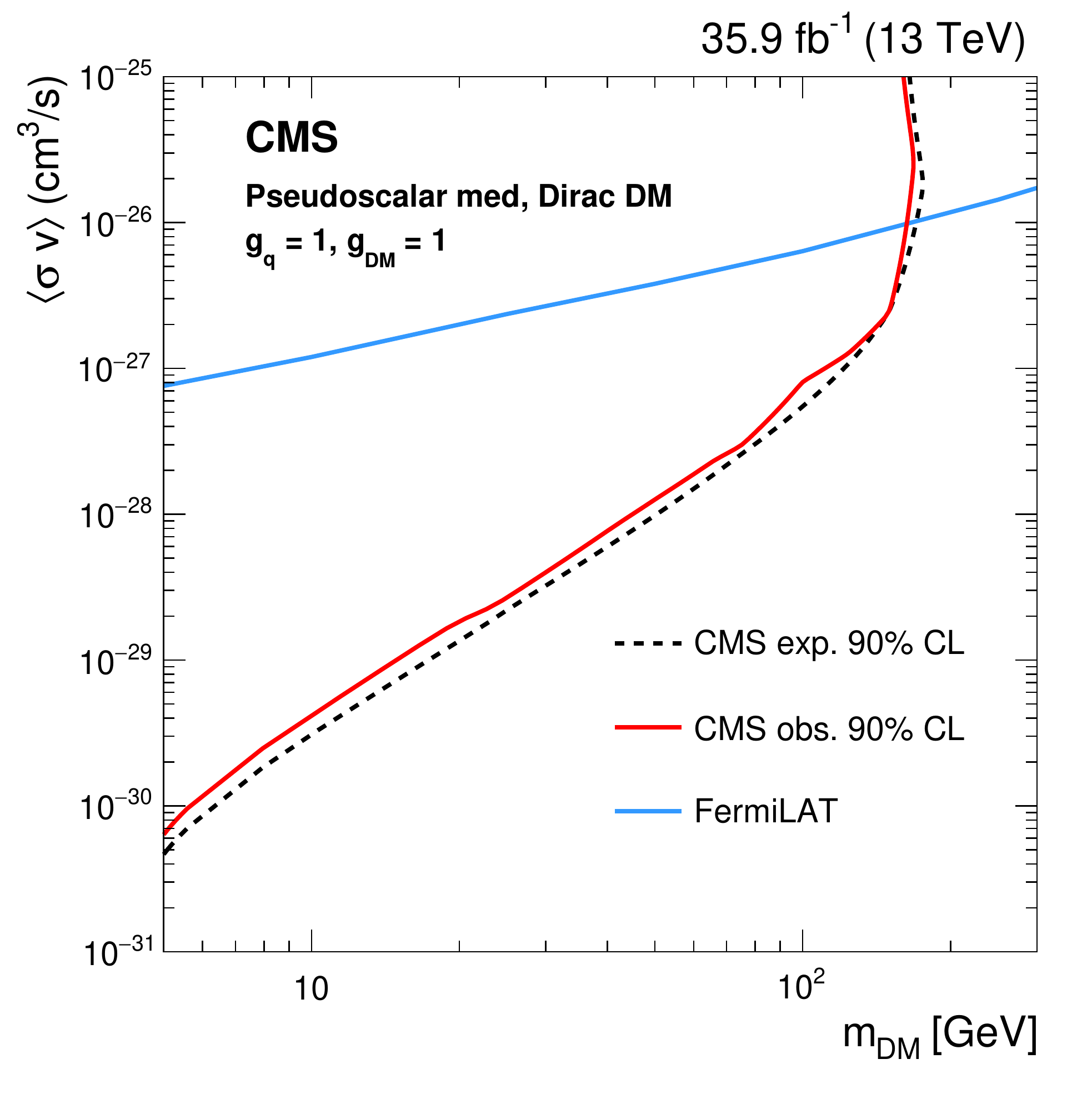}
\caption{
For the pseudoscalar mediator, limits are compared to the the velocity averaged DM
annihilation cross section upper limits from Fermi-LAT~\cite{Ackermann:2015zua}.
There are no comparable limits from direct detection experiments, as the scattering cross section
between DM particles and SM quarks is suppressed at nonrelativistic
velocities for a pseudoscalar mediator~\cite{Haisch:2012kf,Berlin:2014tja}.
}
\label{fig:nucleon2}
\end{figure*}

The exclusion contours obtained from the simplified DM models are translated to
90\%\CL upper limits on the spin-independent/spin-dependent ($\sigma_{\textrm{SI/SD}}$)
DM-nucleon scattering cross sections
using the approach outlined in Refs.~\cite{Buchmueller:2014yoa,Harris:2015kda,Boveia:2016mrp}.
The results for the vector and axial-vector mediators are compared with the results of direct searches in Fig.~\ref{fig:nucleon}.
This search provides the most stringent constraints for vector mediators, for DM particle masses below
5\GeV. For axial-vector mediators, the sensitivity achieved in this search
provides stronger constraints up to a DM particle mass of 550\GeV than those obtained from direct searches.
For pseudoscalar mediators, the 90\%\CL upper limits as shown in Fig.~\ref{fig:nucleon2} are
translated to velocity-averaged DM annihilation cross section ($\langle  \sigma v \rangle$) and are
compared to the indirect detection results from the Fermi-LAT Collaboration~\cite{Ackermann:2015zua}.
The collider results provide stronger constraints for DM masses less than 150\GeV.

\subsubsection{Fermion portal dark matter interpretation}

\begin{figure}[htb]
\centering
\includegraphics[width=\cmsFigWidth]{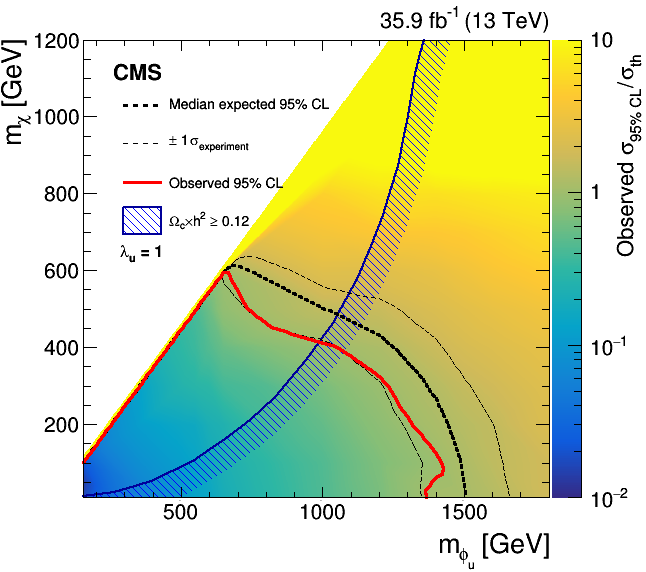}
\caption{
The 95\%\,\CL expected (black dashed line) and observed (red solid line)
upper limits on $\mu=\sigma/\sigma_{\rm{th}}$ in the context of the
fermion portal DM model, for Dirac DM particles with coupling strengths to
the up quark corresponding to $\lambda_{\rm{u}} = 1$ in
the $m_{\phi_{\rm{u}}}$-$m_\chi$ plane.
Constraints from the Planck satellite experiment~\cite{Ade:2015xua} are shown as dark blue contours;
in the shaded area DM is overabundant.}
\label{fig:fermionportal_exlusion}
\end{figure}

The total production cross section in the fermion portal DM model has an exponential (linear)
dependence on the mass of the new scalar mediator $m_{\phi_{\rm{u}}}$ (mass of the DM candidate $m_\chi$).
The middle diagram shown in Fig.~\ref{fig:fermionportal_FeynDiag} represents the main production mechanism for
small $m_{\phi_{\rm{u}}}$ values, whereas the right diagram contributes to the total
cross section for $m_{\phi_{\rm{u}}} > 1\TeV$. The region where $m_{\phi_{\rm{u}}} < m_\chi$
is not considered in the search, because of the reduced production cross section of the model.
The upper limits on the signal strength are set as a function of $m_{\phi_{\rm{u}}}$ and $m_\chi$.
Figure~\ref{fig:fermionportal_exlusion} shows the exclusion contours in the $m_{\phi_{\rm{u}}}$-$m_\chi$ plane, for
which the coupling strength $\lambda_{\rm{u}}$ of the interaction between the scalar mediator and up-type quarks
is fixed at unity. The results are also compared to constraints from the observed cosmological relic density of DM, obtained by the Planck satellite experiment,
for the allowed values of $m_{\phi_{\rm{u}}}$ and $m_\chi$ \cite{fermionportal_ref1}. In this search, mediator (dark matter) masses
up to 1.4 (0.6)\TeV are excluded.

\subsubsection{Nonthermal dark matter interpretation}

\begin{figure*}[h!]
\centering
\includegraphics[width=\cmsFigWidth]{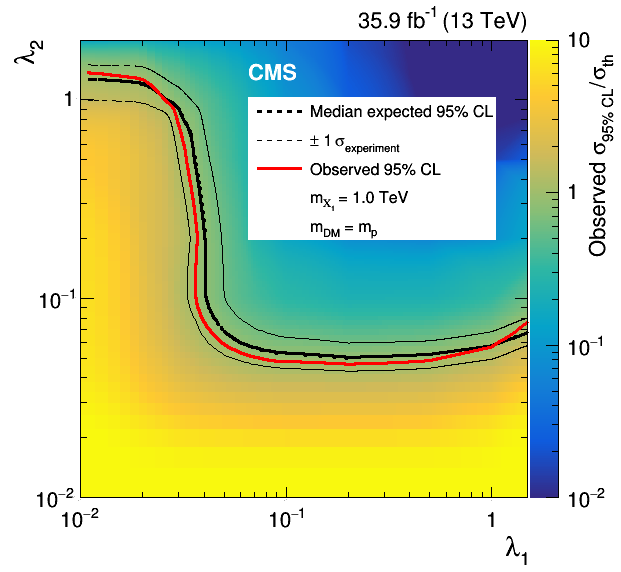}\hfil
\includegraphics[width=\cmsFigWidth]{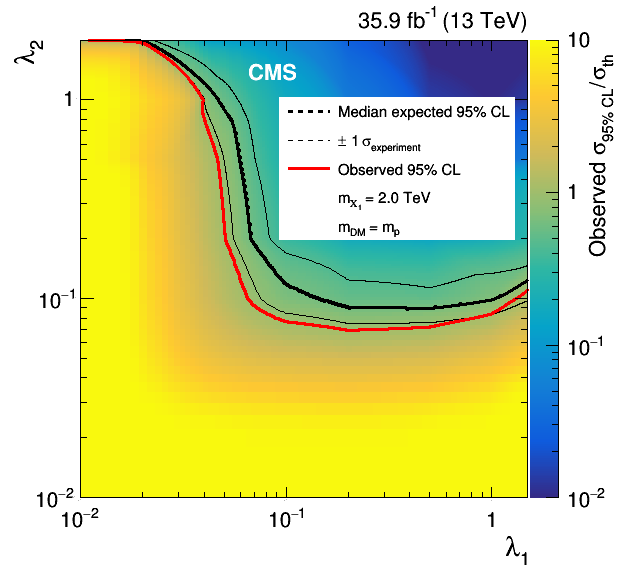}
\caption{
Expected (black line) and observed (red line) 95\%\,\CL upper limits on the signal strength
$\mu=\sigma/\sigma_{\text{th}}$, in the context of a nonthermal dark matter model. Results are reported in
the $\lambda_1$-$\lambda_2$ plane, which represents the coupling strength of the interaction of the new scalar
mediator with down-type quarks and DM with up-type quarks, respectively. Limits are shown for $m_\mathrm{X_1}$ of 1\TeV (left) and 2\TeV (right).}
\label{fig:NonThDM_limit2}
\end{figure*}

This search is also interpreted in the context of the nonthermal DM model where
the DM candidate is not parity protected and therefore could be singly produced. Such production leads
to signatures with an energetic jet and large \ptmiss~whose distribution is
characterized by a Jacobian-like shape, which exhibits a peak at half of the mediator mass.
Therefore, multiple mediator mass points have been studied. The search is restricted
to a coupling range of 0.01--1.5 for $\lambda_1$ and 0.01--2.0 for $\lambda_2$ to ensure
the mediator width is less than about 30\% of its mass. Within these bounds, no significant excesses
were found and limits are reported as a function of coupling strength parameters $\lambda_1$ and $\lambda_2$
for two reference mediator masses $m_\mathrm{X_1}$ of 1 and 2\TeV.
Figure~\ref{fig:NonThDM_limit2} shows the exclusion contours in the $\lambda_1$-$\lambda_2$ plane.

\subsection{Invisible decays of the Higgs boson interpretation}

\begin{figure*}[hbtp]
\begin{center}
\includegraphics[width=\cmsFigWidth]{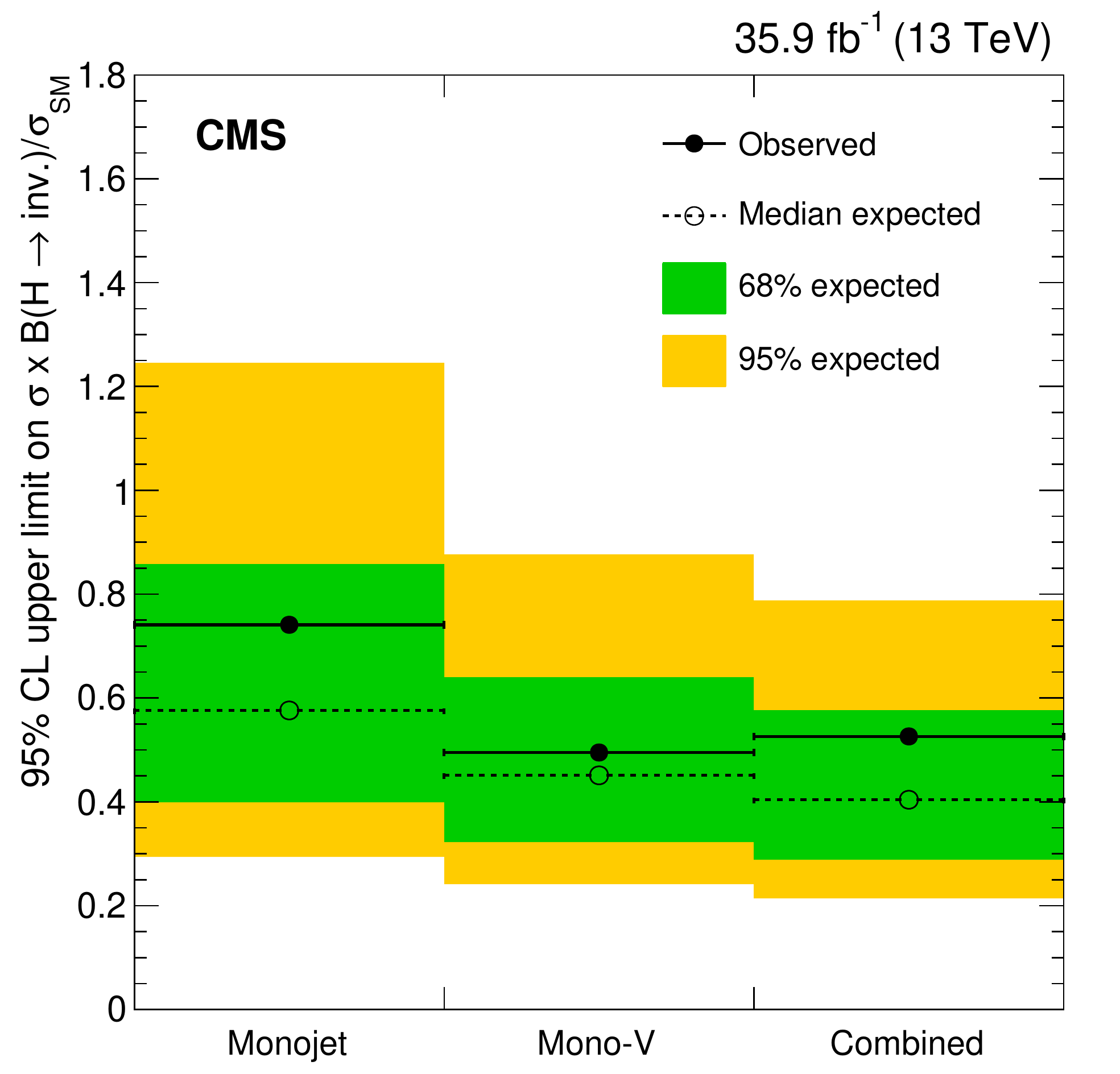}
\caption{Expected (dotted line) and observed (solid line)
95\%\,\CL upper limits on the invisible branching fraction of the 125\GeV
SM-like Higgs boson. Limits are shown for the monojet and mono-$\PV$ categories
separately, and also for their combination.}
\label{fig:HinvLimitsPlot}
\end{center}
\end{figure*}

The results of this search are further interpreted in terms of an upper limit on the production cross section and branching fraction, \brhinv,
where the Higgs boson is produced through gluon fusion (ggH) along with a jet; or in association with a vector boson ($\PZ\PH$, $\PW\PH$);
or through vector boson fusion (VBF). The predictions for the Higgs boson production cross section
and the corresponding theoretical uncertainties are taken from the recommendations of the LHC Higgs
cross section working group~\cite{Heinemeyer:2013tqa}.
The observed (expected) 95\%\,\CL upper limit on the invisible branching fraction of the
Higgs boson, $\sigma \times \brhinv / \sigma_{\textrm{SM}}$,
is found to be 53\% (40\%). The limits are summarized in Fig.~\ref{fig:HinvLimitsPlot}, while
Table~\ref{tab:hinvlimits} shows the individual limits for the monojet and mono-$\PV$ categories.

\begin{table*}[htb]
\topcaption{Expected and observed 95\%\,\CL upper limits on the invisible branching fraction of the Higgs boson. Limits are tabulated for the monojet and mono-$\PV$ categories separately, and for their combination. The one standard deviation uncertainty range in the expected limits is listed. The expected composition of the production modes of a Higgs boson with a mass of 125\GeV is summarized, assuming SM production cross sections.}
\begin{center}
\renewcommand{\arraystretch}{1}
\ifthenelse{\boolean{cms@external}}{\footnotesize}{}
{
\begin{scotch}{l c c c}
Category          & Observed (expected) & 68\% expected  & Expected signal composition \\
\hline\\[-2ex]
Monojet           & 0.74 (0.57)         & 0.40--0.86    & \begin{tabular}[c]{@{}c@{}}72.8\% ggH, 21.5\% VBF,\\ 3.3\% $\PW\PH$, 1.9\% $\PZ\PH$, 0.6\% $\Pg\Pg\PZ\PH$\end{tabular} \\[2ex]
mono-$\PV$            & 0.49 (0.45)         & 0.32--0.64    & \begin{tabular}[c]{@{}c@{}}38.7\% ggH, 7.0\% VBF, \\32.9\% $\PW\PH$, 14.6\% $\PZ\PH$, 6.7\% $\Pg\Pg\PZ\PH$\end{tabular}  \\[2ex]
Combined          & 0.53 (0.40)         & 0.29--0.58    & \NA \\[1ex]
\end{scotch}
}
\label{tab:hinvlimits}
\end{center}
\end{table*}

\subsection{The ADD model interpretation}

\begin{figure*}[hbtp]
\begin{center}
\includegraphics[width=\cmsFigWidth]{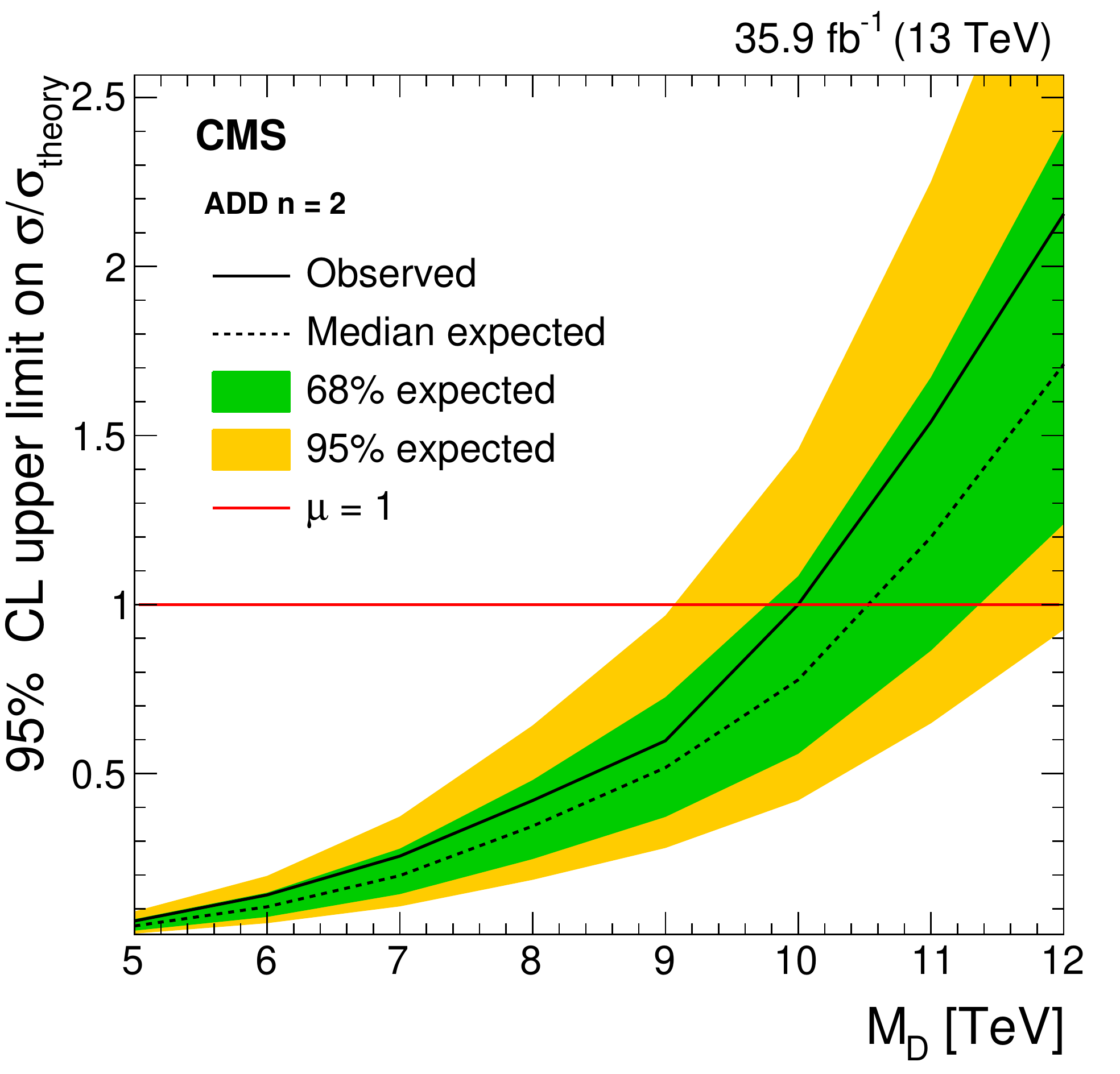}\hfil
\includegraphics[width=\cmsFigWidth]{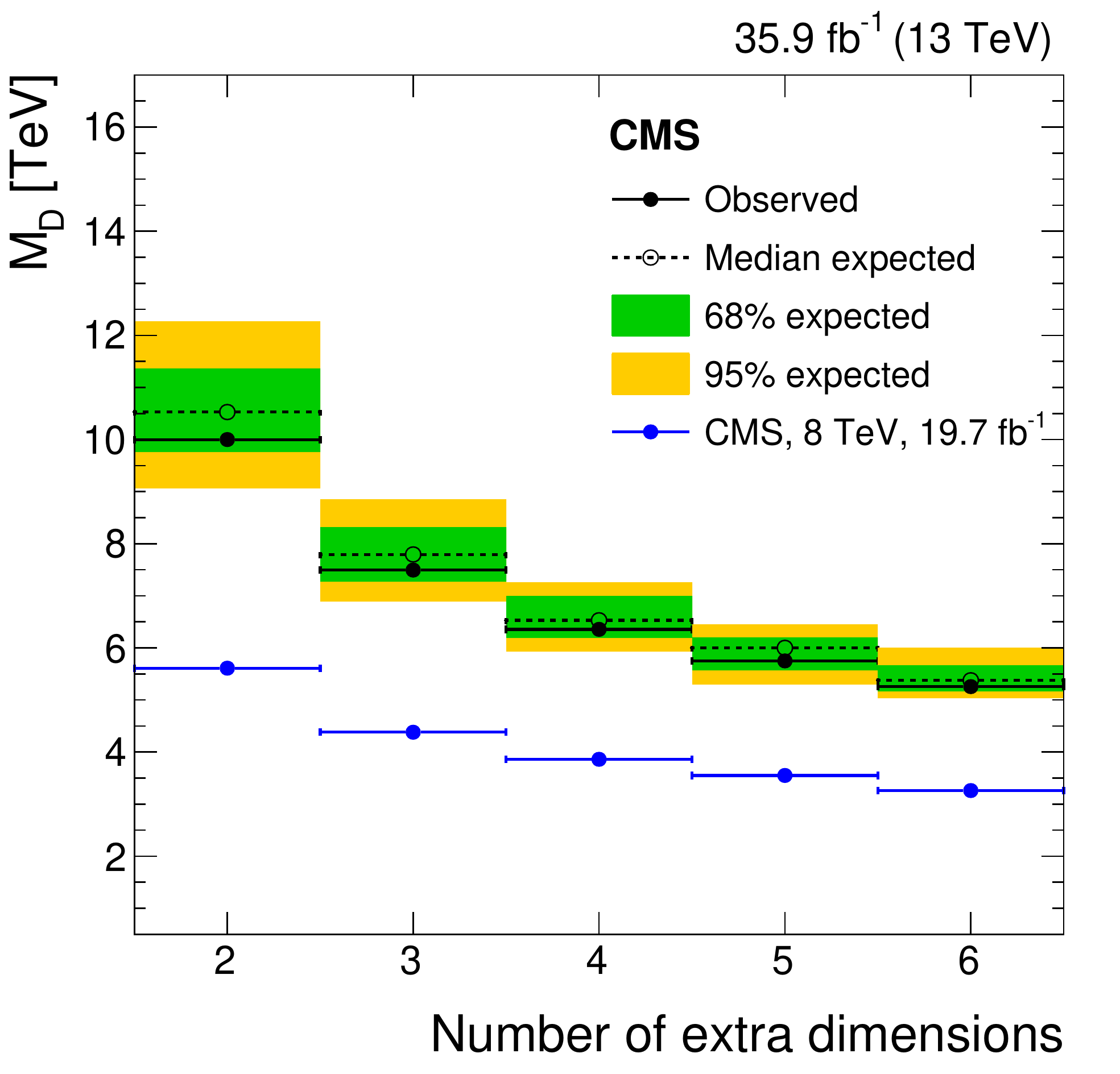}
\end{center}
\caption{
The 95\%\,\CL expected (dotted) and observed (solid) upper limits on the signal strength $\mu=\sigma/\sigma_{\rm{th}}$ for ADD
graviton production (left), as a function of fundamental Planck scale ($M_{\mathrm{D}}$) for $n=2$, where $n$ is the number of extra spatial dimensions.
The 95\%\,\CL expected (dotted) and observed (solid) lower limits (right) on $M_{\mathrm{D}}$ as a function of $n$ in the ADD model.
The results are also compared to earlier ones obtained by the CMS Collaboration with data corresponding to an integrated luminosity
of 19.7\fbinv at a centre-of-mass energy of 8\TeV~\cite{Khachatryan:2014rra} (blue points).
}
\label{fig:ADDLimits}
\end{figure*}

The 95\%\,\CL lower limits on the fundamental Planck scale $M_{\mathrm{D}}$ of the ADD model
are presented as a function of the number of extra spatial dimensions $n$.
The efficiency of the full event selection in the monojet (mono-$\PV$) category for this model ranges between 15 (1)\% and
20 (1.5)\% depending on the values of the parameters $M_{\mathrm{D}}$ and $n$.
An upper limit on the signal strength $\mu=\sigma/\sigma_{\rm{th}}$ is
presented for the ADD graviton production for $n=2$ EDs, as a function of $M_{\mathrm{D}}$ in Fig.~\ref{fig:ADDLimits}.
In addition, Fig.~\ref{fig:ADDLimits} shows the observed exclusion on $M_{\mathrm{D}}$ which varies from 9.9\TeV for
$n=2$ to 5.3\TeV for $n=6$. The results of this search are also compared to earlier ones obtained by the
CMS Collaboration with Run 1 data corresponding to an integrated luminosity of 19.7\fbinv
at a centre-of-mass energy of 8\TeV~\cite{Khachatryan:2014rra}.
The upper limits on the signal production cross section and $M_{\mathrm{D}}$ exclusions are also provided
in Table~\ref{tab:addxsec} as a function of the number of extra dimensions.
Compared to previous CMS publications in this channel, the lower limits on $M_{\mathrm{D}}$ show
a factor of 2 improvement.

\begin{table}[htb]
\topcaption{Upper limits on the signal  production cross section in the ADD model and lower limits on $M_{\mathrm{D}}$, both as functions of the number of extra spatial dimensions ($n$).}
\begin{center}
\renewcommand{\arraystretch}{1}
\ifthenelse{\boolean{cms@external}}{\footnotesize}{}
{
\begin{scotch}{l c c}
$n$ & \begin{tabular}[c]{@{}c@{}} Observed (expected) \\cross section exclusion {[}pb{]}\end{tabular} & \begin{tabular}[c]{@{}c@{}} Observed (expected) \\$M_{\mathrm{D}}$ exclusions [TeV] \end{tabular} \\
\hline
2 & 0.28 (0.22)                                                           &  9.9 (10.5) \\
3 & 0.18 (0.15)                                                            & 7.5 (7.8)  \\
4 & 0.15 (0.13)                                                            & 6.3 (6.5)  \\
5 & 0.13 (0.11)                                                            & 5.7 (6.0)  \\
6 & 0.13 (0.10)                                                            & 5.3 (5.4)  \\
\end{scotch}
}
\label{tab:addxsec}
\end{center}
\end{table}

\section{Summary} \label{sec:summary}

A search for dark matter (DM) particles, invisible decays of a
standard-model-like (SM-like) Higgs boson, and extra spatial dimensions is presented using events with one or more energetic jets
and large missing transverse momentum in proton-proton collisions recorded at $\sqrt{s} = 13\TeV$,
using  a sample of data corresponding to an integrated
luminosity of 35.9\fbinv.
Events are categorized based on whether jets are produced directly in hard scattering as
initial-state radiation or originate from merged quarks from a decay of a highly Lorentz-boosted $\PW$ or $\PZ$ boson.
No excess of events is observed compared to the SM background expectations in either of these two categories.

Limits are computed on the DM production cross section using
simplified models in which DM production is mediated by spin-1
and spin-0 particles. Vector and axial-vector (pseudoscalar) mediators with
masses up to 1.8 (0.4)\TeV are excluded at 95\% confidence level.
Similarly, limits are also presented for the parameters of
the fermion portal DM model and an exclusion up to 1.4\TeV on the mediator
mass is observed at 95\% confidence level.
The first limits on the DM production at a particle collider in
the nonthermal DM model are obtained and presented in the coupling
strength plane. Furthermore, an observed (expected) 95\% confidence level
upper limit of 0.53 (0.40) is set for the invisible branching fraction of an
SM-like 125\GeV Higgs boson, assuming the SM production cross section.
Lower limits are also computed on the fundamental Planck scale
$M_{\mathrm{D}}$ in the context of the Arkani-Hamed, Dimopoulos, and Dvali
model with large extra spatial dimensions, which varies from 9.9\TeV for
$n=2$ to 5.3\TeV for $n=6$ at 95\% confidence level, where $n$ is the number of extra spatial dimensions.
These limits provide the most stringent direct constraints on the fundamental
Planck scale to date.

\begin{acknowledgments}

We congratulate our colleagues in the CERN accelerator departments for the excellent performance of the LHC and thank the technical and administrative staffs at CERN and at other CMS institutes for their contributions to the success of the CMS effort. In addition, we gratefully acknowledge the computing centres and personnel of the Worldwide LHC Computing Grid for delivering so effectively the computing infrastructure essential to our analyses. Finally, we acknowledge the enduring support for the construction and operation of the LHC and the CMS detector provided by the following funding agencies: BMWFW and FWF (Austria); FNRS and FWO (Belgium); CNPq, CAPES, FAPERJ, and FAPESP (Brazil); MES (Bulgaria); CERN; CAS, MoST, and NSFC (China); COLCIENCIAS (Colombia); MSES and CSF (Croatia); RPF (Cyprus); SENESCYT (Ecuador); MoER, ERC IUT, and ERDF (Estonia); Academy of Finland, MEC, and HIP (Finland); CEA and CNRS/IN2P3 (France); BMBF, DFG, and HGF (Germany); GSRT (Greece); OTKA and NIH (Hungary); DAE and DST (India); IPM (Iran); SFI (Ireland); INFN (Italy); MSIP and NRF (Republic of Korea); LAS (Lithuania); MOE and UM (Malaysia); BUAP, CINVESTAV, CONACYT, LNS, SEP, and UASLP-FAI (Mexico); MBIE (New Zealand); PAEC (Pakistan); MSHE and NSC (Poland); FCT (Portugal); JINR (Dubna); MON, RosAtom, RAS, RFBR and RAEP (Russia); MESTD (Serbia); SEIDI, CPAN, PCTI and FEDER (Spain); Swiss Funding Agencies (Switzerland); MST (Taipei); ThEPCenter, IPST, STAR, and NSTDA (Thailand); TUBITAK and TAEK (Turkey); NASU and SFFR (Ukraine); STFC (United Kingdom); DOE and NSF (USA).

\hyphenation{Rachada-pisek} Individuals have received support from the Marie-Curie programme and the European Research Council and Horizon 2020 Grant, contract No. 675440 (European Union); the Leventis Foundation; the A. P. Sloan Foundation; the Alexander von Humboldt Foundation; the Belgian Federal Science Policy Office; the Fonds pour la Formation \`a la Recherche dans l'Industrie et dans l'Agriculture (FRIA-Belgium); the Agentschap voor Innovatie door Wetenschap en Technologie (IWT-Belgium); the Ministry of Education, Youth and Sports (MEYS) of the Czech Republic; the Council of Science and Industrial Research, India; the HOMING PLUS programme of the Foundation for Polish Science, cofinanced from European Union, Regional Development Fund, the Mobility Plus programme of the Ministry of Science and Higher Education, the National Science Center (Poland), contracts Harmonia 2014/14/M/ST2/00428, Opus 2014/13/B/ST2/02543, 2014/15/B/ST2/03998, and 2015/19/B/ST2/02861, Sonata-bis 2012/07/E/ST2/01406; the National Priorities Research Program by Qatar National Research Fund; the Programa Clar\'in-COFUND del Principado de Asturias; the Thalis and Aristeia programmes cofinanced by EU-ESF and the Greek NSRF; the Rachadapisek Sompot Fund for Postdoctoral Fellowship, Chulalongkorn University and the Chulalongkorn Academic into Its 2nd Century Project Advancement Project (Thailand); and the Welch Foundation, contract C-1845.

\end{acknowledgments}

\bibliography{auto_generated}

\clearpage

\appendix

\section{Additional material}\label{sec:app}

Another important cross-check of the application of \pt-dependent NLO QCD and EW corrections
is represented by the agreement between data and simulation in the ratio of
\Zjets events to both \phojets events
and \Wjets events in the control samples as a function of boson \pt.

\begin{figure*}[htbp]
\centering
\includegraphics[width=\cmsFigWidth]{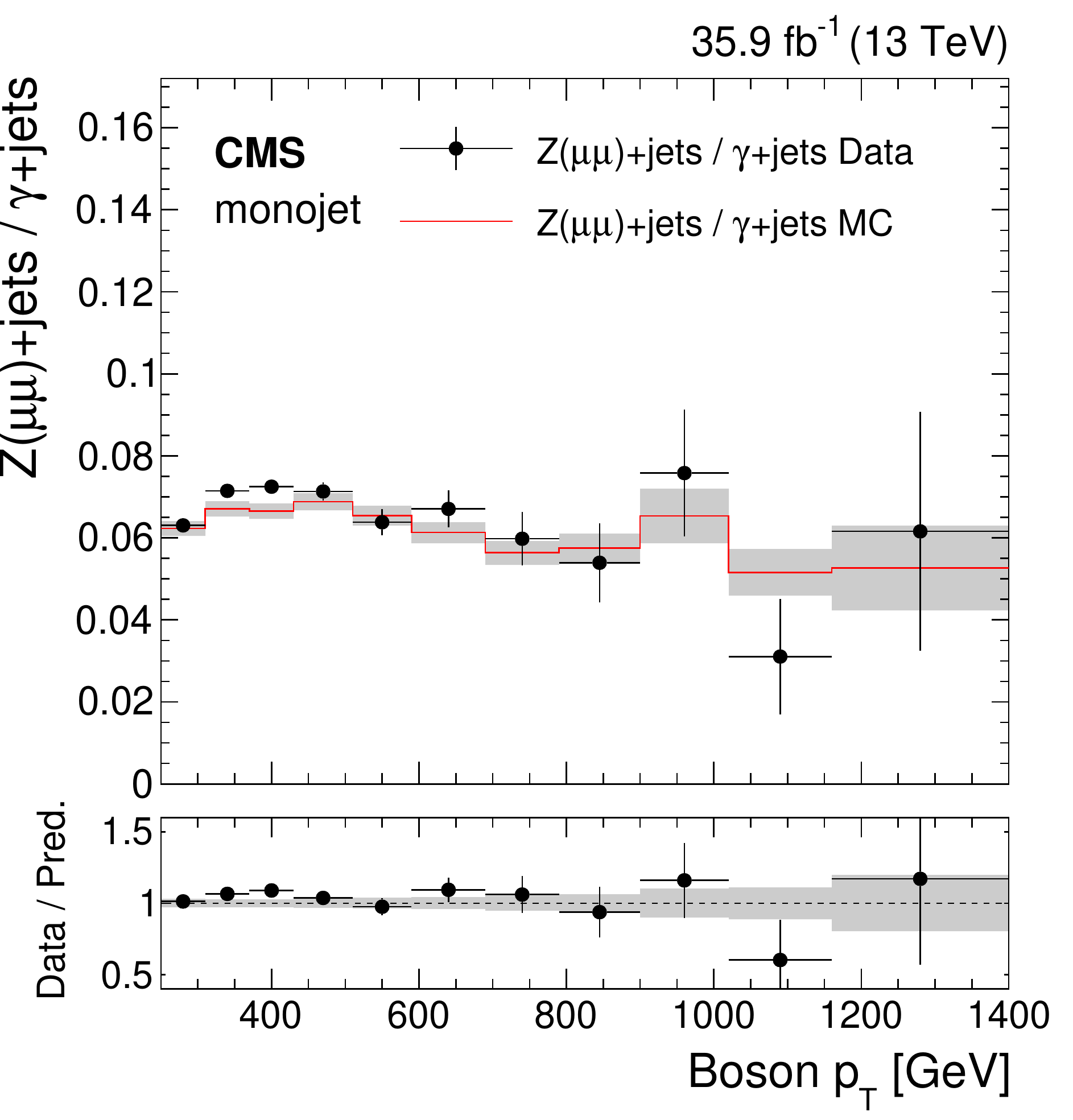}\hfil
\includegraphics[width=\cmsFigWidth]{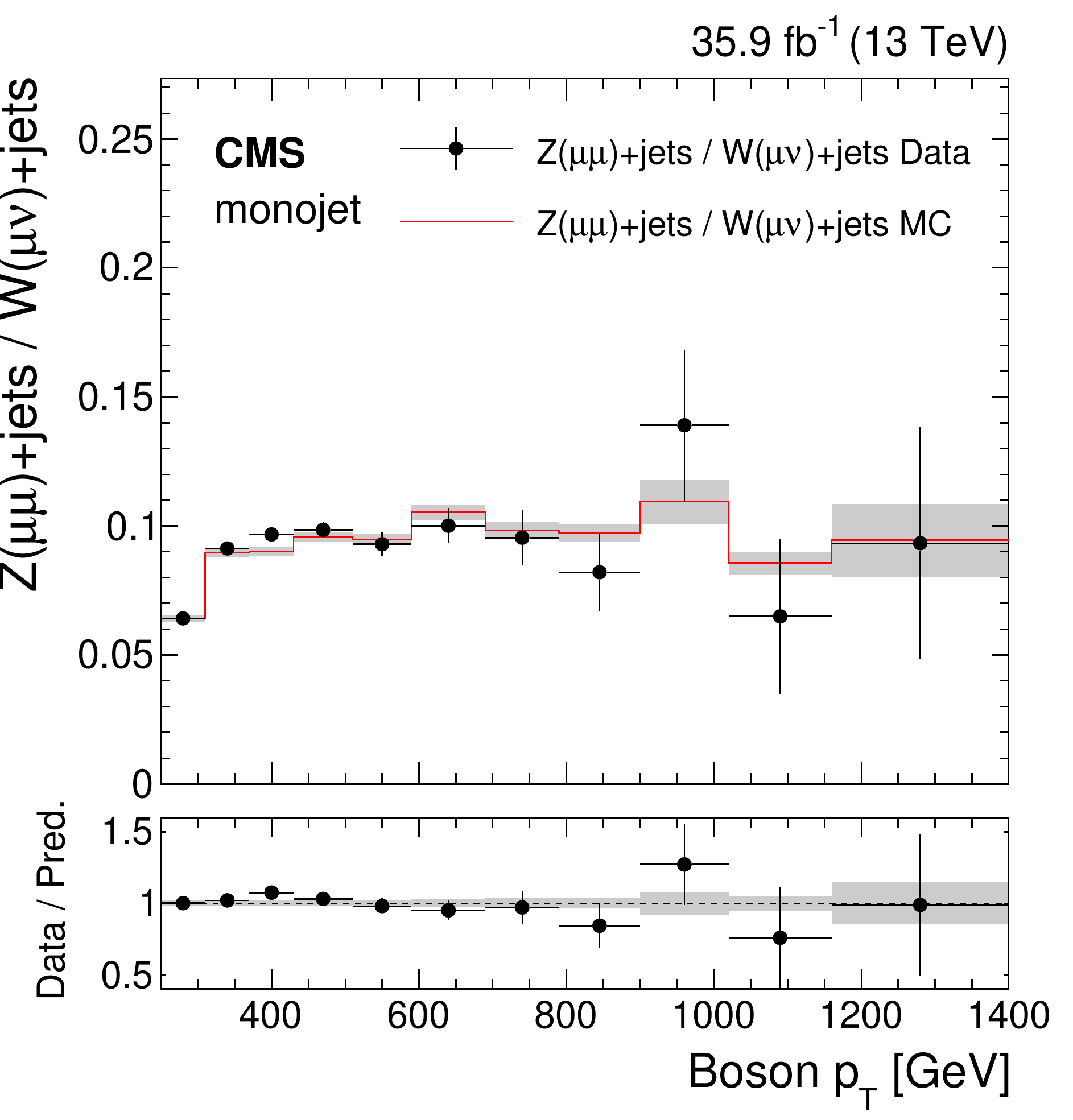}
\\
\includegraphics[width=\cmsFigWidth]{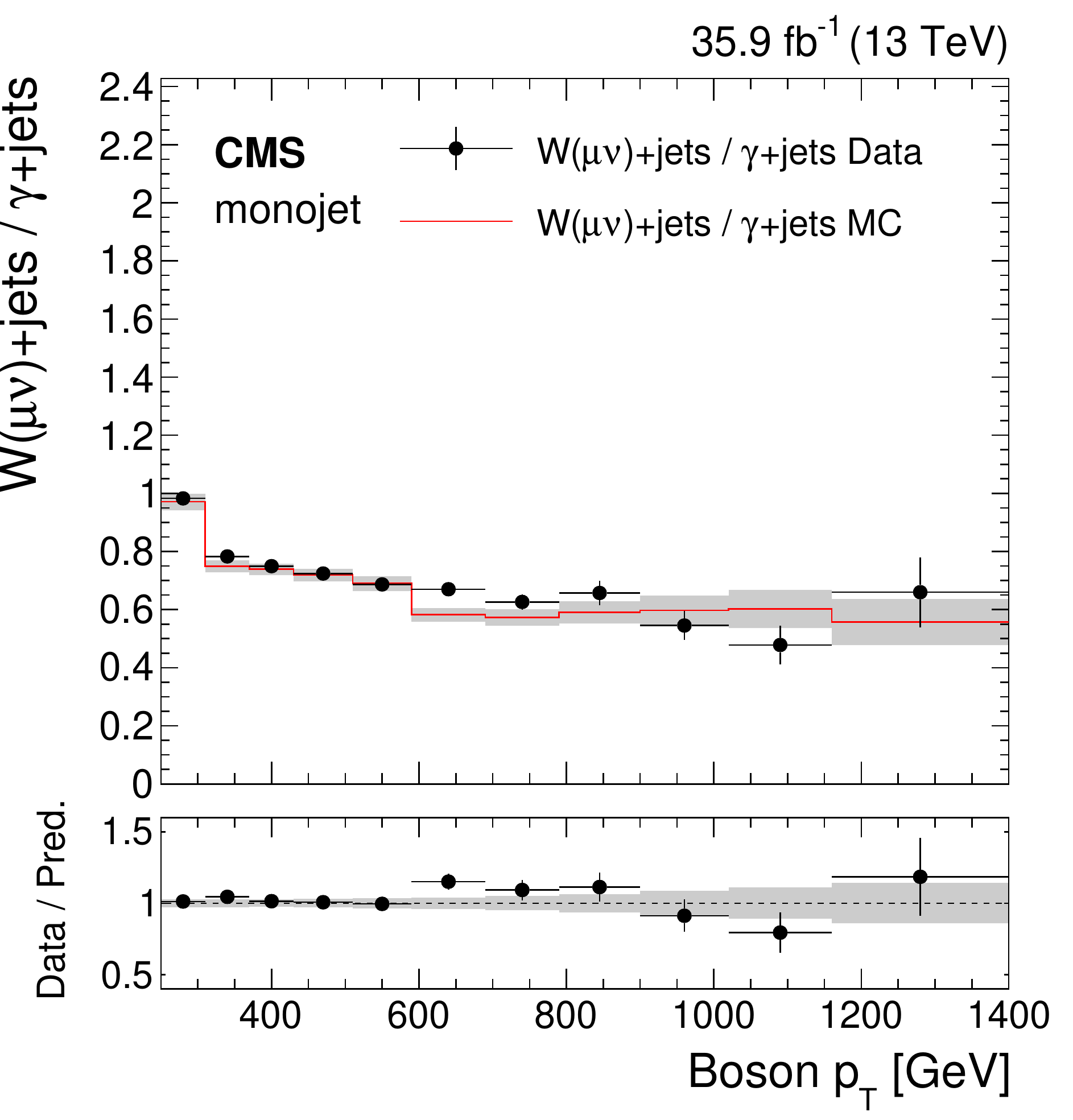}
\caption{Comparison between data and Monte Carlo simulation of the  $\PZ(\PGm\PGm)$/\phojets,
$\PZ(\PGm\PGm)$/$\PW(\PGm\PGn)$ and $\PW(\PGm\PGn)$/\phojets ratios, as a function
of boson \pt, in the monojet category.
In the ratio panel, ratios of data with the pre-fit background prediction are shown.
The gray bands include both the pre-fit systematic uncertainties and the statistical
uncertainty in the simulation.}
\label{fig:Ratio_pt}
\end{figure*}

Figure~\ref{fig:Ratio_pt} shows the ratio between \Zmmjets and \phojets, and the ratio of \Zmmjets and \Wmvjets events as a
function of the boson \pt, for the monojet category. While we do not explicitly use
a \Wmvjets/\phojets constraint in the analysis, the two cross sections
are connected through the \Zjets/\phojets and \Zjets/\Wjets constraints. Therefore, it is instructive to examine
the data-to-simulation comparison for the \Wmvjets/\phojets ratio. This is shown in the same figure.
Good agreement is observed between data and simulation after the application of NLO corrections.

\begin{figure*}[hbtp]
\begin{center}
\includegraphics[width=0.70\textwidth]{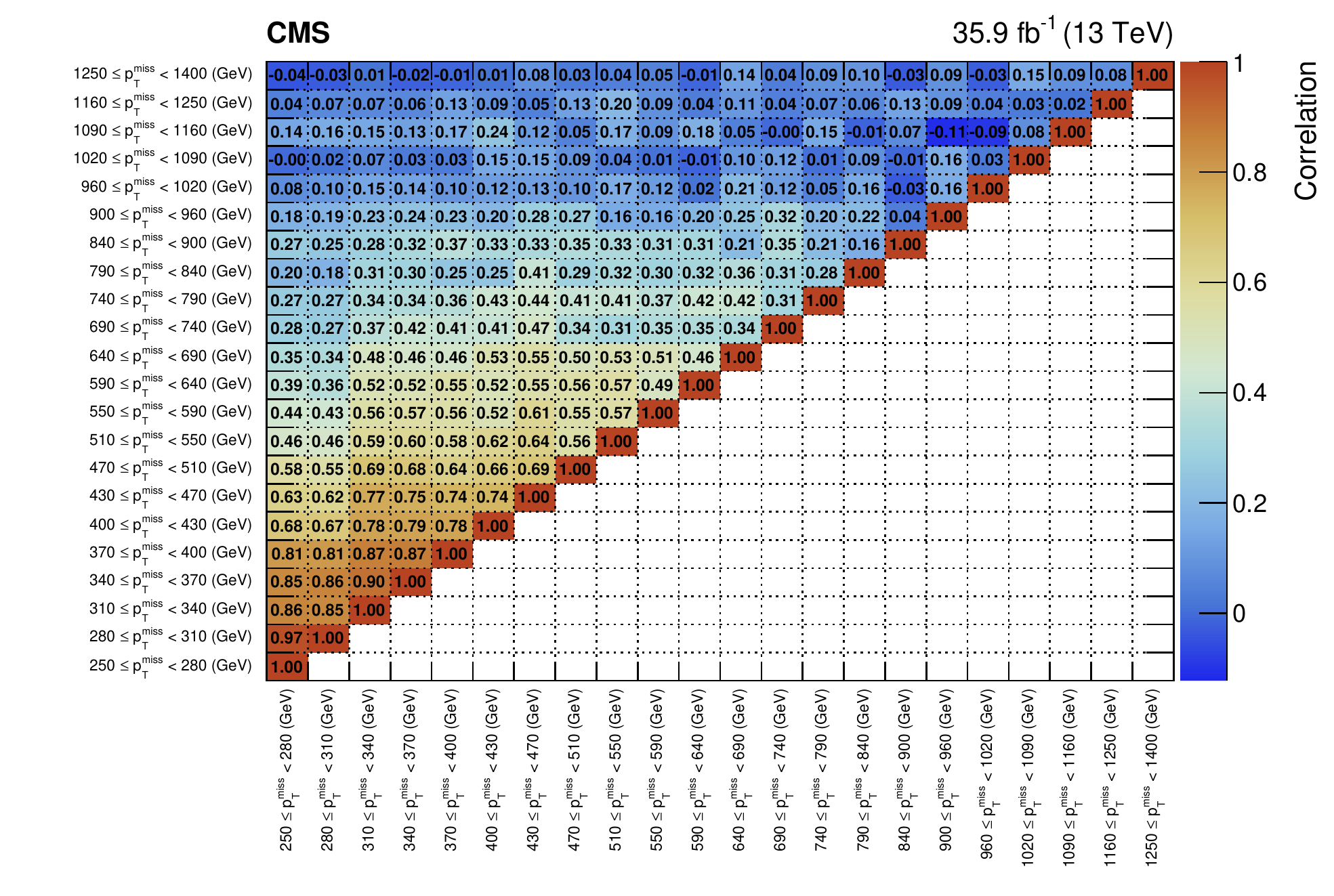}
\caption{Correlations between the predicted background yields in all the \ETm bins of the monojet signal region.
The boundaries of the \ETm bins, expressed in GeV, are shown at the bottom and on the left.}
\label{fig:correlation_matrix_monojet}
\end{center}
\end{figure*}

\begin{figure*}[hbtp]
\begin{center}
\includegraphics[width=0.60\textwidth]{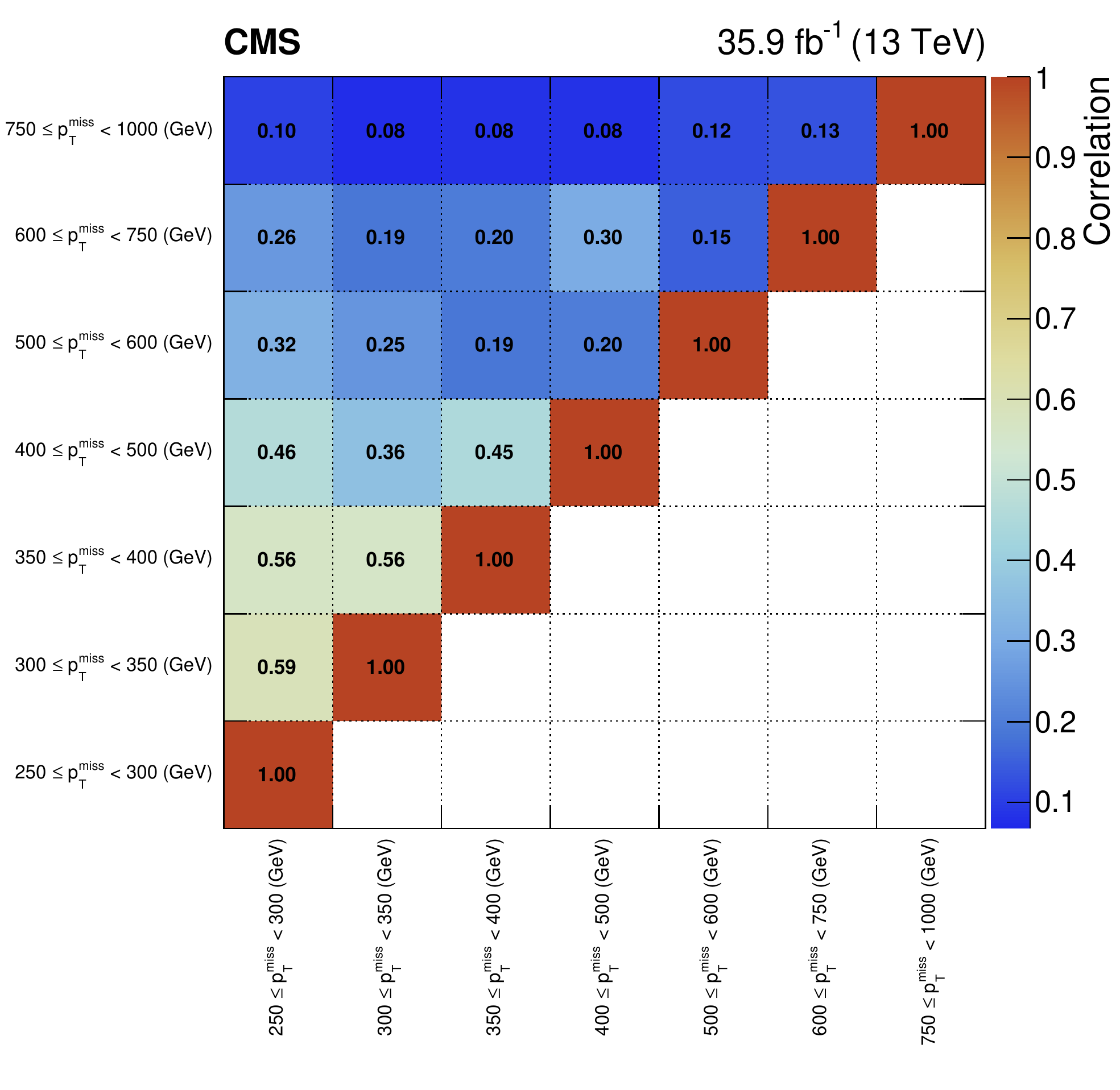}
\caption{Correlations between the predicted background yields in all the \ETm bins of the mono-$\PV$ signal region.
The boundaries of the \ETm bins, expressed in GeV, are shown at the bottom and on the left.}
\label{fig:correlation_matrix_monov}
\end{center}
\end{figure*}

The correlations between the predicted background yields across all the \ptmiss bins in the two signal
regions are shown in Figs.~\ref{fig:correlation_matrix_monojet} and~\ref{fig:correlation_matrix_monov}.
These results can be used with the simplified likelihood approach detailed
in Ref.~\cite{CMS-NOTE-2017-001} for reinterpretations in terms of models not studied in this paper.

\begin{figure*}[hbtp]
\begin{center}
\includegraphics[width=\cmsFigWidth]{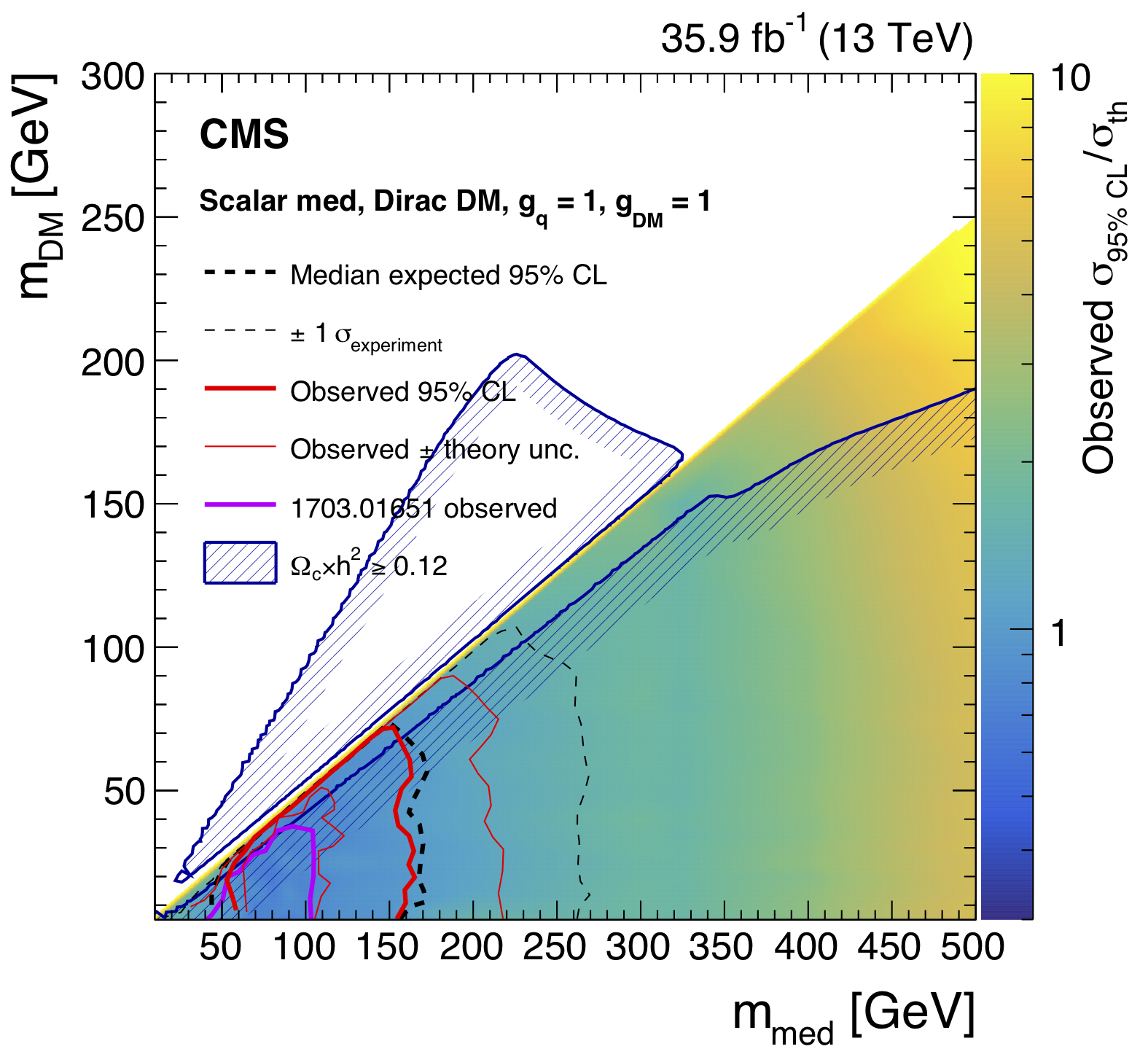}\hfil
\includegraphics[width=\cmsFigWidth]{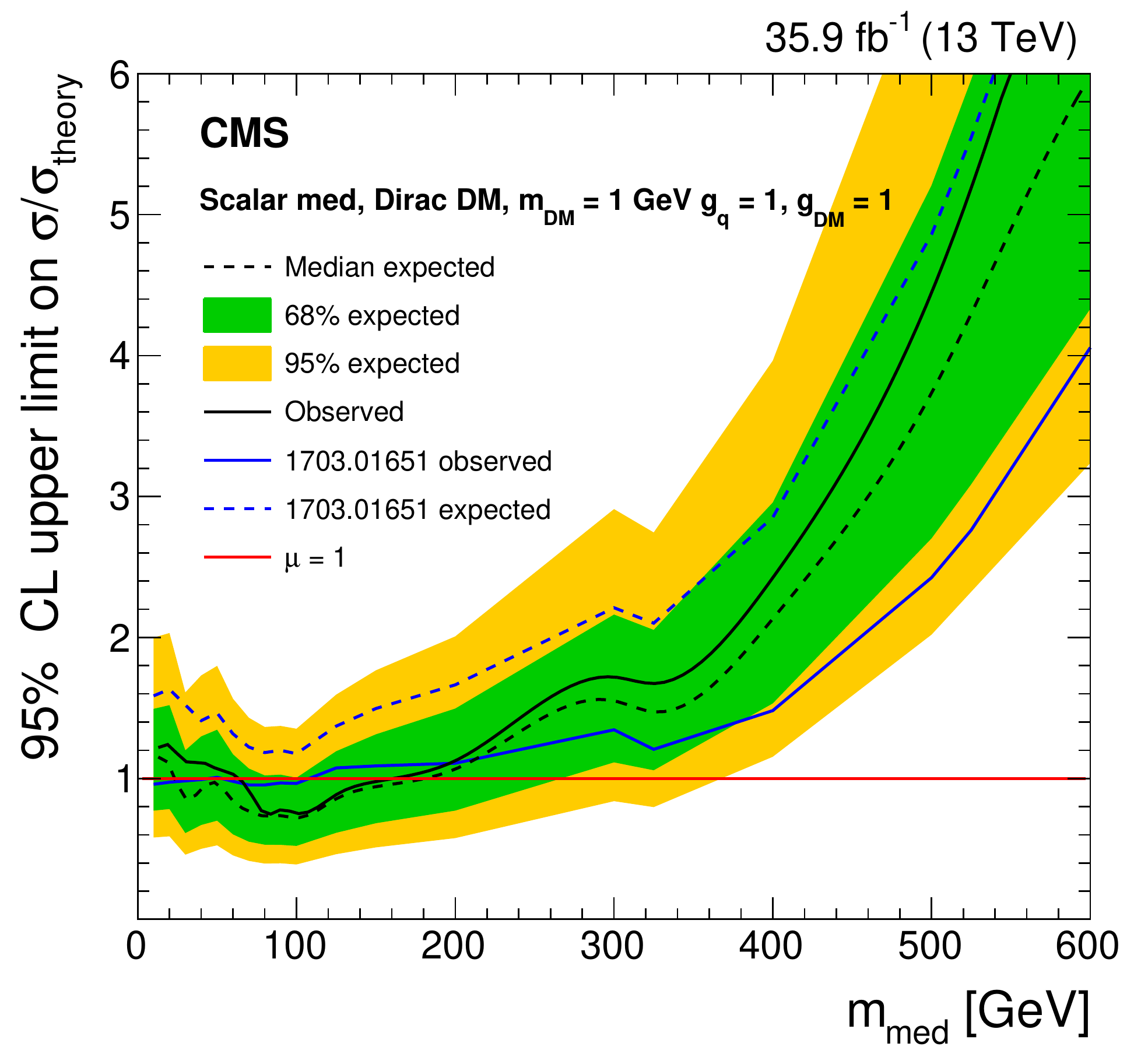}
\caption{
Exclusion limits at 95\%\,\CL on $\mu=\sigma/\sigma_{\textrm{th}}$ in the
$m_{\textrm{med}}$-$m_{\textrm{DM}}$ plane assuming scalar mediators (left) allowing for
vector boson couplings simulated at LO in QCD.
The solid (dotted) red (black) line shows the contour for the observed (expected) exclusion.
The solid contours around the observed limit and the dashed contours
around the expected limit represent one standard deviation due to theoretical uncertainties
in the signal cross section and the quadratic sum of the statistical
and experimental systematic uncertainties, respectively.
Expected and observed sensitivity of the previous CMS publication \cite{paper-exo-037}~are also presented.
Results of the Planck satellite
experiment ~\cite{Ade:2015xua} are shown as dark blue contours.
In the shaded area DM is overabundant.
Expected (dotted black line) and observed (solid black line)
95\%\,\CL upper limits on the signal strength $\mu$ as a function of the mediator mass for the spin-0 models (right).
}
\label{fig:scan_comparision}
\end{center}
\end{figure*}

\begin{figure*}[hbtp]
\begin{center}
\includegraphics[width=\cmsFigWidth]{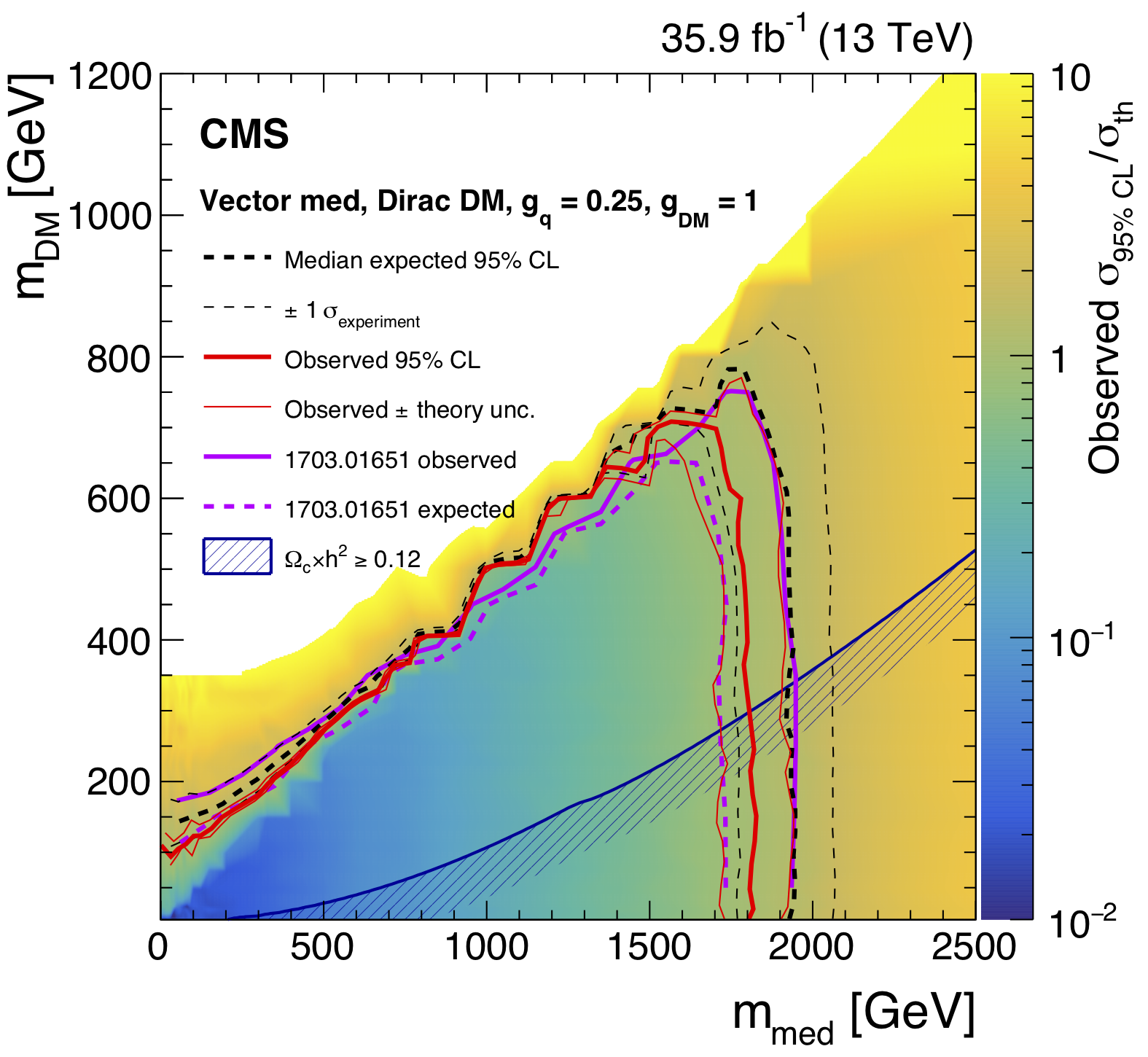}\hfil
\includegraphics[width=\cmsFigWidth]{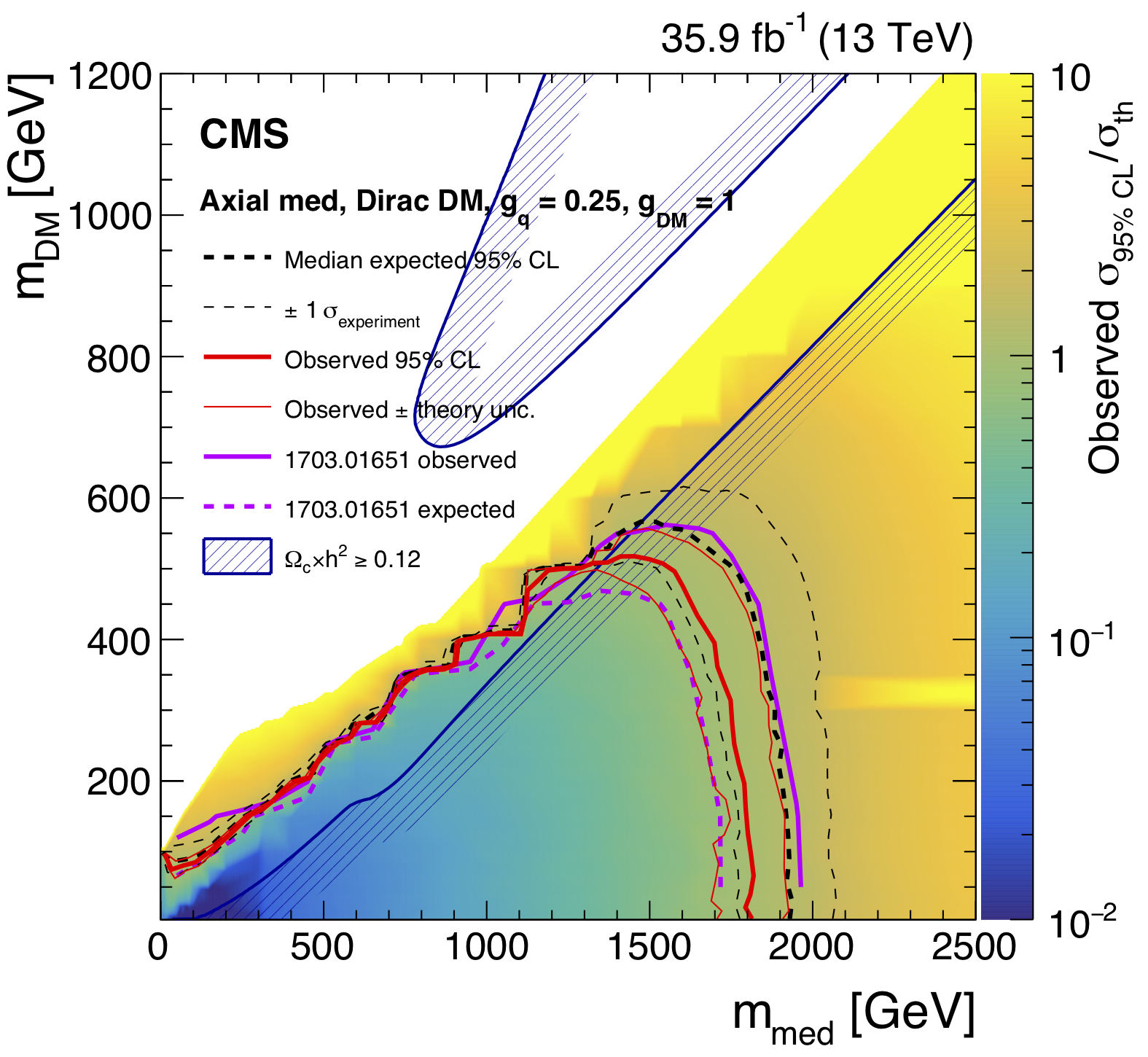}
\caption{
Exclusion limits at 95\%\,\CL on $\mu=\sigma/\sigma_{\textrm{th}}$ in the
$m_{\textrm{med}}$-$m_{\textrm{DM}}$ plane assuming vector (left) and axial-vector (right) mediators
where the the mono-$\PV$ signal is simulated at LO in QCD.
The solid (dotted) red (black) line shows the contour for the observed (expected) exclusion.
The solid contours around the observed limit and the dashed contours
around the expected limit represent one standard deviation due to theoretical uncertainties
in the signal cross section and the quadratic sum of the statistical
and experimental systematic uncertainties, respectively.         Planck satellite
experiment~\cite{Ade:2015xua} are shown as dark blue contours.
In the shaded area DM is overabundant.
}
\label{fig:scan_spin1_dmf}
\end{center}
\end{figure*}

To allow for a direct comparison with the results of Ref.~\cite{paper-exo-037}
for simplified DM models, the results are presented for scalar mediators allowing for vector boson couplings
simulated at LO in QCD, as shown in Fig.~\ref{fig:scan_comparision}.
Similarly, results for spin-1 mediators are also presented in Fig.~\ref{fig:scan_spin1_dmf}, where
the mono-$\PV$ signal is simulated at LO in QCD. The comparison of MC generators are also provided in Table~\ref{tab:MC_Generators}.

\begin{table*}[!htb]
\topcaption{Monte Carlo generators and perturbative order in QCD used for simulating various signal processes studied in this work, and in Ref.~\cite{paper-exo-037}}
\begin{center}
\renewcommand{\arraystretch}{1}
\ifthenelse{\boolean{cms@external}}{\footnotesize}{\resizebox{\textwidth}{!}}
{
\begin{scotch}{l l l }
Process                                         & Monte Carlo generator           & Monte Carlo generator   \\
& (Perturbative order in QCD)     & (Perturbative order in QCD) \\
& Ref.~\cite{paper-exo-037}       & this work \\
\hline
Monojet (spin-1 med.)                & \POWHEG 2.0 (NLO)            & \MADGRAPH{}5\_a\MCATNLO 2.2.3 (NLO)     \\
Monojet (spin-0 med.)                & \POWHEG 2.0 (LO)             & \MADGRAPH{}5\_a\MCATNLO 2.2.3 (NLO)      \\
mono-$\PV$ (spin-1 med.)                & \MADGRAPH{}5\_a\MCATNLO 2.2.3 (LO)   & \MADGRAPH{}5\_a\MCATNLO 2.2.3 (NLO)      \\
mono-$\PV$ (spin-0 med.)                & \textsc{JHUGenerator} 5.2.5          & Not used      \\
\end{scotch}
}
\end{center}
\label{tab:MC_Generators}
\end{table*}

\cleardoublepage \section{The CMS Collaboration \label{app:collab}}\begin{sloppypar}\hyphenpenalty=5000\widowpenalty=500\clubpenalty=5000\input{EXO-16-048-authorlist.tex}\end{sloppypar}
\end{document}

%% file: EXO-16-048-authorlist.tex
\textbf{Yerevan Physics Institute,  Yerevan,  Armenia}\\*[0pt]
A.M.~Sirunyan, A.~Tumasyan
\vskip\cmsinstskip
\textbf{Institut f\"{u}r Hochenergiephysik,  Wien,  Austria}\\*[0pt]
W.~Adam, F.~Ambrogi, E.~Asilar, T.~Bergauer, J.~Brandstetter, E.~Brondolin, M.~Dragicevic, J.~Er\"{o}, A.~Escalante Del Valle, M.~Flechl, M.~Friedl, R.~Fr\"{u}hwirth\cmsAuthorMark{1}, V.M.~Ghete, J.~Grossmann, J.~Hrubec, M.~Jeitler\cmsAuthorMark{1}, A.~K\"{o}nig, N.~Krammer, I.~Kr\"{a}tschmer, D.~Liko, T.~Madlener, I.~Mikulec, E.~Pree, N.~Rad, H.~Rohringer, J.~Schieck\cmsAuthorMark{1}, R.~Sch\"{o}fbeck, M.~Spanring, D.~Spitzbart, A.~Taurok, W.~Waltenberger, J.~Wittmann, C.-E.~Wulz\cmsAuthorMark{1}, M.~Zarucki
% ...
\vskip\cmsinstskip
% ...
1:~~Also at Vienna University of Technology, Vienna, Austria\\
% ...
